\documentclass[aps,amsmath,amssymb,twocolumn,floatfix,pra,footinbib,superscriptaddress,longbibliography]{revtex4-2}

\usepackage{graphicx}
\usepackage{epstopdf}

\usepackage[applemac]{inputenc}
\usepackage[T1]{fontenc}
\usepackage{lmodern}
\usepackage[english]{babel}
\usepackage{ae}
\usepackage{color}
\usepackage{url}

\usepackage[colorlinks]{hyperref}
\hypersetup{%
	plainpages=true,
	breaklinks=true,%not default in dvips mode, so we must specify
	hypertexnames=false,%not ideal, but needed when pagenums duplicate (`i' vs. `1')
	pageanchor=true,
	colorlinks=true,
	linkcolor={[RGB]{32,119,180}},
	citecolor={[RGB]{255,127,15}},
	urlcolor={[RGB]{32,119,180}},
	anchorcolor={black}
}

\usepackage{amsmath,amssymb,natbib,bm}
\usepackage{psfrag}
\usepackage{amssymb}
\usepackage{amsthm}
\usepackage{mathrsfs}
\usepackage{siunitx}
\usepackage{physics}
\usepackage{mathtools}
\usepackage[boxruled]{algorithm2e}

\usepackage{cancel}

\usepackage{mleftright} %to get rid of the extra space around parentheses with \left and \right: write \mleft, \mright instead

\usepackage[export]{adjustbox}

\newcommand{\rd}{\ensuremath{\mathrm{d}}}
\newcommand{\id}{\ensuremath{\,\rd}}

\newcommand{\abssq}[1]{\mleft| #1 \mright|^2}

\renewcommand*{\i}{\mathrm{i}}      % Imaginary unit
\newcommand*{\e}{\mathrm{e}}        % Euler's number

\newcommand{\figref}[1]{\mbox{Fig.~\ref{#1}}}

\newcommand{\secref}[1]{\mbox{Section~\ref{#1}}}

\newcommand{\appref}[1]{\mbox{Appendix~\ref{#1}}}
\renewcommand{\eqref}[1]{\mbox{Eq.~(\ref{#1})}}
\newcommand{\algref}[1]{\mbox{Algorithm~\ref{#1}}}

\newcommand{\figpanel}[2]{Fig.~\hyperref[#1]{\ref*{#1}(#2)}}
\newcommand{\figpanels}[3]{Fig.~\hyperref[#1]{\ref*{#1}(#2)--(#3)}}
\newcommand{\figpanelNoPrefix}[2]{\hyperref[#1]{\ref*{#1}(#2)}}
\begin{document}

\title{Comprehensive explanation of ZZ coupling in superconducting qubits}

\author{Simon Pettersson Fors}
\email{simon.pettersson.fors@chalmers.se}
\affiliation{Department of Microtechnology and Nanoscience, Chalmers University of Technology, 412 96 Gothenburg, Sweden}

\author{Jorge Fern\'andez-Pend\'as}
\affiliation{Department of Microtechnology and Nanoscience, Chalmers University of Technology, 412 96 Gothenburg, Sweden}

\author{Anton Frisk Kockum}
\email{anton.frisk.kockum@chalmers.se}
\affiliation{Department of Microtechnology and Nanoscience, Chalmers University of Technology, 412 96 Gothenburg, Sweden}

\date{\today}

\begin{abstract}

A major challenge for scaling up superconducting quantum computers is unwanted couplings between qubits, which lead to always-on ZZ couplings that impact gate fidelities by shifting energy levels conditional on qubit states. To tackle this challenge, we introduce analytical and numerical techniques, including a diagrammatic perturbation theory and a state-assignment algorithm. Together, these tools enable us to explain the emergence of ZZ coupling in three linked pictures, where each picture tells us more about the underlying mechanisms creating the ZZ coupling. These pictures generalize previous efforts, which focused on specific setups and a single mechanism. The deeper understanding that we provide of the mechanisms behind the ZZ coupling facilitate finding parameter regions of weak and strong ZZ coupling. We showcase our techniques for a system consisting of two fixed-frequency transmon qubits connected by a flux-tunable transmon coupler. There, we find three types of parameter regions with zero or near-zero ZZ coupling, all of which are accessible with current technology. We furthermore find regions of strong ZZ coupling nearby, which may be used to implement adiabatic controlled-phase gates and quantum simulations. Our framework is applicable to many types of qubits and opens up for the design of large-scale quantum computers with improved gate fidelities.

\end{abstract}

\maketitle

\tableofcontents

%%%%%%%%%%%%%%%%%%%%%%%%%%%%%%%%%%%%%%%%%%%%%%%

\section{Introduction}

Quantum computers are currently being scaled up to hundreds of qubits~\cite{Madsen2022, Kim2023, Bluvstein2024, Acharya2024} and beyond in the pursuit of machines that can unlock significant speedups over classical computers for important problems in many areas, e.g., quantum physics and chemistry, optimization, and machine learning~\cite{Georgescu2014, Montanaro2016, Wendin2017, Preskill2018, McArdle2020, Bauer2020, Cerezo2021, Cerezo2022, Dalzell2023}. However, to ensure that such scaled-up machines can achieve fault tolerance through quantum error correction~\cite{Terhal2015}, it is crucial to maintain or increase high fidelities for individual operations (gates, readout, reset, etc.) on the quantum computer while increasing its qubit count~\cite{Nielsen2000}. Those fidelities are limited by several problematic challenges, such as unwanted couplings of the qubits to surrounding environments causing decoherence~\cite{abadUniversalFidelityReduction2022, krantzQuantumEngineerGuide2019, Siddiqi2021}, unwanted couplings between the qubits themselves~\cite{Chow2011, McKay2019, berkeTransmonPlatformQuantum2022}, and control-signal crosstalk~\cite{Dai2021, Spring2022, Lawrie2023, Kosen2024}. All of these challenges may be exacerbated when scaling up, since the number of elements that can couple to each other then increases.

In this article, we focus on static couplings between qubits, the effect of which generally can be viewed as resulting in a ZZ coupling (cf.~\secref{sec:ZZ}). The ZZ coupling manifests as an energy shift for the qubits that depends on which states the qubits are in. As such, the ZZ coupling can be used to implement a controlled-phase (CPHASE) or controlled-Z (CZ) gate if it is strong~\cite{dicarloDemonstrationTwoqubitAlgorithms2009a, barendsSuperconductingQuantumCircuits2014, xuHighFidelityHighScalabilityTwoQubit2020, Collodo2020, stehlikTunableCouplingArchitecture2021a, sungRealizationHighFidelityCZ2021, longUniversalQuantumGate2021, Chu2021, negirneacHighFidelityControlledGate2021, Xiong2022, Huang2024}, but it constitutes an unwanted coherent error when operating other gates~\cite{Chow2011, McKay2019, Ganzhorn2020, caiImpactSpectatorsTwoQubit2021a, Chen2023} (and thus, by extension, quantum algorithms~\cite{Lacroix2020, karamlouAnalyzingPerformanceVariational2021}), and in those cases it is therefore desirable to minimize the ZZ coupling.

As coherence times for qubits have increased and ZZ coupling has become one of the most prominent errors to deal with in superconducting quantum computers, a host of architectures have been proposed and tried to suppress or cancel the ZZ coupling between two superconducting qubits. The setups considered seem to include virtually every imaginable combination of coupling elements and qubits: transmon qubits~\cite{kochChargeinsensitiveQubitDesign2007} coupled by a resonator~\cite{Kandala2021, zhaoMitigationQuantumCrosstalk2023a}, by a fixed-frequency transmon qubit~\cite{zhaoSuppressionStaticZZ2021}, by a tunable coupler~\cite{liTunableCouplerRealizing2020a, Collodo2020, stehlikTunableCouplingArchitecture2021a, Chu2021, xuParasiticFreeGateErrorProtected2023}, or by various driven coupling elements (a driven resonator~\cite{Huang2024}, a driven qubit~\cite{niScalableMethodEliminating2022, Shirai2023}, an unconventional parametric coupler~\cite{jinVersatileParametricCoupling2023}, a qubit coupled to a driven resonator~\cite{lerouxSuperconductingCouplerExponentially2021b}); off-resonant drives on both qubits~\cite{Mitchell2021, weiHamiltonianEngineeringMulticolor2022, Xiong2022}; two coupling elements (two transmons~\cite{gotoDoubleTransmonCouplerFast2022, kuboFastParametricTwoqubit2023a, campbellModularTunableCoupler2023, aokiControlZZCoupling2024}, a transmon and a resonator~\cite{mundadaSuppressionQubitCrosstalk2019}, two resonators~\cite{Wang2024}); floating qubits for a negative direct coupling~\cite{seteFloatingTunableCoupler2021a}; a floating coupler~\cite{Marxer2023}; qubits of opposite anharmonicity~\cite{zhaoHighContrastZZInteraction2020a, kuSuppressionUnwantedZZ2020a, xuZZFreedomTwoQubit2021, Ciani2022}; a coupler of different anharmonicity than the qubits~\cite{zhaoSwitchableNextNearestNeighborCoupling2020, heunischTunableCouplerFully2023a}, e.g., a transmon coupler between fluxonium qubits~\cite{Moskalenko2022, dingHighFidelityFrequencyFlexibleTwoQubit2023a}; direct~\cite{Ficheux2021} or tunable inductive~\cite{zhangTunableInductiveCoupler2024} coupling between fluxonium qubits; coupled two-mode qubits~\cite{finckSuppressedCrosstalkTwoJunction2021a}. There are also several approaches for limiting the impact of ZZ coupling, including pulse shaping~\cite{longUniversalQuantumGate2021, Xie2022, watanabeZZInteractionfreeSinglequbitgate2024a}, dynamical decoupling~\cite{Tripathi2022, Zhou2023}, and more~\cite{Liang2024}.

The wealth of specific proposed setups for controlling ZZ coupling shows that there is a jungle of parameters (types of qubits and couplers, their coupling topology, their frequencies, drive frequencies and amplitudes, coupling strengths between elements, etc.) to explore. To guide such exploration, and to evaluate whether existing proposals have exhausted the possible ways to design ZZ coupling, it would be valuable to have a general picture of the mechanisms that give rise to ZZ coupling. For specific setups, the explanations that have been put forward so far are mainly based on level repulsions~\cite{xuHighFidelityHighScalabilityTwoQubit2020, sungRealizationHighFidelityCZ2021, zhaoSuppressionStaticZZ2021, Chu2021, zhaoQuantumCrosstalkAnalysis2022a}, i.e., avoided level crossings pushing apart energy levels, some of which contribute to the ZZ coupling. This picture finds support in and is complemented by many analytical calculations of ZZ coupling performed for specific setups using perturbation theory~\cite{liTunableCouplerRealizing2020a, sungRealizationHighFidelityCZ2021, Chu2021, liNonperturbativeAnalyticalDiagonalization2022, dipaoloExtensibleCircuitQEDArchitecture2022, dingHighFidelityFrequencyFlexibleTwoQubit2023a, heunischTunableCouplerFully2023a} and other methods~\cite{solgunDirectCalculationZZ2022a}, as well as numerical calculations~\cite{stehlikTunableCouplingArchitecture2021a, dipaoloExtensibleCircuitQEDArchitecture2022, dingHighFidelityFrequencyFlexibleTwoQubit2023a, zhangTunableInductiveCoupler2024}. However, it remains unclear to what extent level repulsion is a sufficient explanation for ZZ coupling more generally.

To truly enable both harnessing ZZ coupling for high-fidelity gates and reducing its harmful influence on other operations, a more general and detailed treatment of the problem thus appears warranted. Such a treatment can provide a deeper understanding of the mechanisms that give rise to this coupling and a unified picture explaining results in specific setups. Here, we therefore present a hierarchy of analytical and numerical methods yielding increasingly detailed pictures of ZZ coupling. We showcase these methods for two fixed-frequency transmon qubits and a flux-tunable transmon coupler, which is the most common configuration in experiments and covers many of the specific setups previously considered. Our analytical methods include the introduction of a diagrammatic perturbation theory to clarify the mechanisms behind the ZZ coupling. To support our approximations in the perturbation theory, and the results emerging from it, our numerical modeling considers the Hamiltonian for the transmon qubits from a low level and leverages an improved algorithm to identify eigenstates in the system.

We find that the qubit frequencies, anharmonicities, and coupling strengths in our considered system can be chosen to create three types of parameter regions with zero or near-zero ZZ coupling that can be accessed by current technology without major redesigns. To the best of our knowledge, some of these parameter regions have not been pointed out in previous works. Through our diagrammatic perturbation theory we are able to explain the primary mechanisms (both level repulsions and some higher-order mechanisms) for the existence of all these regions, and we can use these explanations to understand how to engineer the ZZ coupling in experiments. Furthermore, we are able to combine the understanding of these mechanisms with combinatorial arguments for the configurations of energy levels to show that there are no other parameter regions than these that exhibit zero or near-zero ZZ coupling for the considered system.

Our results thus open up both for improving gate speeds for CPHASE and CZ gates, and for improving fidelities of other gates, which are negatively affected by ZZ coupling. Through the analytical and numerical methods we introduce, system parameters and architectures can be constrained to a more manageable search space. Indeed, our methods are not limited to the three-transmon setup we study here as a paradigmatic example; we expect them to find applications in investigations of larger systems (including ZZZ and higher-order couplings), in setups with other types of superconducting qubits (e.g., with other anharmonicities than transmon qubits), and possibly also in other quantum-computing systems where ZZ coupling constitutes a challenge, e.g., semiconductor qubits~\cite{Buterakos2018}.

This article is organized as follows. In \secref{sec:ZZ}, we provide further background and motivation for the importance of the ZZ coupling, showing how it emerges generally in systems of coupled qubits. By comparing the strength of the ZZ coupling with typical timescales for decoherence, we further estimate both how strong the ZZ coupling needs to be to implement a good CZ gate and how weak the ZZ coupling should be to not be the main limitation for high-fidelity operation of other gates. In \secref{sec:EffectiveHamiltonian}, we give the details of the system we study, where two fixed-frequency transmon qubits are connected through a flux-tunable coupler. We set up the circuit Hamiltonian that we later use for numerical computations and derive an effective Hamiltonian that enables our analytical calculations. 

From there, we proceed to present increasingly detailed pictures of the ZZ coupling. First, in \secref{sec:intuitive_picture}, we give an intuitive picture based on level repulsions and use it to predict where regions of zero or strong ZZ coupling can be expected in a parameter space defined by the qubit transition frequencies. Then, in \secref{sec:SW_diagrammatics}, we introduce a diagrammatical technique for the Schrieffer--Wolff transformation to enable reasoning about mechanisms for the ZZ coupling in a more detailed analytical picture. We apply this technique in \secref{sec:analytical_predictions} to expand our understanding of the mechanisms beyond level repulsion giving rise to the ZZ coupling and thereby refine our predictions of parameter regions for zero and high ZZ coupling as well as our understanding of how to control these regions. In \secref{sec:numerical_predictions}, we complete the picture of the ZZ coupling with numerical computations (utilizing an algorithm for stable matching to label dressed eigenstates of the system optimally) that we compare to the analytical predictions from the preceding sections. We conclude in \secref{sec:Conclusions} with a summary and conclusions, as well as an outlook for future work and applications.

Some details of our calculations and results are relegated to appendices. In \appref{app:normal_ordering}, we provide more details about the normal-ordering of the transmon Hamiltonian used to derive an effective model for our system in \secref{sec:EffectiveHamiltonian} and \appref{app:SWT_transmon} similarly contains more information about the Schrieffer--Wolff transformation used to arrive at that effective model. In \appref{app:diagram_evaluations}, \appref{app:4th_order_non_EC_diagrams}, and \appref{app:5th_order_SWT}, we give further information on our diagrammatic approach to the Schrieffer--Wolff transformation from \secref{sec:SW_diagrammatics}, including evaluations of more diagrams. Appendix~\ref{app:ZZ_other_parameters} includes an extended discussion of the effect of changing the values of several system parameters. Finally, we provide the details of our numerical computations in \appref{app:simulation_details}.

%%%%%%%%%%%%%%%%%%%%%%%%%%%%%%%%%%%

\section{The importance of ZZ coupling}
\label{sec:ZZ}
Here, we show how the ZZ coupling emerges as a fundamental property of coupled qubits. We then estimate what levels of ZZ coupling reduce gate fidelities to the same extent as decoherence and what levels of ZZ coupling are needed to implement a fast CZ gate.

%%%%%%%%%%%%%%%%%%%%%%%%%%%%%%%%%

\subsection{Emergence of ZZ coupling in coupled qubits}
\label{sec:DefinitionZZ}

Consider a pair of two-level systems (qubits) coupled via some mediating interaction. Such two-level systems can be realized in many physical systems, e.g., superconducting circuits, trapped ions, and neutral atoms~\cite{Preskill2018}. The total Hamiltonian of the two qubits is $H = H_0 + V$, where $H_0$ is the bare Hamiltonian of the uncoupled qubits and $V$ is the interaction Hamiltonian. In the eigenbasis $\{ \ket{00}, \ket{01}, \ket{10}, \ket{11} \}$ of the bare Hamiltonian, the uncoupled Hamiltonian has a matrix representation ($\hbar = 1$; the following ordering according to eigenstates is used also in subsequent matrices)
\begin{equation}
    H_0 =
    \mqty(\dmat[0]{0 , \omega_2, \omega_1, \omega_1 + \omega_2})
    \;  \mqty{\scriptstyle \ket{00} \\ \scriptstyle \ket{01} \\ \scriptstyle \ket{10} \\ \scriptstyle \ket{11}} 
    \, ,
    \label{eq:TLSbare}
\end{equation}
where $\omega_1$ and $\omega_2$ are the respective excitation energies for the qubits and we have set the energy of the $\ket{00}$ state to 0. Importantly, the energy needed to excite the state $\ket{11}$ is equal to the energy of exciting the individual states $\ket{01}$ and $\ket{10}$: $\omega_1 + \omega_2 $. 

The above equality is usually broken by the coupling $V$. Diagonalizing the total Hamiltonian $H$ yields the dressed Hamiltonian
\begin{equation}
    H' = \mqty(\dmat[0]{0 , \omega'_2, \omega'_1, \omega'_1 + \omega'_2 + \zeta})
    ,
\label{eq:TLSdressed}
\end{equation}
where $ \omega'_1 = E'_{10} - E'_{00}$ and $ \omega'_2 = E'_{01} - E'_{00}$ are defined from the eigenenergies $E'_{00}$, $E'_{01}$, $E'_{10}$, and $E'_{11}$ of $H$. The energy of the state $\ket{11}$ is in many cases, contrary to \eqref{eq:TLSbare}, not equal to the individual energies of the eigenmodes: $ \omega'_1 + \omega'_2 + \zeta \neq \omega'_1 + \omega'_2$. The discrepancy
\begin{equation}
    \zeta = E'_{11} - E'_{10} - E'_{01} + E'_{00}.
    \label{eq:ZZ}
\end{equation}
is called the ZZ coupling, or cross-Kerr term, where the name originates from the fact that the discrepancy in \eqref{eq:TLSdressed} can be expressed with a term proportional to the Pauli-Z matrices $\sigma_z \otimes \sigma_z $. The ZZ coupling is caused by the coupling between the qubits, and normally has a significant additional contribution from higher-excited states. A consequence of the ZZ coupling is that the energy of one qubit is conditional on the state of the other qubit, and vice versa. In the case of a time-dependent Hamiltonian, the eigenenergies in \eqref{eq:ZZ} are the instantaneous eigenenergies giving a dynamic ZZ coupling $\zeta = \zeta(t)$.

\subsection{Estimation of impact on gate fidelities}
\label{sec:estimate_ZZ}
To see the importance of ZZ coupling for achieving high-fidelity quantum gates, we consider a system implementing an iSWAP gate while subject to a ZZ coupling. The average gate fidelity can be computed as~\cite{nielsenSimpleFormulaAverage2002, willschSupercomputerSimulationsTransmon2020}
\begin{align}
    F &= \int \id \ket{\psi} \bra{\psi}U^\dag M \ket{\psi} \bra{\psi}M^\dag U \ket{\psi} \\
    &= \frac{\abssq{\tr{M U^\dag}} + \tr{M^\dag M}}{d(d+1)}, 
    \label{eq:fidelity}
\end{align}
where $U$ is the ideal (iSWAP) gate that we aim to implement, $M$ is the actual implemented gate, and $d$ is the dimension of the computational subspace, i.e., $d = 2^2 = 4$ for the two-qubit system. 

The effect of an isolated ZZ coupling is a phase accumulation in the state $\ket{11}$. By going to a frame rotating with the dressed frequencies $\omega'_1$ and $\omega'_2$ and noting that phases generated from dressed frequencies during the gate operation can be undone with virtual Z gates~\cite{krantzQuantumEngineerGuide2019}, the unitary evolution generated by \eqref{eq:TLSdressed} is
\begin{equation}
    U_\zeta =
    \mqty(\dmat[0]{1 , 1, 1, \e^{- \i \phi_\zeta}})
    ,
    \label{eq:TLSunitary}
\end{equation}
where $\phi_\zeta(t_g) = \int_0^{t_g} \zeta(t')\dd{t}' \equiv \Bar{\zeta} t_g $, $t_g$ is the gate time, and $\Bar{\zeta}$ is the time-averaged ZZ coupling.

Using \eqref{eq:fidelity} for the iSWAP gate $U = U_\text{iSWAP}$, and approximating the ZZ-coupling-affected gate with $M \approx U_\zeta U_\text{iSWAP}$ yields the average gate fidelity
\begin{equation}
    F = 1 - \frac{3}{10} \left[1 - \cos{(\Bar{\zeta} t_g)}  \right] .
    \label{eq:ZZ_fidelity}
\end{equation}
Here, the approximation $M \approx U_\zeta U_\text{iSWAP}$ can be shown to be exact if the generators, i.e., Hamiltonians of $U_\zeta$ and $U_\text{iSWAP}$ commute. In the limit of weak average ZZ coupling and short gate times, the average gate fidelity becomes $F = 1 - (3/20) (\Bar{\zeta} t_g)^2 + \mathcal{O}{[(\Bar{\zeta} t_g)^4]}$. 

To understand to what extent the ZZ coupling needs to be mitigated in this example, it is illuminating to compare the result in \eqref{eq:ZZ_fidelity} to the reduction in average gate fidelity from incoherent errors. To first order in gate time and decoherence rates, the average gate fidelity with uncorrelated energy relaxation and pure dephasing on $N$ qubits is~\cite{abadUniversalFidelityReduction2022}
\begin{equation}
    F = 1 - \frac{d}{2(d+1)} t_g \sum_{k=1}^N \mleft( \Gamma_1^{(k)} + \Gamma_\phi^{(k)} \mright),
    \label{eq:decoherence_fidelity}
\end{equation}
where $\Gamma_1^{(k)} = 1 / T_1^{(k)}$ ($\Gamma_\phi^{(k)} = 1 / T_\phi^{(k)}$) is the relaxation (pure dephasing) rate of qubit $k$ and $T_1^{(k)}$ ($T_\phi^{(k)}$) is the relaxation (pure dephasing) time. 

Comparing to relaxation (ignoring for the moment pure dephasing; $\Gamma_\phi^{(k)} = 0$), we find by equating \eqref{eq:ZZ_fidelity} and \eqref{eq:decoherence_fidelity} that the average ZZ coupling which to leading order reduces the average gate fidelity by the same amount is
\begin{equation}
    \Bar{\zeta} = \sqrt{\frac{16}{3 t_g T_1}} .
    \label{eq:ZZ_equal_T1}
\end{equation}
For current superconducting-circuit technology with realistic gate times and relaxation times on the order of $t_g = \SI{100}{\nano \second}$ and $T_1 = \SI{100}{\micro \second}$~\cite{Burnett2019, Kjaergaard2020}, \eqref{eq:ZZ_equal_T1} tells us that we need an average ZZ coupling below $\Bar{\zeta} = 2\pi \times \SI{100}{\kilo \hertz}$ for relaxation to be the dominant error source. Note that a similar estimate holds for pure dephasing since \eqref{eq:decoherence_fidelity} is linear in $\Gamma_1^{(k)}$ and $\Gamma_\phi^{(k)}$.

\subsection{CZ gate based on ZZ coupling}
\label{sec:CZ}

The ZZ coupling generates a phase for the state $\ket{11}$, as shown in \eqref{eq:TLSunitary}. For the iSWAP gate, this additional phase is an error that needs to be mitigated. However, the generated phase on its own can be used to create a CPHASE gate instead of an iSWAP gate, if the ZZ coupling can be turned on and off in a well-controlled manner. For a CZ gate, a CPHASE gate with phase $\pi$, the gate time of such a gate is given by $t_g = \pi / \Bar{\zeta}$. To implement, e.g., a $\SI{100}{\nano \second}$ CZ gate thus requires an average ZZ coupling of $2 \pi \times \SI{5}{\mega \hertz}$. Combining this result with \eqref{eq:ZZ_equal_T1} gives that we need to be able to tune the ZZ coupling between at least $2 \pi \times \SI{100}{\kilo \hertz}$ and $2 \pi \times \SI{5}{\mega \hertz}$ to implement a $\SI{100}{\nano \second}$ CZ gate with coherence-limited fidelity. This type of gate, typically referred to as an adiabatic CZ gate \cite{martinisFastAdiabaticQubit2014}, has been experimentally implemented in Refs.~\cite{dicarloDemonstrationTwoqubitAlgorithms2009a, barendsSuperconductingQuantumCircuits2014, xuHighFidelityHighScalabilityTwoQubit2020, Collodo2020, stehlikTunableCouplingArchitecture2021a, sungRealizationHighFidelityCZ2021, longUniversalQuantumGate2021, Chu2021, negirneacHighFidelityControlledGate2021, Xiong2022, Huang2024}.

\section{Effective Hamiltonian model for transmon architectures}
\label{sec:EffectiveHamiltonian}

The starting point for analyzing the ZZ coupling is a system Hamiltonian, as shown in the simple example in \secref{sec:DefinitionZZ}. The Hamiltonian is used to compute the eigenenergies defining the ZZ coupling. To concretize our analysis of the ZZ coupling in this paper, we apply it to a three-transmon system of two fixed-frequency transmon qubits and a flux-tunable transmon coupler, which is the most common architecture in experiments. In this section, we derive an effective Hamiltonian model for the low-energy subspace of the three-transmon system. In particular, we consider the low-energy properties of the capacitive couplings in the circuit Hamiltonian for this setup. Our derivation yields a model that enables a consistent analytical computation, up to a chosen approximation precision, of the primary mechanisms that generate the ZZ coupling. These primary mechanisms are the main instrument used to predict the ZZ coupling in \secref{sec:analytical_predictions}.

\subsection{Circuit Hamiltonian}
\label{sec:circuit_hamiltonian}

\begin{figure}[hbtp]
    \centering
    \includegraphics[width=\linewidth]{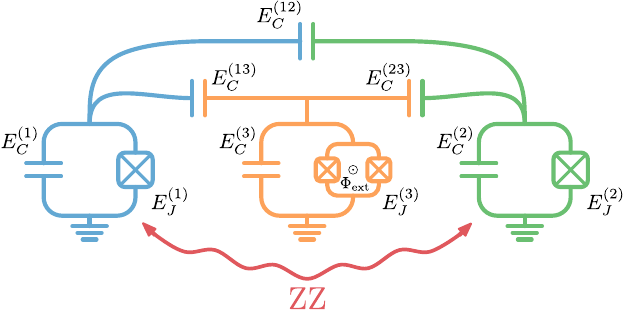}
    \caption{Circuit diagram for two fixed-frequency qubits (blue and green) coupled through both a direct capacitive coupling and a flux-tunable coupler (orange). The qubits and the coupler are implemented with transmon circuits, where the nonlinear inductances are the Josephson junctions (crossed boxes). Two Josephson junctions in a loop create a superconducting quantum interference device (SQUID) with an effective inductance tunable by the external magnetic flux through the loop. The direct capacitive coupling and the coupler together generate an effective ZZ coupling (red arrows) between the qubits.}
    \label{fig:three_qubit_circuit}
\end{figure}

Figure~\ref{fig:three_qubit_circuit} shows the circuit diagram for the three-transmon setup, which consists of two qubit circuits and one coupler circuit connecting them. The two qubits are fixed-frequency transmons. In addition to being coupled through the flux-tunable coupler, here a frequency-tunable transmon, they are also coupled through a direct capacitance. The three-transmon system Hamiltonian is the sum of a bare Hamiltonian $H_0$ and capacitive couplings $V$:
\begin{align}
    H &= H_0 + V, \label{eq:circuit_Hamiltonian} \\
    H_0 &= \sum_{i=1}^3 \underbrace{ \mleft[ 4 E^{(i)}_C \mleft(\Hat{n}_i - n_g^{(i)} \mright)^2 - E^{(i)}_J \cos{\Hat{\phi}_i} \mright] }_{\textstyle H_i} ,  \label{eq:transmon_Hamiltonian} \\
    V &=  \sum_{i<j}^3 4E^{(ij)}_C \Hat{n}_i \Hat{n}_j, \label{eq:interaction_Hamiltonian}
\end{align}
where $\Hat{n}_i$ ($\Hat{\phi}_i$) is the charge (phase) operator of transmon $i$ with offset charge $n_g^{(i)}$, $E^{(i)}_C$ ($E^{(i)}_J$) is the charging (Josephson) energy of transmon $i$, $E^{(ij)}_C$ is the mutual charging energy between transmons $i$ and $j$, and $H_i$ is the bare Hamiltonian of transmon $i$ \cite{kochChargeinsensitiveQubitDesign2007}. We label the qubits with $i = 1, 2$ and the coupler with $i = 3$. Unlike the qubits, the coupler has a flux-tunable Josephson energy $E^{(3)}_J(\Phi_\text{ext})$, where $\Phi_\text{ext}$ is the external flux through the coupler's SQUID loop. We will for the remainder of this paper neglect the offset charges since their impact on the lowest eigenenergies is exponentially suppressed in the transmon regime $E_C^{(i)} / E_J^{(i)} \ll 1$ \cite{kochChargeinsensitiveQubitDesign2007}.

For each of the three bare transmon Hamiltonians in \eqref{eq:transmon_Hamiltonian}, we can solve the eigenproblem $H_i \ket*{\Psi^{(i)}_{m}} = E^{(i)} \ket*{\Psi^{(i)}_{m}}$ exactly. The eigenstates $\ket*{\Psi^{(i)}_{m}}$ in the phase representation are the Mathieu functions. These states form an orthogonal eigenbasis spanning a Hilbert space $\mathcal{H}$ of states that are $2 \pi$-periodic in the phase variable $\phi_i$~\cite{kochChargeinsensitiveQubitDesign2007}. The $2 \pi$-periodicity is a consequence of the phase operator acting on a compact domain $\phi_i \sim \phi_i + 2 \pi$ with periodic boundary conditions. The transmon eigenenergies $E^{(i)}_{m}(E^{(i)}_C, E^{(i)}_J)$ can be computed from Mathieu's characteristic values.

Although we can handle a bare transmon, it is challenging to obtain analytical results directly from the full circuit Hamiltonian in \eqref{eq:circuit_Hamiltonian}; in particular, to solve for the eigenenergies. The reason for this difficulty is that there does not exist closed-form expressions for the Mathieu functions, meaning that the bare-transmon solutions cannot easily be generalized to exact expressions for the coupled system.

Instead of the phase representation, charge states are an alternative basis for $\mathcal{H}$. The charge states are the eigenstates of the charge operator: $\Hat{n}_i \ket*{n^{(i)}} = n_i \ket*{n^{(i)}}$. These states are the eigenstates of the transmon Hamiltonian in the limit $E_J^{(i)} / E_C^{(i)} \to 0$. However, the transmons are operated in the transmon regime $E_C^{(i)} / E_J^{(i)} \ll 1$,  which causes a matrix representation of the Hamiltonian in terms of charge states to be non-perturbative when solving for the eigenenergies. The charge basis is thus not well-suited for an analytical approach, but apt for numerical methods.

\subsection{Anharmonic-oscillator approximation}
\label{sec:anharmonic_oscillator}
A common approach to circumvent the challenges with using the transmon eigenstates or the charge states for analytical calculations is to approximate the transmon Hamiltonians in \eqref{eq:transmon_Hamiltonian} with some variation of anharmonic oscillators. Here, we review this approach and consider its limitations, setting the stage for the approximate Hamiltonian we will derive in \secref{sec:EffectiveHamiltonianModel} below and use for calculating the ZZ coupling in \secref{sec:analytical_predictions}.

To reformulate the bare transmon Hamiltonian on a form more reminiscent of an anharmonic oscillator, we express the charge and phase operators in Hermitian adjoint operators $\Hat{a}_i$ and $\Hat{a}_i^\dagger$ similar to creation and annihilation operators:
\begin{align}
    \Hat{\phi}_i &= \sqrt{ \frac{\lambda_i}{2} } \mleft( \Hat{a}_i + \Hat{a}_i^\dagger \mright), \label{eq:a} \\
    \Hat{n}_i &= - \i \sqrt{ \frac{1}{2\lambda_i} } \mleft( \Hat{a}_i - \Hat{a}_i^\dagger \mright), \label{eq:a_dagger}
\end{align}
where $\lambda_i$ is a free parameter that we choose such that terms proportional to $\Hat{a}_i^2$ and $(\Hat{a}_i^\dagger)^2$ cancel in the Hamiltonian (see \appref{app:normal_ordering} for details).

With this particular choice of $\lambda_i$, we expand the cosine potential in \eqref{eq:transmon_Hamiltonian} in its power series and normal-order the expansion~\cite{petrescuAccurateMethodsAnalysis2023, leibNetworksNonlinearSuperconducting2012, mikhailovOrderingBosonOperator1983a}, giving
\begin{equation}
    H_i = \omega_0^{(i)} \Hat{a}_i^\dagger \Hat{a}_i + 2 \alpha_0^{(i)} \hspace{-2.46mm} \sum_{m,n \in M} \mleft( \frac{4\alpha_0^{(i)}}{\omega_0^{(i)}} \mright)^{\frac{m+n-4}{2}} \frac{ (\Hat{a}_i^\dagger)^m \Hat{a}_i^n}{m! \, n!},
    \label{eq:normal_Hamiltonian}
\end{equation}
where $\omega_0^{(i)}$ is the bare harmonic oscillator frequency, $\alpha_0^{(i)}$ is the bare anharmonicity, and $M = \{ m, n \in \mathbb{Z}^+ \mid m+n \geq 4, \, \text{and} \ m+n \ \text{is even} \}$. Here, $\mathbb{Z}^+$ denotes the natural numbers including zero. The new parametrization is given by
\begin{align}
    \frac{E_C^{(i)}}{E_J^{(i)}} &= \frac{1}{2}\mleft( \frac{\lambda_i}{2} \mright)^2 \e^{-\lambda_i/4} \label{eq:lambda}, \\
    \lambda_i &= - \frac{8 \alpha_0^{(i)}}{\omega_0^{(i)}}, \label{eq:lambda_alpha_omega}\\
    \alpha_0^{(i)} &= - E_C^{(i)} .
\end{align}
Equation~(\ref{eq:lambda}) is a transcendental equation with at most two solutions for $\lambda_i > 0$. We choose the smallest of the two solutions since the transmon Hamiltonian should resemble a weak anharmonic oscillator with $\abs{\alpha_0^{(i)} / \omega_0^{(i)}} \ll 1$. Hence, it holds in the transmon regime that $\lambda_i \ll 1$.

The form of \eqref{eq:normal_Hamiltonian} makes it tempting to promote $\Hat{a}_i^\dagger$ and $\Hat{a}_i$ to proper creation and annihilation operators acting on a Fock space. However, if we introduce harmonic-oscillator eigenstates $\ket*{m^{(i)}}$ with the properties
\begin{align}
    \Hat{a}_i^\dagger \ket*{m^{(i)}} &= \sqrt{m^{(i)} + 1} \ket*{m^{(i)} + 1}, \\ 
    \Hat{a}_i \ket*{m^{(i)}} &= \sqrt{m^{(i)}} \ket*{m^{(i)} - 1},
\end{align}
these states are not orthogonal for the inner product of $\mathcal{H}$, i.e., $\braket*{m^{(i)}}{m'^{(i)}} \neq \delta_{mm'}$. Defining $\braket*{m^{(i)}}{m'^{(i)}} \equiv \Hat{\delta}_{mm'}^{(i)}$, we can show that $\Hat{\delta}^{(i)}$ is a Hermitian matrix that satisfies $\Hat{\delta}_{mm'}^{(i)} = 1$ for $m = m'$, due to normalization, and
\begin{equation}
    \Hat{\delta}_{mm'}^{(i)} = \frac{\sqrt{\lambda_i / 2} }{m' - m} \mleft[\sqrt{m} \psi_{m-1}^{(i)} \psi_{m'}^{(i)} - \sqrt{m'} \psi_m^{(i)} \psi_{m'-1}^{(i)}  \mright]_{\phi=-\pi}^{\phi=+\pi} ,
    %\Hat{\delta}_{mm'}^{(i)} = \frac{\sqrt{\lambda_i / 2} }{m' - m} \left[\Hat{a}_i \psi_m^{(i)} \psi_{m'}^{(i)} - \psi_m^{(i)} \Hat{a}_i \psi_{m'}^{(i)}  \right]_{\phi=-\pi}^{\phi=+\pi} ,
    \label{eq:delta_hat}
\end{equation}
for $m \neq m'$. Here, $\psi_m^{(i)} (\phi) = \braket*{\phi}{m^{(i)}}$ is the normalized $m$th eigenstate of the harmonic oscillator in the phase representation, and the bracket is evaluated at the boundary $\phi_i = \pm \pi$. The nonorthogonality arises from the fact that the eigenstates of the harmonic oscillator violate the periodic boundary condition $\phi_i \sim \phi_i + 2 \pi$. The states are thus not elements of $\mathcal{H}$; nor are they a basis.

Even though the harmonic-oscillator states are not a basis, they can still be used as an approximate basis for the low-energy subspace of $\mathcal{H}$. To see this, first note from \eqref{eq:delta_hat} that the low-energy states with energies below $2 E_J^{(i)}$, i.e., inside of the cosine potential, are approximately orthogonal $\Hat{\delta}_{mm'}^{(i)} \approx \delta_{m m'}$. This approximate orthogonality follows from the fact that the low-energy states have near-zero amplitude at $\phi_i = \pm \pi$. Furthermore, consider for a set $\omega_0^{(i)}$ the limit $E_C^{(i)} / E_J^{(i)} \to 0$ deep in the transmon regime $E_C^{(i)} / E_J^{(i)} \ll 1$ (the reciprocal of the limit where the charge states are eigenstates). For this limit, the height of the cosine potential becomes infinite ($2 E_J^{(i)} \to \infty$), and $H_i \to \omega_0^{(i)} \Hat{a}_i^\dagger \Hat{a}_i$ and $\Hat{\delta}_{mm'}^{(i)} \to \delta_{m m'}$, using Eqs.~(\ref{eq:normal_Hamiltonian}) and (\ref{eq:delta_hat}). Thus, the transmon eigenstates approach the harmonic-oscillator states such that the approximate basis is exact for $E_C^{(i)} / E_J^{(i)} \to 0$. We therefore assume in the transmon regime when $\Hat{\delta}_{mm'}^{(i)} \approx \delta_{m m'}$ that we can use the harmonic-oscillator states as an approximate basis for the low-energy subspace of $\mathcal{H}$.

We note that (a) elevating the harmonic-oscillator states to a basis on the compact phase domain and (b) setting $\Hat{\delta}_{mm'}^{(i)} = \delta_{m m'}$ is equivalent to neglecting the periodic boundary condition and extending the domain to the real line $\phi_i \in \mathbb{R}$. The real line is the proper domain for an anharmonic oscillator, and we hence refer to approximations (a) and (b) as the anharmonic-oscillator approximation. 

\subsection{Effective Hamiltonian model}
\label{sec:EffectiveHamiltonianModel}

We can directly apply the anharmonic-oscillator approximation to \eqref{eq:normal_Hamiltonian} and decide to use the result as an effective Hamiltonian model for the whole coupled three-transmon system. However, this violates the periodic boundary condition and creates a complicated Hamiltonian with a large number of terms. Instead, we mitigate these two problems by only applying the anharmonic-oscillator approximation to dress the capacitive couplings with a Schrieffer--Wolff transformation \cite{schriefferRelationAndersonKondo1966, bravyiSchriefferWolffTransformation2011} (see also \secref{sec:SW_review}). 

To handle the capacitive couplings in this way, we first express the circuit Hamiltonian in the bare eigenbasis $\{\ket*{\Psi^{(i)}_{m}} \}_{m=0}^\infty$ of each transmon $i$:
\begin{align}
    H_0 &\equiv \sum_{i=1}^3 \sum_{m=0}^\infty E^{(i)}_{m} \dyad*{\Psi^{(i)}_{m}} \label{eq:Mathieu_Hamiltonian_H0}, \\
    V &\equiv  \sum_{i<j}^3 4E^{(ij)}_C N_i \otimes N_j, \label{eq:Mathieu_Hamiltonian_V}
\end{align}
where $N_i \equiv \sum_{m,m'=0}^\infty \mel*{\Psi^{(i)}_m}{\Hat{n}_i}{\Psi^{(i)}_{m'}} \dyad*{\Psi^{(i)}_{m}}{\Psi^{(i)}_{m'}}$ and $\mel*{\Psi^{(i)}_m}{\Hat{n}_i}{\Psi^{(i)}_{m'}}$ are the matrix elements of the charge operator for transmon $i$. Thus, expressing the circuit Hamiltonian in the bare eigenbasis is reduced to evaluating $\mel*{\Psi^{(i)}_m}{\Hat{n}_i}{\Psi^{(i)}_{m'}}$.

We use the anharmonic-oscillator approximation to compute these matrix elements perturbatively. The perturbative parameter is $\alpha_0^{(i)} / \omega_0^{(i)}$, which is small when $E_C^{(i)} / E_J^{(i)} \ll 1 $. We perform the perturbative expansion to first order in $\alpha_0^{(i)} / \omega_0^{(i)}$ to later be able to compute the ZZ coupling to a precision of $2\pi \times \SI{100}{\kilo \hertz}$ (see \secref{sec:truncation_scheme} for further explanation). If needed, the precision of the approximation can be improved by including higher-order terms. We approximate the eigenstates with
\begin{equation}
    \ket*{\Psi_m^{(i)}} \approx \e^{-S^{(i)}} \ket*{m^{(i)}},
\end{equation}
where $S^{(i)}$ is an anti-Hermitian generator for the Schrieffer--Wolff transformation. This approximation is only valid for the low-energy subspace with $\Hat{\delta}_{mm'}^{(i)} \approx \delta_{m m'}$. Using the Baker--Campbell--Hausdorff lemma \cite{sakuraiModernQuantumMechanics2020} then yields
\begin{multline}
    \mel*{\Psi^{(i)}_m}{\Hat{n}_i}{\Psi^{(i)}_{m'}} = \\ - \i \sqrt{\frac{1}{2 \lambda_i}} \sum_{p=0}^\infty \frac{1}{p!} \mel*{m^{(i)}}{ [S^{(i)} ,\Hat{a}_i - \Hat{a}_i^\dagger ]^{(p)} }{m'^{(i)}},
    \label{eq:matrix_element_BCH}
\end{multline}
where $[A, B]^{(p)} = [A,[A,...[A,B]...]]$ denotes $p$ nested commutators and $[A, B]^{(0)} \equiv 1$.

The generator $S^{(i)}$ is determined by applying a Schrieffer--Wolff transformation to \eqref{eq:normal_Hamiltonian}. Since we consider the first-order expansion in $\alpha_0^{(i)} / \omega_0^{(i)}$, it is sufficient to split the Hamiltonian into
\begin{align}
    H_\text{bare}^{(i)} &= \omega_0^{(i)} \Hat{a}_i^\dagger \Hat{a}_i, \label{eq:harmonic_Hamiltonian} \\
    H_\text{int}^{(i)} &= 2 \alpha_0^{(i)} \mleft(\frac{1}{4!} (\Hat{a}_i^\dagger)^4 + \frac{1}{3!} (\Hat{a}_i^\dagger)^3 \Hat{a} + \text{H.c.} \mright),
    \label{eq:harmonic_interaction}
\end{align}
and neglect remaining terms (see \appref{app:SWT_transmon} for details). Here, $\text{H.c.}$ denotes the Hermitian conjugate of the preceding terms. Thus, the generator is
\begin{equation}
    S^{(i)} = 2 \alpha_0^{(i)} \mleft( \frac{1}{4!} \frac{1}{4 \omega_0^{(i)}} (\Hat{a}_i^\dagger)^4 + \frac{1}{3!} \frac{1}{2 \omega_0^{(i)}} (\Hat{a}_i^\dagger)^3 \Hat{a}_i - \text{H.c.} \mright).
    \label{eq:SWT_transmon_generator}
\end{equation} 

Having the matrix elements in hand from Eqs.~(\ref{eq:matrix_element_BCH}) and (\ref{eq:SWT_transmon_generator}), we find an effective model for the full circuit Hamiltonian with the capacitive couplings approximated to first order in $\alpha_0^{(i)} / \omega_0^{(i)}$. Using $\Hat{\delta}_{mm'}^{(i)} \approx \delta_{m m'}$, we define creation- and annihilation-like operators
\begin{align}
    a_i^\dagger &\equiv \sum_{m,m'=0}^\infty \mel*{m^{(i)}}{\Hat{a}^\dagger_i}{m'^{(i)}} \dyad*{\Psi^{(i)}_{m}}{\Psi^{(i)}_{m'}}, \\
    a_i &\equiv \sum_{m,m'=0}^\infty \mel*{m^{(i)}}{\Hat{a}_i}{m'^{(i)}} \dyad*{\Psi^{(i)}_{m}}{\Psi^{(i)}_{m'}},
\end{align}
that act on the transmon eigenstates. The charge operators can then be written as
\begin{multline}
    N_i \equiv -\i \sqrt{\frac{1}{2\lambda_i}} \mleft[ \mleft(1 + \frac{\alpha_0^{(i)}}{2 \omega_0^{(i)}} a_i^\dagger a_i \mright)a_i \mright. \\ + \mleft. \frac{\alpha_0^{(i)}}{4 \omega_0^{(i)}} a_i^3 - \text{H.c.} \mright] + \mathcal{O}\mleft[\mleft( \frac{\alpha_0^{(i)}}{\omega_0^{(i)}} \mright)^2 \mright].
    \label{eq:dressed_charge_operator}
\end{multline}
Using these dressed charge operators gives the effective Hamiltonian
\begin{align}
    H_0 &\equiv \sum_{i=1}^3 \sum_{m=1}^{\infty} \frac{\Delta^{(i)}_m }{m!}(a^\dagger_i)^m a_i^m \label{eq:effective_H0}, \\
    V &\equiv - \sum_{i<j}^3 g_{ij} \mleft[ a^\dagger_i  \mleft(1 + \frac{\alpha_0^{(i)}}{2 \omega_0^{(i)}} a_i^\dagger a_i \mright) + \frac{\alpha_0^{(i)}}{4 \omega_0^{(i)}} (a^\dagger_i)^3 - \text{H.c.} \mright] \nonumber \\
    &\hphantom{\equiv \; - \sum_{i<j}^3 \;} \otimes \mleft[ a^\dagger_j  \mleft(1 + \frac{\alpha_0^{(j)}}{2 \omega_0^{(j)}} a_j^\dagger a_j \mright) + \frac{\alpha_0^{(j)}}{4 \omega_0^{(j)}} (a^\dagger_j)^3 - \text{H.c.} \mright] \nonumber \\
    &\hphantom{\equiv \; \sum_{i<j}^3 \;} + \mathcal{O}\mleft[\mleft( \frac{\alpha_0^{(i)}}{\omega_0^{(i)}} \mright)^2 \mright], \label{eq:effective_V}
\end{align}
where the energy difference $\Delta_m^{(i)}$ is solved recursively from $\sum_{m=0}^{n} \binom{n}{m} \Delta_{m}^{(i)} = E_{n}^{(i)}$ and the coupling strength is $g_{ij} = 2 E_C^{(ij)} / \sqrt{\lambda_i \lambda_j}$. We recall from \eqref{eq:lambda_alpha_omega} that $\lambda_i = -8\alpha_0^{(i)} / \omega_0^{(i)}$. In particular, we refer to $\omega_i \equiv \Delta^{(i)}_1$ as the transmon frequency and $\alpha_i \equiv \Delta^{(i)}_2$ as the transmon anharmonicity. Note that if the sum in \eqref{eq:effective_H0} is truncated after $m = 2$, and the corrections at first order in $\alpha_0^{(i)} / \omega_0^{(i)}$ in \eqref{eq:effective_V} are neglected, the effective Hamiltonian simplifies to a Kerr-oscillator model. In that simplified case, the eigenstates of $H_0$ are still the transmon eigenstates.

\subsection{Hamiltonian graph representation}
\label{sec:Hamiltonian_graph}
\begin{figure*}
    \centering
    \includegraphics[]{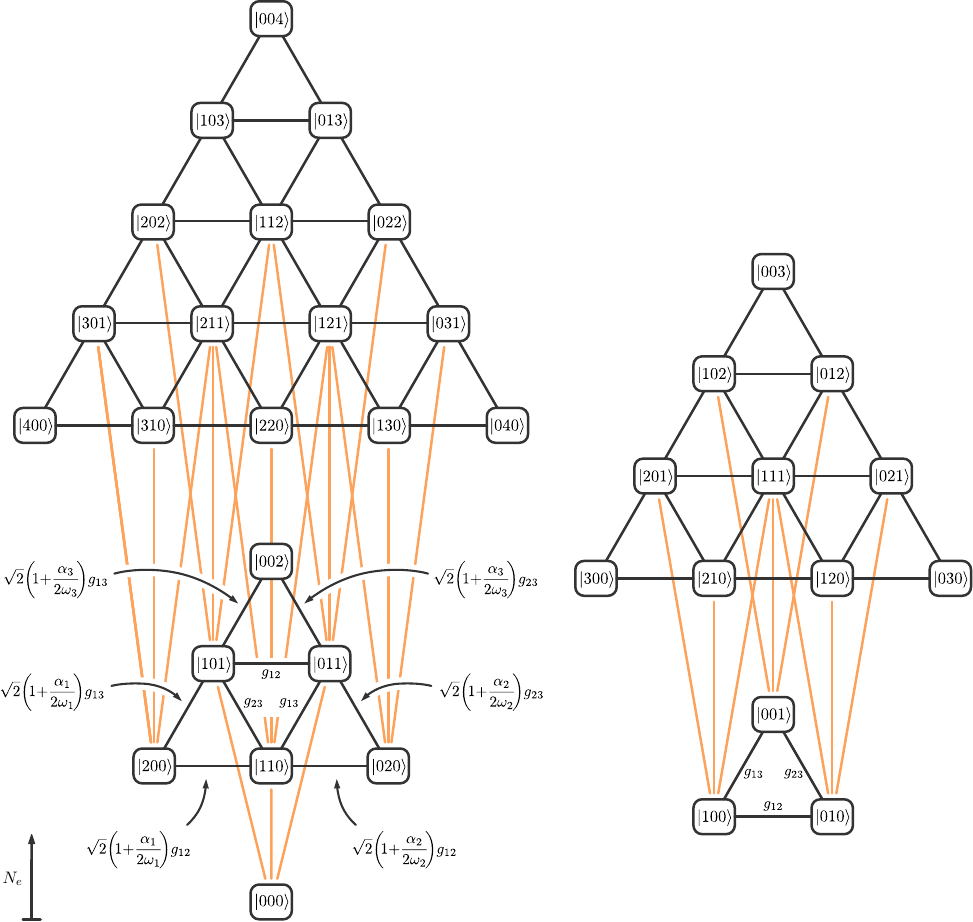}
    \caption{Hamiltonian graph representation of the effective Hamiltonian in Eqs.~(\ref{eq:effective_H0}) and (\ref{eq:effective_V}). The effective Hamiltonian conserves the parity of a state's total excitation number $N_e$. As a result, the graph is decoupled into two subgraphs with states of even (left) and odd (right) total number of excitations. Excitation-conserving edges (black solid lines) connect states into excitation subgraphs. Non-excitation-conserving edges (orange solid lines) connect states between excitation subgraphs. The edges are undirected since the Hamiltonian is Hermitian. The five excitation subgraphs with the lowest excitation number are shown. To simplify the representation, we have removed loop edges and non-excitation-conserving edges that have contributions less than the approximation precision defined in \secref{sec:analytical_predictions} below. We only include the edge weights for the excitation-conserving edges in the three lowest excitation subgraphs; see the main text in \secref{sec:Hamiltonian_graph} for further details.
    }
    \label{fig:Hamiltonian_graph}
\end{figure*}
The effective Hamiltonian in Eqs.~(\ref{eq:effective_H0}) and (\ref{eq:effective_V}) has notable properties that are important for predicting the ZZ coupling. We gain insights into these properties by finding a visual graph representation of the effective Hamiltonian. The graph representation efficiently shows how the capacitive couplings in $V$ couple the eigenstates of $H_0$ in a graph structure particular for the Hamiltonian.

We define the Hamiltonian graph in \figref{fig:Hamiltonian_graph} by letting the vertices of the graph represent the bare eigenstates $\ket{i}$ of $H_0$. The bare eigenstates are the transmon eigenstates from \secref{sec:circuit_hamiltonian} such that $\ket{i} \equiv \ket*{\Psi_{m_1}^{(1)}} \otimes \ket*{\Psi_{m_2}^{(2)}} \otimes \ket*{\Psi_{m_3}^{(3)}}$, where $i = (m_1, m_2, m_3)$ is a composite index. The weighted edges are the elements of the weighted adjacency matrix of the graph. We let the Hamiltonian matrix $H_{ij} \equiv \mel{i}{H}{j}$ be the adjacency matrix giving that the couplings in \eqref{eq:effective_V} between the bare eigenstates are represented by the weighted edges. There are two types of edges: the black solid edges conserve the total excitation number $N_e \equiv m_1 + m_2 + m_3$ between the two states they connect, while the orange solid edges connect states with different excitation numbers. Hence, we refer to the black and orange solid edges as excitation-conserving and non-excitation-conserving edges, respectively. We note that the edges are undirected since the Hamiltonian is Hermitian. Loops representing the self-couplings in \eqref{eq:effective_H0} are omitted in \figref{fig:Hamiltonian_graph} to simplify the representation; they are considered implicit in each vertex.

It is clear from \figref{fig:Hamiltonian_graph} that the effective Hamiltonian decouples the bare eigenstates into two subgraphs. The subgraphs differ with respect to the excitation-number parity of the states such that the left (right) subgraph includes only states with even (odd) $N_e$. This decoupling is a consequence of the fact that the effective Hamiltonian in Eqs.~(\ref{eq:effective_H0}) and (\ref{eq:effective_V}) conserves the parity of $N_e$. Intuitively, the conserved parity can be understood from the fact that the capacitive coupling in \eqref{eq:dressed_charge_operator} only changes the excitation number of a transmon by an odd amount.

We have simplified the graph in \figref{fig:Hamiltonian_graph} to focus on the details that are the most relevant for the ZZ-coupling predictions. First, only states with at most four total excitations are shown, i.e., $N_e \leq 4$. The states with the same number of excitations are typically referred to as excitation manifolds or subspaces. These manifolds form the black triangular blocks in \figref{fig:Hamiltonian_graph} that in turn split the parity subgraphs into distinct smaller subgraphs based on the number of total excitations. We refer to these smaller subgraphs as excitation subgraphs.

Second, the edge weights are not shown beyond the second-excitation subgraph for simplicity. If needed, these edge weights can easily be inferred from \eqref{eq:effective_V}.

Third, only a subset of the non-excitation-conserving edges are presented in \figref{fig:Hamiltonian_graph}. The omitted edges are all either edges created by terms proportional to $(a_i^\dagger)^3$ and $(a_i)^3$ in \eqref{eq:effective_V}, or edges generated by $a_i^\dagger a_j^\dagger$ and $a_i a_j$ that are connected to vertices beyond the nearest-neighbor vertices of the states defining the ZZ coupling in \eqref{eq:ZZ}. The omission of these edges is motivated by the fact that they generate corrections to the bare energies that are beyond the approximation precision considered in \secref{sec:analytical_predictions}.

\section{Intuitive picture for the ZZ coupling}
\label{sec:intuitive_picture}

Having considered different Hamiltonian models for the three-transmon circuit in \figref{fig:three_qubit_circuit}, culminating in the effective Hamiltonian in Eqs.~(\ref{eq:effective_H0}) and (\ref{eq:effective_V}), we are now ready to return to the ZZ coupling. In a similar spirit to the different Hamiltonian models, we predict the static ZZ coupling at different levels of completeness. We start in this section by predicting the ZZ coupling from an intuitive, but incomplete, picture. We incrementally complete this picture throughout the remainder of this paper. To allow the intuitive picture to be the central element in this section, we limit the mathematical computations to a minimum. Instead, we develop the intuitive picture from considering the effects of level repulsions, and in how many ways we can arrange the levels involved in these repulsions.

The level repulsions depend on the detuning between the energy levels. Starting in this section, and continuing in the remainder of this paper, we use $\Delta_{ij}$ to denote both the transmon-frequency detuning $\Delta_{ij} = \omega_i - \omega_j$ and the bare-energy detuning $\Delta_{ij} = E_{0, i} - E_{0, j}$, since $\hbar = 1$. Here, $E_{0, i}$ is the bare eigenenergy given by $H_0 \ket{i} = E_{0,i} \ket{i}$; recall that $i$ is a composite index. The two cases will be easily distinguishable by the context, or otherwise explicitly stated. 

The idea of using level repulsion to explain the static ZZ coupling has previously been used by Sung \emph{et al.} \cite{sungRealizationHighFidelityCZ2021}. We extend the level-repulsion picture in this section by adding the mentioned aspect of level arrangement, and by considering which of these arrangements that are likely to yield a zero or non-zero ZZ coupling.

\begin{figure*}
    \centering
    %\hspace*{-1.5cm}
    \includegraphics[width=\textwidth]{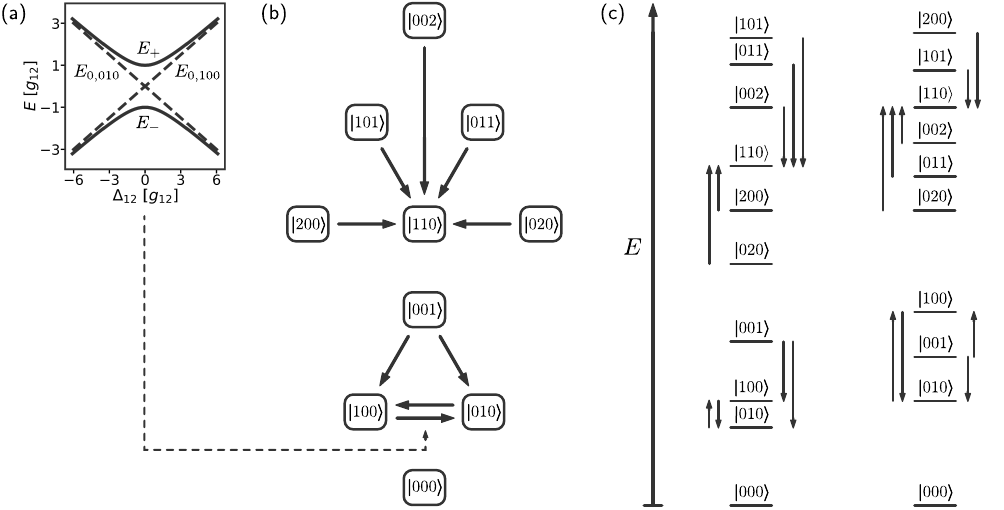}
    \caption{The nine main level repulsions on the energies defining the ZZ coupling. (a) The energy spectrum of the two-level system of $\ket{010}$ and $\ket{100}$ as a function of the bare detuning. Here, we neglect effects from other couplings outside of the two-level system and position the energy levels around $E = 0$. The energy spectrum shows the avoided level crossing of the eigenenergies $E_\pm$ (black solid lines) which deviate from the bare energies (black dashed lines) as a result of (fictive) level repulsions. Similar avoided level crossings are present for the other eight level repulsions that are represented with black arrows towards the states associated with the energies. (b) The main level repulsions following the edges of the Hamiltonian graph in \figref{fig:Hamiltonian_graph}. The black dashed line highlights the level repulsions shown in detail in (a). We note that the level repulsion from the energy level of the $\ket{002}$ is mediated by the edges through the states $\ket{101}$ and $\ket{011}$. (c) The level-diagram representation of the Hamiltonian graph with level repulsions in (b). We show two examples of level configurations that have balanced level repulsions. The balanced configurations predict likely parameter regions with zero ZZ coupling. Note that the level repulsions in the first-excitation subgraph have an opposite contribution to the ZZ coupling.}
    \label{fig:level_repulsion_diagram}
\end{figure*}

\subsection{The level-repulsion picture}
\label{sec:level_repulsion_picture}

Eigenenergies of coupled states generally exhibit the phenomenon of avoided level crossings \cite{krantzQuantumEngineerGuide2019}. Avoided level crossings refer to the observation that eigenvalues from coupled states do not cross one another subject to variations in system parameters, i.e., the eigenvalues are non-degenerate. These non-degeneracies in an energy spectrum appear as if there are repelling forces between the energy levels. Even though the repelling forces are fictive, they provide an intuitive understanding of how the shape of an energy spectrum depends on the couplings between bare states. We show a simple energy spectrum with an avoided level crossing in \figpanel{fig:level_repulsion_diagram}{a}.

From \figpanel{fig:level_repulsion_diagram}{a}, we note that the magnitude of a level repulsion between the energy levels assigned to two bare states $\ket{i}$ and $\ket{j}$ depends on two factors: the coupling strength $g_{ij}$ and the bare energy distance between the states, i.e., the bare detuning $\Delta_{ij} = E_{0, i} - E_{0, j}$. The repulsions grow with larger coupling strengths and smaller detunings. As such, the effects of level repulsions on an energy spectrum are the most significant close to resonances.

We consider which repulsions are expected to be the main contributors to the ZZ coupling. Recall the Hamiltonian graph in \figref{fig:Hamiltonian_graph}, which gives a direct overview of the possible couplings between states. These couplings, represented by the edges, show the possible level repulsions on the energies associated with each state. Since the detunings are larger between states in different excitation subgraphs than between states in the same excitation subgraph, we expect that the level repulsions from excitation-conserving edges dominate over the repulsions from non-excitation-conserving edges. Hence, we focus on the level repulsions from excitation-conserving edges to keep the intuitive picture as minimalistic as possible.

We show the level repulsions mediated by excitation-conserving edges on the energies defining the ZZ coupling [the energies of $\ket{000}$, $\ket{010}$, $\ket{100}$, and $\ket{110}$; cf.~\eqref{eq:ZZ}] in \figpanel{fig:level_repulsion_diagram}{b}. Even if $\ket{002}$ has no direct coupling to $\ket{110}$ in \figref{fig:Hamiltonian_graph}, we note that there is still a (higher-order) level repulsion mediated by, e.g., the couplings to $\ket{101}$ and $\ket{011}$. Thus, we identify nine main level repulsions that perturb the energies of the computational states.

For these nine level repulsions, it is difficult to draw any conclusions from the intuitive picture about how they affect one another. For instance, we might imagine different shielding and amplification effects. To refrain from these complications, we assume that the level repulsions do not interact and hence are additive in the intuitive picture. This additivity leads to the questions of how many ways there are to add up the level repulsions and how many of these that can yield a zero ZZ coupling. We answer these questions in the following subsections by considering how many configurations there are for the energy levels.

\subsection{Energy-level configurations}
\label{sec:configurations}

The direction of a level repulsion depends on the ordering of the levels involved. If the energy level $E_{0, i}$ is higher than the energy level $E_{0, j}$, the repulsion on $E_{0, i}$ is positive, i.e., directed upwards. This ordering corresponds to a positive detuning $\Delta_{ij}$ such that the direction of the repulsion is given by the sign of the detuning. Consequently, the number of different ways to add up the level repulsions is determined by the number of ways to order the energy levels. We refer to an arrangement of a set of energy levels as a configuration. We show two examples of configurations in \figpanel{fig:level_repulsion_diagram}{c}.

We consider the total number of configurations. To focus on distinct and experimentally relevant configurations for transmons, we make three assumptions: (1) $\omega_1 > \omega_2$, (2) the anharmonicities $\alpha_1 = \alpha_2 = \alpha_3 \equiv \alpha < 0$ of the transmons are equal and negative, and (3) the transmon frequencies $\omega_1 \sim \omega_2 \sim \omega_3 \gg \abs{\alpha}$ are comparable and much larger than the anharmonicities. We make assumption (1) without loss of generality since the configurations for $\omega_1 < \omega_2$ are symmetric under relabeling the qubits $1 \leftrightarrow 2$. 

Assumption (2) reduces the configuration space by omitting extra configurations arising due to small variations in the anharmonicites. These extra configurations can be neglected since the energies of the computational states do not include any anharmonicites, such that small variations in the anharmonicites do not change the directions of the particular level repulsions in \figpanel{fig:level_repulsion_diagram}{a}. 

Assumption (3) implies that the excitation subspaces are well separated in energy, i.e., states from different subspaces are far off resonance. This assumption is needed to be consistent with the assumption that the level repulsions from excitation-conserving couplings dominate. The assumption also implies that it is sufficient to only consider configurations of the three lowest excitation subspaces, since the direction of the level repulsions from the three- and higher-excitation subspaces are not expected to depend on the internal configurations in these higher subspaces.

Making the above assumptions, we find 24 possible configurations of the energy levels in the three lowest excitation manifolds. We find these configurations by first noting that there are two sets of configurations: 11 with $\abs{\Delta_{12}} < \abs{\alpha}$ and 13 with $\abs{\Delta_{12}} > \abs{\alpha}$, where $\Delta_{12} = \omega_1 - \omega_2 = \Delta_{100,010}$. These relations between $\abs{\Delta_{12}}$ and $\abs{\alpha}$ change the ordering of the states $\ket{200}$, $\ket{110}$, and $\ket{020}$; we display both arrangements in \figpanel{fig:level_repulsion_diagram}{c}. We note that only these two arrangements exist for the states $\ket{200}$, $\ket{110}$, and $\ket{020}$ since we assume negative anharmonicites. Each of the 24 configurations is given by assuming one of the relations between $\abs{\Delta_{12}}$ and $\abs{\alpha}$ and then varying the bare frequency of the coupler. One instructive way to visualize the transitions between all of the 24 configurations is to vary the bare frequency from low to high frequencies such that the frequency of the coupler goes from below to above the frequencies, i.e., transitioning through all the orderings of the first-excitation subspace. We stress that showing that there is only a finite small set of configurations is an important conclusion since it reduces the need of studying the ZZ coupling in an infinite and multidimensional parameter space to a finite set of parameter regions.

Furthermore, we note that there is still a finite number of configurations if we weaken assumption (2) by allowing positive anharmonicities. In this scenario, for example, in setups with fluxonium \cite{manucharyanFluxoniumSingleCooperPair2009} instead of transmons, there are additional configurations to the 24 ones focused on here. These additional configurations are rearrangements of the second-excitation subgraph.  However, there are no new level repulsions in the case of positive anharmonicities compared to the main nine ones visualized in \figpanel{fig:level_repulsion_diagram}{a}. As such, the intuitive picture is extendable to positive-anharmonicity qubits and couplers by simply adding the additional arrangements of the second-excitation subgraph. 

\subsection{Predictions from the intuitive picture}
\label{sec:intuitive_picture_predictions}

\begin{figure}
    \centering
    \includegraphics[width=\columnwidth]{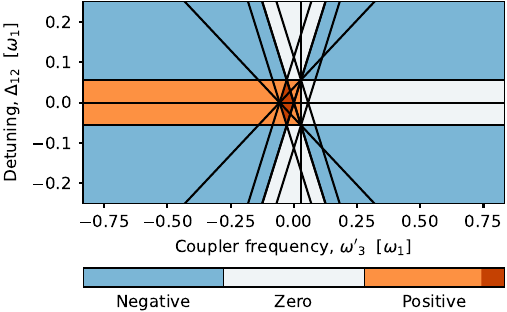}
    \caption{Predictions for the static ZZ coupling from the intuitive picture. The predictions are given in the parameter space of the shifted coupler frequency relative to the mean qubit frequency $\omega'_3 = \omega_3 - (\omega_1 + \omega_2)/2$ and the qubit detuning $\Delta_{12} = \omega_1 - \omega_2$. The parameter space is partitioned according to the 24 configurations, resulting in 24 bounded parameter regions for $\Delta_{12} > 0$. We note that the figure is symmetric under $\Delta_{12} \to -\Delta_{12}$ due to the freedom to label the qubits. The partition lines (solid black) are the borders between configurations; they represent the conditions for two or more energy levels permuting. We predict nine configurations for $\Delta_{12} > 0$ where it is likely to find a zero ZZ coupling. We note that we do not expect zero ZZ coupling everywhere in the white regions. The regions colored blue and orange are predicted to have negative and positive ZZ coupling, respectively. The dark orange region is predicted to yield the strongest ZZ coupling.}
    \label{fig:configurations}
\end{figure}

Here, we predict which of the 24 configurations above that are likely to give a zero ZZ coupling. We also predict the sign of the ZZ coupling in the different configurations and where the ZZ coupling is likely to be the strongest. The predictions are given in \figref{fig:configurations}, where the two-dimensional parameter space of qubit detuning and coupler frequency is partitioned according to the configurations. For $\Delta_{12} > 0$, we note the 24 parameter regions corresponding to the configurations discussed above. The parameter regions are bounded by the resonance conditions for states $\ket{i}$ and $\ket{j}$ ($\Delta_{ij} = 0$; black solid lines), across which the configurations rearrange. For each of these 24 configurations, the predictions are made from considering the balance between the level repulsions.

Intuitively, the ZZ coupling is predicted from the balance between the level repulsions due to the earlier assumption that the level repulsions are additive. As such, we require that the level repulsions counteract each other and result in a net-zero repulsion to obtain a zero ZZ coupling. Likewise, a non-zero ZZ coupling emerges when there is an imbalance between the level repulsions. As a consequence of the imbalance, the ZZ coupling increases in strength with the number of level repulsions that align. Note that the level repulsions in the first-excitation subgraph have an opposite contribution to the ZZ coupling due to the negative signs for the energies $E_{010}$ and $E_{100}$ in the definition in \eqref{eq:ZZ}.

We consider which of the 24 configurations that support balanced or imbalanced level repulsions. We find nine configurations with potential balance between the level repulsions; these configurations correspond to the white regions in \figref{fig:configurations} (for $\Delta_{12} > 0$). Two of the balanced configurations are shown in \figpanel{fig:level_repulsion_diagram}{c}. In all the nine balanced configurations, the balance is explained by first fixating the level repulsions from the energy levels of the qubit states $\ket{200}$ and $\ket{020}$ by the relation between $\abs{\Delta_{12}}$ and $\abs{\alpha}$, and then counteracting these repulsions with the level repulsions from states including at least one excitation in the coupler: $\ket{001}$, $\ket{011}$, $\ket{101}$, and $\ket{002}$. Breaking this balance results in the remaining 15 imbalanced configurations (for $\Delta_{12} > 0$) that we display as the 15 orange and blue regions in \figref{fig:configurations}. In these regions, having a majority of positive level repulsions accumulates to a positive ZZ coupling, and vice versa.

We predict that the most imbalanced configurations creates the strongest ZZ coupling. We construct the most imbalanced configurations by aligning as many level repulsions as possible. By choosing $\abs{\Delta_{12}} < \abs{\alpha}$ , we align the level repulsions from the energy levels of the qubit states $\ket{200}$ and $\ket{002}$. Then, we align the level repulsions from the energy levels of the qubit states with the level repulsions from the energy levels of the states $\ket{011}$, $\ket{101}$, and $\ket{002}$. In other words, we choose a coupler frequency less than the frequencies of the qubits: $\Delta_{13}, \Delta_{23} < 0$. We highlight the predicted parameter region with the strongest ZZ coupling in \figref{fig:configurations} with dark orange.

To conclude this section, we note the three main limiting factors of the intuitive picture and its resulting predictions. First, we assume in the intuitive picture that the level repulsions are non-interacting and hence additive. We do not expect that this holds in general when repulsions from multiple energy levels are closely involved. Second, the intuitive picture does not take into account that different level repulsions can have different magnitudes. To consider the magnitudes, we need to move beyond the qualitative description in the intuitive picture to a more quantitative method. Third, we paid little attention to the level repulsions caused by non-excitation-conserving couplings. Once again, we need quantitative methods to consider the magnitude of the non-excitation-conserving contributions. With these limiting factors in mind, we move on in the next section to introducing the method we use to analytically predict the static ZZ coupling. This results in a more complete description that addresses the limitations of the intuitive picture. 

\section{Introduction to Schrieffer--Wolff diagrammatics}
\label{sec:SW_diagrammatics}

Motivated by the limitations of the intuitive picture, we proceed to methods that quantitatively predict the static ZZ coupling. To obtain exact quantitative results, numerical methods are preferred, and we indeed give numerical predictions later in \secref{sec:numerical_predictions}. However, in stark contrast to the intuitive picture, the numerical methods provide little in way of explanations, but rather offer a direct route to the end result. This lack of explainability is problematic when we consider scaling up to larger systems, where naive numerical approaches fail because of computational load. For large systems, building intuition is vital to run efficient numerical simulations. 

Incentivized by the limitations of numerical methods, we introduce in this section a diagrammatical technique for the Schrieffer--Wolff transformation. The purpose of this technique is to make extensive analytical computations with the Schrieffer--Wolff transformation efficient and interpretable. The interpretability is achieved by capturing the intuitive picture in the technique. We develop the technique by combining the Schrieffer-Wolff transformation with the Hamiltonian graph representation in \figref{fig:Hamiltonian_graph}, which results in a small set of diagrammatic rules. In particular, we introduce the technique from the viewpoint of analytically computing the eigenenergies of the effective Hamiltonian in Eqs.~(\ref{eq:effective_H0}) and (\ref{eq:effective_V}) in order to extract the ZZ coupling.

\subsection{The Schrieffer--Wolff transformation}
\label{sec:SW_review}

For our ends, the Schrieffer--Wolff (from here on: SW) transformation \cite{schriefferRelationAndersonKondo1966, bravyiSchriefferWolffTransformation2011} is a perturbative method for solving the eigenproblem $H \ket{\Psi_i} = E_i \ket{\Psi_i}$, where $H = H_0 + V$. The solution is obtained by perturbatively constructing the transformation $U$, which transforms the bare eigenstates $\ket{i}$ of $H_0$ into the dressed eigenstates $\ket{\Psi_i}$ of $H$, in orders of the perturbation $V$. Recall that the bare eigenstates are the transmon eigenstates.

The SW transformation is written $U = \e^{-S}$, where $S$ is its generator. The generator is commonly expanded in a power series $S = \sum_n S_n$, where the partial generators are of order $\mathcal{O}(S_n) = \mathcal{O}(V^n)$ \cite{bravyiSchriefferWolffTransformation2011, colemanIntroductionManyBodyPhysics2015}. However, we do not follow this common approach, but instead partition the transformation as $U = \prod_n e^{-S_n}$. The rationale is that this partitioning reduces the number of commutators that need to be evaluated in the SW transformation. The partial generators are still of order $\mathcal{O}(S_n) = \mathcal{O}(V^n)$, and equivalent to the power-series generators subject to the Baker--Campbell--Hausdorff formula.

For the ZZ coupling in the three-transmon system, we estimate in \secref{sec:truncation_scheme} that the energy corrections of the bare energies need to be computed to fourth order in $V$ to achieve a sufficient truncation error. We therefore compute the second-order SW transformation, since it yields up to the fourth-order energy corrections. In comparison, the third-order transformation introduce corrections first at sixth order (see \appref{app:5th_order_SWT} for details). We thus truncate the partitioning after $e^{-S_2}$ such that the second-order SW transformation to the eigenstates is
\begin{equation}
    \ket{\Psi_i} = \e^{-S_1} \e^{-S_2} \ket{i} .
\end{equation}
Since the SW transformation is unitary, the partial generators are anti-Hermitian: $S_n^\dagger = - S_n$.

We construct the partial generators by diagonalizing $H$ in orders of $V$. The diagonalization proceeds by shifting the transformation to the Hamiltonian to obtain the equivalent eigenproblem 
\begin{equation}
    \e^{S_2} \e^{S_1} H \e^{-S_1} \e^{-S_2} \ket{i} = E_i \ket{i}.
    \label{eq:bare_eigenproblem}
\end{equation}
Perturbatively expanding the transformation with the Baker--Campbell--Hausdorff lemma \cite{sakuraiModernQuantumMechanics2020} and imposing that $S_1$ ($S_2)$ diagonalizes $H$ to first (second) order in $V$, the generators become
\begin{align}
    \mel{i}{S_1}{j} &= \frac{\mel{i}{V}{j}}{\Delta_{ij}}, \quad \text{for} \quad i \neq j, \label{eq:S1} \\
    \mel{i}{S_2}{j} &= \frac{\mel{i}{[S_1, V]}{j}}{2 \Delta_{ij}}, \quad \text{for} \quad i \neq j, \label{eq:S2}
\end{align}
where $\Delta_{ij} = E_{0,i} - E_{0,j}$ is the bare detuning and $E_{0,i}$ is the bare, i.e., zeroth-order, energy given by $H_0 \ket{i} = E_{0,i} \ket{i}$. It also holds that $\mel{i}{S_1}{i} = \mel{i}{S_2}{i} = 0$ for all $i$. We note that the generators diverge in the limit $\Delta_{ij} \to 0$ for which the perturbative expansion is invalid.

Using Eqs.~(\ref{eq:S1}) and (\ref{eq:S2}) for the generators in \eqref{eq:bare_eigenproblem} yields the second-, third-, and fourth-order corrections to the bare energies:
\begin{align}
    E_{2,i} &= \mel{i}{\frac{1}{2} [S_1,V]}{i}, \label{eq:E2} \\
    E_{3,i} &= \mel{i}{\frac{1}{3} [S_1,[S_1,V]]}{i}, \label{eq:E3} \\
    E_{4,i} &= \mel{i}{\mleft( \frac{1}{8} [S_1,[S_1,[S_1,V]]] + \frac{1}{4} [S_2,[S_1,V]] \mright)}{i}. \label{eq:E4}
\end{align}
The first-order correction is $E_{1,i} = 0$ by defining (without loss of generality) $V$ to be off-diagonal in the bare eigenbasis. This is automatically the case for the effective Hamiltonian in \eqref{eq:effective_V}. 

Computing the corrections to the bare energies in Eqs.~(\ref{eq:E2})--(\ref{eq:E4}) relies heavily on evaluating nested commutators of $V$, $S_1$, and $S_2$ in-between the inner product of $\ket{i}$. This is a convoluted task if we start from the matrix representations in Eqs.~(\ref{eq:S1}) and (\ref{eq:S2}), which result in long expressions that are difficult to interpret for the effective Hamiltonian in Eqs.~(\ref{eq:effective_H0}) and (\ref{eq:effective_V}). Instead, we combine the Schrieffer-Wolff transformation with the Hamiltonian graph representation in \secref{sec:Hamiltonian_graph} to more efficiently compute and interpret the energy corrections.

\subsection{Schrieffer--Wolff transformation on Hamiltonian graphs}
\begin{figure}
    \centering
    \includegraphics{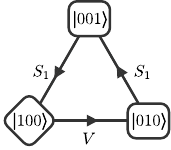}
    \caption{One of the closed paths contributing to the third-order energy correction in \eqref{eq:E3} for the initial state $\ket{100}$. The initial state is highlighted by making its vertex a diamond. The closed path is generated by the commutator $[S_1,[S_1,V]]$ acting on the initial state $\ket{100}$, and it visits all vertices in the first-excitation subgraph in \figref{fig:Hamiltonian_graph}. The sum of all possible paths similar to the one here forms the diagram in \eqref{eq:tri}.}
    \label{fig:closed_path}
\end{figure}

We interpret the energy corrections in Eqs.~(\ref{eq:E2})--(\ref{eq:E4}) on the Hamiltonian graph in \figref{fig:Hamiltonian_graph}. The key observation is that the energy corrections only have non-zero contributions from closed paths on the Hamiltonian graph. A closed path is a path that starts and ends in the same vertex, as exemplified in \figref{fig:closed_path}. To see why only closed paths contribute, we dissect the inner products in Eqs.~(\ref{eq:E2})--(\ref{eq:E4}). By expanding the nested commutators, we note that the operators $V$, $S_1$, and $S_2$ act on the state $\ket{i}$, i.e., the vertex. Acting with $V$ transitions the state along the edges of the vertex in \figref{fig:Hamiltonian_graph} to a superposition of its nearest-neighbor states. $S_1$ generates similar transitions along the edges, but weighted with the detuning, due to the construction of $S_1$ in \eqref{eq:S1}. Since $S_2$ is constructed from a product of $V$ and $S_1$ in \eqref{eq:S2}, $S_2$ transitions states along two connected edges in the Hamiltonian graph. For a sequence of these transitions to yield a contribution to the inner product, the transitions need to loop back to the initial state. This is equivalent to closed paths with initial state $\ket{i}$.

\subsection{Rules for Schrieffer--Wolff diagrammatics}
\label{sec:rules}

In order to compute the energy corrections in Eqs.~(\ref{eq:E2})--(\ref{eq:E4}), we here give diagrammatic rules for how to evaluate the nested commutators from closed paths in Hamiltonian graphs. Since Eqs.~(\ref{eq:E2})--(\ref{eq:E4}) are up to fourth order in $V$, there is only a small set of possible closed paths. The possible paths that give corrections of order $n$ follow from the fact that the correction order $n$ is the same as the number of transitions along a path. Note that $S_2$ gives two transitions along two connected edges.

We define diagrams that represent the sum of all possible closed paths with initial vertex $\ket{i}$. In a diagram, every edge must be transitioned at least once by the paths it represents. With this definition, all energy corrections up to fourth order are given by diagrams that have at most four edges. Below, we show all diagrams needed for calculations to this order.

Starting with the simplest case without edges, the trivial diagram consists of a single vertex:
\begin{equation}
    \includegraphics[valign=c]{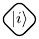} = \omega_i .
    \label{eq:x}
\end{equation}
It represents the bare energy of the initial state $\ket{i}$. We use diamond vertices to emphasize the initial state in \eqref{eq:x} and in all following diagrams.

By adding a single edge and a vertex $\ket{j}$ to \eqref{eq:x}, we create the first diagram that gives an energy correction to the bare energy of $\ket{i}$. Since each transition adds an additional order to the correction, only the diagram with a single edge supports second-order energy corrections. Evaluating \eqref{eq:E2} and the first term of \eqref{eq:E4} on the possible closed paths gives
\begin{equation}
    \includegraphics[valign=c]{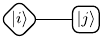} = \frac{g_{ij}^2}{\Delta_{ij}} \mleft[1 - \mleft( \frac{g_{ij}}{\Delta_{ij}} \mright)^2 \mright] ,
    \label{eq:x-}
\end{equation}
where $g_{ij}$ is the coupling strength and $\Delta_{ij}$ is the detuning between states $\ket{i}$ and $\ket{j}$. The same notation is used in the following diagrams.

It is illuminating to contemplate what kind of mechanism the diagram in \eqref{eq:x-} represents. For positive detuning (and sufficiently large, $\Delta_{ij} > \abs{g_{ij}}$, such that the SW transformation is valid), the diagram evaluates to a positive energy correction for the state $\ket{i}$. Similarly, by exchanging $i \leftrightarrow j$, we obtain the corresponding energy correction for $\ket{j}$. We note that this correction has the same magnitude as for $\ket{i}$, but is instead negative, meaning that the energy levels are equally repelled from each other. These are the properties for level repulsion from the intuitive picture in \secref{sec:intuitive_picture}. We therefore interpret the diagram in \eqref{eq:x-} as level repulsion. The second-order term $g_{ij}^2/\Delta_{ij}$ is the leading-order expression for the repulsion, and the fourth-order term $-g_{ij}^2/\Delta_{ij} \times (g_{ij}/\Delta_{ij})^2$ corrects the overestimation at leading order.

We create two additional diagrams by adding a vertex $\ket{k}$ and a single edge to \eqref{eq:x-}. The new vertex can either be connected to the $\ket{i}$ or $\ket{j}$ vertex, and the expressions follow from the first term in \eqref{eq:E4}:
\begin{align}
    \includegraphics[valign=c]{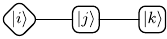} &= - \frac{1}{4} \frac{g_{ij}^2}{\Delta_{ij}} \mleft( \frac{g_{jk}}{\Delta_{jk}} \mright)^2 + \frac{3}{4} \frac{g_{jk}^2}{\Delta_{jk}} \mleft( \frac{g_{ij}}{\Delta_{ij}} \mright)^2 , \label{eq:x--} \\
    \includegraphics[valign=c]{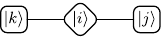} &= - \frac{g_{ij}^2}{\Delta_{ij}} \mleft( \frac{g_{ik}}{\Delta_{ik}} \mright)^2 - \frac{g_{ik}^2}{\Delta_{ik}} \mleft( \frac{g_{ij}}{\Delta_{ij}} \mright)^2 . \label{eq:-x-}
\end{align}
The terms in Eqs.~(\ref{eq:x--}) and (\ref{eq:-x-}) have the same form as the correction term of the level repulsion in \eqref{eq:x-}, but involve the added vertex. The added vertex thus creates an additional level repulsion that interacts with the original repulsion. Therefore, we interpret both diagrams as level-repulsion corrections caused by interacting repulsions from surrounding states. 

Building upon Eqs.~(\ref{eq:x--}) and (\ref{eq:-x-}), we create a new diagram by closing them into a loop with a third edge:
\begin{equation}
    \includegraphics[valign=c]{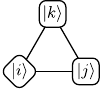} = \frac{2 g_{ij} g_{jk} g_{ik}}{\Delta_{ij} \Delta_{ik}} .
    \label{eq:tri}
\end{equation}
This is the only diagram that gives third-order energy corrections from \eqref{eq:E3}, which is a consequence of the fact that only a triangle closes a path with three edges. Note that the diagram is linear in all the coupling strengths such that we cannot factor out a leading-order repulsion term, e.g., $g_{ij}^2/\Delta_{ij}$, in \eqref{eq:tri}. This inhibits an interpretation of the diagram as a level repulsion. Therefore, we instead refer to the diagram as a three-loop mechanism. We note that the three-loop mechanism is not captured in the intuitive picture, where we only consider non-interacting level repulsions.

Another loop diagram is created by introducing a fourth vertex $\ket{l}$. Evaluating again the first term in \eqref{eq:E4} gives
\begin{equation}
    \begin{aligned}
        \includegraphics[valign=c]{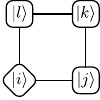} = &\frac{1}{4} \frac{ g_{ij} g_{jk} g_{kl} g_{il} }{\Delta_{jk} \Delta_{kl} \Delta_{il}} + \frac{1}{4} \frac{ g_{ij} g_{jk} g_{kl} g_{il} }{\Delta_{ij} \Delta_{jk} \Delta_{kl}} \\
        - &\frac{3}{4} \frac{ g_{ij} g_{jk} g_{kl} g_{il} }{\Delta_{ij} \Delta_{kl} \Delta_{il}} + \frac{3}{4} \frac{ g_{ij} g_{jk} g_{kl} g_{il} }{\Delta_{ij} \Delta_{jk} \Delta_{il}} .
    \end{aligned}
\label{eq:square}
\end{equation}
Similar to \eqref{eq:tri}, this diagram is linear in all the coupling strengths and cannot be interpreted as level repulsions. We therefore refer to the diagram in \eqref{eq:square} as a four-loop mechanism.

Up until this point, all diagrams have been generated by the leftmost commutators in Eqs.~(\ref{eq:E2})--(\ref{eq:E4}), where the only generator appearing is $S_1$. The last diagram involves the generator $S_2$, and it is created from the second term in \eqref{eq:E4}:
\begin{equation}
    \includegraphics[valign=c]{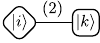} = \frac{g_{2,ik}^2}{\Delta_{ik}},
    \label{eq:x-2}
\end{equation}
where $g_{2, ik} = \sum_j \frac{1}{2} \mleft( \frac{g_{ij}}{\Delta_{ij}} g_{jk} - g_{ij} \frac{g_{jk}}{\Delta_{jk}} \mright)$, and the sum is evaluated over all intermediate states $\ket{j}$. Identically to \eqref{eq:x-}, we interpret this diagram as level repulsion, but at second order via the intermediate states $\ket{j}$. The seven kinds of diagrams enumerated above exhaust the list of possible diagrams up to fourth order.

\subsection{Diagram contractions}
\label{sec:diagram_contractions}

Having listed the possible diagrams up to fourth order, we consider if this set can be reduced. A potential reduction is valuable since it both shortens computations and simplifies their interpretations. In particular, the simplest possible interpretation is preferable. With this motivation in mind, we find two related cases where the fourth-order diagrams in Eqs.~(\ref{eq:x--}), (\ref{eq:square}), and (\ref{eq:x-2}) can be contracted to simpler forms.

We remark that the three-state diagram in \eqref{eq:x--} always is accompanied by the diagram in \eqref{eq:x-2} describing second-order level repulsion. Since their contributions are connected in this sense, we sum the two diagrams to find a simpler contracted form:
\begin{equation}
    \begin{aligned}
        \includegraphics[valign=c, trim=0 0 0 1.05mm]{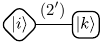} &\equiv \includegraphics[valign=c]{x--.pdf} + \includegraphics[valign=c, trim=0.3mm 0 0 0]{x-2.pdf} \\ &= \frac{1}{\Delta_{ik}} \mleft( \frac{g_{ij} g_{jk}}{\Delta_{ij}} \mright)^2 .
    \end{aligned}
    \label{eq:x-2_contract}
\end{equation}
Note that the sum for $g_{2, ik}$ in \eqref{eq:x-2} in this case only involves a single intermediate state. Similar to \eqref{eq:x-2}, we interpret this contracted diagram as second-order level repulsion, but now with the different coupling strength $g_{ij} g_{jk} / \Delta_{ij}$. We note that the new coupling strength is not invariant under exchange of $i \leftrightarrow k$ in contrast to \eqref{eq:x-2}, unless $\abs{\Delta_{ij}} = \abs{\Delta_{jk}}$. This variance implies that the two contracted level repulsions in-between the energy levels of $\ket{i}$ and $\ket{k}$ in general are different. As such, the contracted level repulsions do not fully adhere to the standard property of a level repulsion, where the coupling strength is symmetric between the two states. 

Using the contraction in \eqref{eq:x-2_contract}, we find that the four-loop diagram in \eqref{eq:square} can be contracted with two diagrams of the type in \eqref{eq:x--} and the diagram in \eqref{eq:x-2}:
\begin{equation}
    \begin{aligned}
    \includegraphics[valign=c]{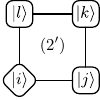} &\equiv \includegraphics[valign=c]{CD.pdf} + \includegraphics[valign=c]{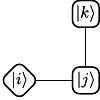} + \includegraphics[valign=c]{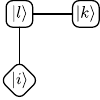} \\ &+ \includegraphics[valign=c]{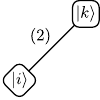} =  \frac{1}{\Delta_{ik}} \mleft( \frac{g_{ij} g_{jk}}{\Delta_{ij}} + \frac{g_{il} g_{kl}}{\Delta_{il}} \mright)^2 .
    \end{aligned}
    \label{eq:square_contract}
\end{equation}
This contraction is a consequence of the observation that \eqref{eq:square_contract} can be split into two of the contractions in \eqref{eq:x-2_contract} separately involving the states in the left branch $\ket{i}$--$\ket{j}$--$\ket{k}$ and the right branch $\ket{i}$--$\ket{l}$--$\ket{k}$ plus additional cross terms. These cross terms then complete the square. We interpret \eqref{eq:square_contract} as second-order level repulsion identically to \eqref{eq:x-2_contract} with coupling strength $g_{ij} g_{jk} / \Delta_{ij} + g_{il} g_{kl} / \Delta_{il}$. We note that this coupling strength is the sum of the contracted coupling strengths along the left and right branches following \eqref{eq:x-2_contract}. Again, this coupling strength is in general not symmetric between the energy levels of $\ket{i}$ and $\ket{k}$.

Importantly, we remark that the contraction in \eqref{eq:square_contract} removes the four-loop mechanism in \eqref{eq:square}. After the contractions, we are left with diagrams that are interpreted as mechanisms of level repulsion and the three-loop mechanism.  

With the diagrams and their contractions in hand, we can reduce the computation of the eigenenergies of the effective Hamiltonian in Eqs.~(\ref{eq:effective_H0}) and (\ref{eq:effective_V}) to expansions in diagrams that are immediate to evaluate with the rules derived in Eqs.~(\ref{eq:x})--(\ref{eq:square_contract}). To illustrate, the diagram expansion up to third order of the eigenenergy assigned to the state $\ket{100}$ is
\begin{equation}
    E_{100} = \includegraphics[valign=c]{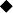}
    + \includegraphics[valign=c]{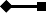} \,
    + \includegraphics[valign=c]{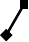}
    + \includegraphics[valign=c]{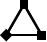} \, ,
    \label{eq:example_diagram_expansion}
\end{equation}
where we have only considered closed paths constrained to the first-excitation subgraph in \figref{fig:closed_path}. In \eqref{eq:example_diagram_expansion}, we have also simplified the diagram notation by dropping the vertex labels which instead are uniquely inferred from positioning the diagram in the Hamiltonian graph in \figref{fig:Hamiltonian_graph} given the initial state denoted with the diamond. We refer to the expansion and evaluation of the SW transformation in the diagrams and their rules in Eqs.~(\ref{eq:x})--(\ref{eq:square_contract}) as Schrieffer-Wolff diagrammatics. The mechanism interpretation of the diagrammatics enables an intuitive understanding of the eigenenergies and relates the SW transformation to the intuitive picture in \secref{sec:intuitive_picture}. We return in the next section to predicting the static ZZ coupling and to improve on the intuitive picture by applying SW diagrammatics.

\section{Analytical predictions for the ZZ coupling}
\label{sec:analytical_predictions}

In the intuitive picture in \secref{sec:intuitive_picture}, we used level repulsion to make basic predictions for the static ZZ coupling. These predictions are limited in the sense that they leave three questions unanswered: 
\begin{itemize}
    \item Are there other relevant mechanisms than level repulsion?
    \item How do the relevant mechanisms compare in strength?
    \item Can the non-excitation-conserving contributions be neglected?
\end{itemize}
To address the questions left open by the intuitive picture, we here analytically predict the static ZZ coupling using the Schrieffer-Wolff diagrammatics introduced in \secref{sec:SW_diagrammatics}. The analytical predictions add necessary details to the intuitive picture while still maintaining the interpretability. Through the analytical predictions, we identify the primary mechanisms causing the static ZZ coupling up to the limitations of perturbation theory. We investigate the strength of these mechanisms in the energy-level configurations from \secref{sec:configurations} for the three-transmon system to infer all possible parameter regions with zero or strong ZZ coupling that the primary mechanisms support. Using this mechanism picture, we also explain why the regions with zero or strong ZZ coupling exist and show that the found parameter regions with zero ZZ coupling belong to two different types.

The ZZ coupling has previously been computed with Rayleigh--Schr\"{o}dinger perturbation theory by Li \emph{et al.}~in Ref.~\cite{liTunableCouplerRealizing2020a}, Sung \emph{et al.}~in Ref.~\cite{sungRealizationHighFidelityCZ2021}, and Zhao \emph{et al.}~in Ref.~\cite{zhaoSuppressionStaticZZ2021}.
Although the Rayleigh-Schr\"{o}dinger approach is as viable, we find that the SW transformation has two advantages for the problem at hand: (1) it is easier to systematically extend beyond fourth-order perturbation theory, and (2) its components are more straightforward to interpret. For these reasons, we use the SW transformation to extend the perturbation theory of Li \emph{et al.},~Sung \emph{et al.},~and Zhao \emph{et al.}~beyond fourth order and beyond the weak-qubit-coupling-strength assumption: $\abs{g_{12}} \ll \abs{g_{13}}, \abs{g_{23}}$.

\subsection{Estimations of energy corrections and truncation scheme}
\label{sec:truncation_scheme}

We here estimate the eigenenergy corrections from the SW transformation in Eqs.~(\ref{eq:E2})--(\ref{eq:E4}). These estimates are our starting point to determine if we can neglect the non-excitation-conserving contributions. The estimates also have another key function: we need them to construct a consistent truncation scheme for the diagram expansions of the eigenenergies. We recall from \secref{sec:estimate_ZZ} that we want to predict the ZZ coupling to a precision of at least $2\pi \times \SI{100}{\kilo \hertz}$. As such, we consider the needed truncation scheme to achieve a truncation error at the scale of $2\pi \times \SI{100}{\kilo \hertz}$ in regions where the SW transformation is valid.

The diagrammatic rules in Eqs.~(\ref{eq:x})--(\ref{eq:x-2}) highlight that the SW transformation gives energy corrections that are multivariate polynomials in a small set of perturbative ratios. The relevant ratios are expressed in the system parameters of the effective Hamiltonian in Eqs.~(\ref{eq:effective_H0}) and (\ref{eq:effective_V}). Assuming that the transmon frequencies are comparable ($\omega_1 \sim \omega_2 \sim \omega_3$), the relevant ratios are: $g_{ij} / \Delta_{ij} $, $g_{ij} / \Sigma_{ij} $, and $\alpha_i / 2 \omega_i $, where $\Delta_{ij} = \omega_i - \omega_j$ and $\Sigma_{ij} \equiv \omega_i + \omega_j$. Here, ($g_{ij} / \Sigma_{ij} $) $g_{ij} / \Delta_{ij}$ originates from the (non-) excitation-conserving edges in \figref{fig:Hamiltonian_graph}. For the estimates, we assume currently conventional experimental parameters for transmons: $ \abs{g_{ij}} \sim \abs{\alpha_i} \sim 2\pi \times \SI{100}{\mega \hertz}$, $\abs{\Delta_{ij}} \sim 2\pi \times \SI{1}{\giga \hertz}$, and $\Sigma_{ij} \sim 2 \omega_i \sim 2\pi \times \SI{10}{\giga \hertz}$, which implies
\begin{equation}
    \abs{ \frac{ g_{ij} } { \Delta_{ij} } } \sim \frac{1}{10}, \quad \abs{\frac{\alpha_i}{2\omega_i}} \sim \abs{ \frac{ g_{ij} } { \Sigma_{ij} } } \sim \frac{1}{100} .
    \label{eq:perturbative_ratios}
\end{equation}

We use \eqref{eq:perturbative_ratios} to make a series of estimates in comparison to the sought precision of $2\pi \times \SI{100}{\kilo \hertz}$. First, the ratio $\abs{\alpha_i / 2 \omega_i}$ explains why we include the first-order corrections of the charge operators in the effective Hamiltonian. This first-order correction combined with a second-order excitation-conserving energy correction in \eqref{eq:x-} has a noticeable contribution: $\abs{\alpha_i / 2 \omega_i \times g_{ij}^2 / \Delta_{ij} } \sim 2\pi \times \SI{100}{\kilo \hertz}$. We neglect the second-order corrections for the charge operators for the same reason. Note that we refer to corrections from only excitation-conserving edges as excitation-conserving corrections, and else as non-excitation-conserving corrections.

Second, we estimate that the non-excitation-conserving contributions are non-negligible. For example, the second-order non-excitation-conserving energy corrections are $g_{ij}^2 / \Sigma_{ij} \sim 2\pi \times \SI{1}{\mega \hertz}$.

Third, we estimate which orders in the SW transformation that need to be taken into account to achieve a truncation error at the scale of $2\pi \times \SI{100}{\kilo \hertz}$. This scale is reached at fourth order for the excitation-conserving corrections with $g_{ij}^4 / \Delta_{ij}^3 $ and at third order for the non-excitation-conserving corrections $g_{ij}^3 / \Delta_{ij} \Sigma_{ij}$. The next order of magnitude of $2\pi \times \SI{10}{\kilo \hertz}$ is reached with $g_{ij}^5 / \Delta_{ij}^4$ for the excitation-conserving corrections, and with $g_{ij}^4 / \Delta_{ij}^2 \Sigma_{ij}$ for the non-excitation-conserving corrections, corresponding to fifth and fourth order, respectively. 

However, the truncation error of the diagram expansions depends not only on the scale, but also on the number of neglected corrections, i.e., diagrams that constructively add up. In particular, for the ZZ coupling in \eqref{eq:ZZ}, which is a linear combination of eigenenergies, we note that the diagram expansions subtract, making it difficult to estimate the number of diagrams that add up to contribute. With these complications in mind, it is not obvious that the hundreds of corrections of scale $2\pi \times \SI{10}{\kilo \hertz}$ can be neglected to achieve a truncation error at the scale of $2\pi \times \SI{100}{\kilo \hertz}$.

Taking the above estimates into consideration, we conclude that we need a truncation scheme that at least includes the (third-) fourth-order diagrams in the (non-) excitation-conserving diagram expansions. It is then necessary to investigate if the (fourth-) fifth-order (non-) excitation-conserving diagrams are negligible. With this truncation scheme, we can only capture mechanisms estimated to contribute at the scale of $2\pi \times \SI{100}{\kilo \hertz}$. To go below the scale of $2\pi \times \SI{100}{\kilo \hertz}$, we resort to numerical methods in \secref{sec:numerical_predictions}. 

With a truncation error at the scale of $2\pi \times \SI{100}{\kilo \hertz}$, we emphasize that it is still possible to analytically predict system-parameter regions with zero ZZ coupling. These predictions are possible in regions that fulfill: (1) the perturbative approximation is valid and continuous, i.e., in regions without poles $\Delta_{ij} = 0$, and (2) there are at least one positive and one negative ZZ-coupling point in the region that have absolute values greater than the truncation error. It then follows from the intermediate-value theorem of real analysis \cite{abbottUnderstandingAnalysis2015} that there exists a parameter point in the region with zero ZZ coupling.

With the truncation scheme in hand, we are ready to analytically predict the ZZ coupling. From the estimates in \eqref{eq:perturbative_ratios}, we expect that the excitation-conserving corrections will contribute more than the non-excitation-conserving corrections. As such, we first focus on the excitation-conserving corrections in Sections~\ref{sec:mechanism_description}--\ref{sec:ec_predictions} to predict the ZZ coupling. Then we separately investigate in \secref{sec:nec_corrections} how the non-excitation-conserving and fifth-order corrections modify the predictions. Following this separation, we write the eigenenergies as $E_i = E_{\Delta, i} + E_{\Sigma, i}$, where ($E_{\Sigma, i}$) $E_{\Delta, i}$ is the (non-) excitation-conserving contribution assigned to the bare state $\ket{i}$.

\subsection{The mechanism picture}
\label{sec:mechanism_description}

The immediate result of the SW transformation is a perturbative series for the ZZ coupling in the system parameters: $\omega_i$, $\alpha_i$, and $g_{ij}$. This series is involved for the three-transmon system at hand and difficult to use to in detail explain the emergence of the ZZ coupling. To give structure to the perturbative series, we coarse grain the series into what we refer to as mechanisms. The mechanisms are subseries that share the same parameter dependence, e.g., by common factors. We isolate the mechanisms by applying SW diagrammatics in the excitation-conserving subgraphs. All this results in a mechanism picture, which is a refined analogue of the level-repulsion picture in \secref{sec:level_repulsion_picture}.

To construct the mechanism picture, we first consider the excitation-conserving diagram expansions of the eigenenergies $E_{\Delta, i}$ that define the ZZ coupling in \eqref{eq:ZZ}. In these expansions, we neglect the non-excitation-conserving edges in the Hamiltonian graph in \figref{fig:Hamiltonian_graph}. Under this assumption, the excitation subgraphs decouple, which significantly reduces the number of possible closed paths and thus the number of diagrams in the expansions. Using the notation in \eqref{eq:example_diagram_expansion}, the excitation-conserving diagram expansions are
\begin{widetext}
\begin{align}
    E_{\Delta, 000} = \includegraphics[valign=c]{E100_x.pdf} \,  \hphantom{+}& \hspace{-2.5mm},
    \label{eq:E000=} \\[1.5ex]
    E_{\Delta, 010} = \includegraphics[valign=c]{E100_x.pdf} \, +& \mleft( \includegraphics[valign=c]{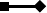} +  \includegraphics[valign=c]{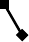} \, \mright) + \includegraphics[valign=c]{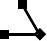} + \mleft( \, \includegraphics[valign=c]{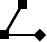}  + \includegraphics[valign=c]{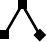} \, \mright) + \includegraphics[valign=c]{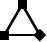} + \mleft( \includegraphics[valign=c]{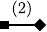} + \includegraphics[valign=c]{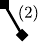} \, \mright) \, , 
    \label{eq:E010=} \\[1.5ex]
    E_{\Delta, 100} = \includegraphics[valign=c]{E100_x.pdf} \,   +& \mleft( \includegraphics[valign=c]{E100_x-.pdf} +  \includegraphics[valign=c]{E100_-x.pdf} \, \mright) + \includegraphics[valign=c]{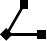} + 
    \mleft( \, \includegraphics[valign=c]{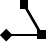}  + \includegraphics[valign=c]{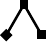} \, \mright) 
    + \includegraphics[valign=c]{E100_xCI.pdf} + 
    \mleft( \includegraphics[valign=c]{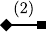} + 
    \includegraphics[valign=c]{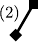} \, \mright) \, , 
    \label{eq:E100=} \\[1.5ex]
    E_{\Delta, 110} = \includegraphics[valign=c]{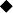} \, +& 
    \mleft( \includegraphics[valign=c]{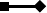} + \includegraphics[valign=c]{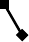} + \includegraphics[valign=c]{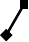} + \includegraphics[valign=c]{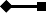} \mright)
    + \mleft( \, \includegraphics[valign=c]{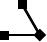} + \includegraphics[valign=c]{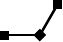} + \includegraphics[valign=c]{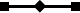} + \includegraphics[valign=c]{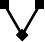}
    + \includegraphics[valign=c]{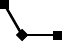} + \includegraphics[valign=c]{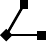} \, \mright) \nonumber \\ 
    +& \mleft( \includegraphics[valign=c]{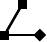} + \includegraphics[valign=c]{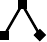} + \includegraphics[valign=c]{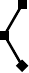}
    + \, \includegraphics[valign=c]{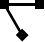} + \includegraphics[valign=c]{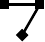} + \includegraphics[valign=c]{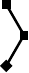} + \includegraphics[valign=c]{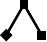} + \includegraphics[valign=c]{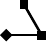} \mright)
    + \mleft( \, \includegraphics[valign=c]{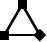} + \includegraphics[valign=c]{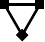} + \includegraphics[valign=c]{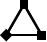} \, \mright ) \nonumber \\ 
    +& \mleft( \includegraphics[valign=c]{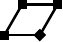} + \includegraphics[valign=c]{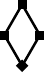} +
    \includegraphics[valign=c]{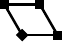} \mright)
    + \mleft( \includegraphics[valign=c]{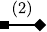} + \includegraphics[valign=c]{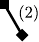} + \includegraphics[valign=c]{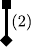} +
    \includegraphics[valign=c]{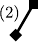} +
    \includegraphics[valign=c]{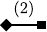} \mright) \, ,
    \label{eq:E110=}
\end{align}
\end{widetext}
where we have collected the individual diagrams in brackets after their kinds. Note that the diagram expansion for $E_{\Delta, 010}$ in \eqref{eq:E010=} is equivalent to the one for $E_{\Delta, 100}$ in \eqref{eq:E100=}. The expansion for $E_{\Delta, 010}$ is given by mirroring each diagram for $E_{\Delta, 100}$ horizontally in the vertical axis. This mirroring is identical to relabeling the qubits $1 \leftrightarrow 2$ in Eqs.~(\ref{eq:effective_H0}) and (\ref{eq:effective_V}). Even though the two expansions are equivalent, they in general give different energy corrections due to the presence of the coupler state $\ket{001}$.

We isolate the mechanisms in the diagram expansions in Eqs.~(\ref{eq:E000=})--(\ref{eq:E110=}). This division is achieved by first using the contractions in Eqs.~(\ref{eq:x-2_contract}) and (\ref{eq:square_contract}), and then regrouping the diagrams based on their mechanism interpretations in \secref{sec:rules}. Starting with the simplest case, the dressed ground-state energy in \eqref{eq:E000=} is the trivial diagram in \eqref{eq:x}, and thus not corrected: $E_{\Delta , 000} = 0$.

In contrast, the first-excitation energies $E_{\Delta, 010}$ and $E_{\Delta, 100}$ have corrections to the bare energies. Since the two diagram expansions are equivalent, we focus on the expansion for $E_{\Delta, 100}$ in \eqref{eq:E100=}. We contract using \eqref{eq:x-2_contract}, resulting in new collections based on the mechanism interpretations that we delimit with brackets:

\begin{widetext}
\begin{equation}
\begin{aligned}
    E_{\Delta, 100} 
    &=
    \includegraphics[valign=c, scale=0.9]{E100_x.pdf} + 
    \mleft( 
    \includegraphics[valign=c, scale=0.9]{E100_x-.pdf} +
    \includegraphics[valign=c, scale=0.9]{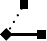} +
    \includegraphics[valign=c, scale=0.9]{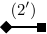} \,
    \mright)
    +
    \mleft( \,
    \includegraphics[valign=c, scale=0.9]{E100_-x.pdf} +
    \includegraphics[valign=c, scale=0.9]{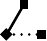} +
    \includegraphics[valign=c, scale=0.9]{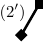} \,
    \mright)
    +
    \mleft(
    \includegraphics[valign=c, scale=0.9]{E100_xCI.pdf} \,
    \mright)
    = \omega_1 
    + 
    \frac{g_{12}^2}{\Delta_{12}} \mleft[1 - \mleft( \frac{g_{12}}{\Delta_{12}} \mright)^2 - \mleft( \frac{g_{13}}{\Delta_{13}} \mright)^2 \mright] 
    \\
    &\hphantom{= \includegraphics[valign=c, scale=0.9]{E100_x.pdf} }~ + \frac{1}{\Delta_{12}} \mleft( \frac{g_{13} g_{23}}{\Delta_{13}} \mright)^2
    + \frac{g_{13}^2}{\Delta_{13}} \mleft[1 - \mleft( \frac{g_{13}}{\Delta_{13}} \mright)^2 - \mleft( \frac{g_{12}}{\Delta_{12}} \mright)^2 \mright] + \frac{1}{\Delta_{13}} \mleft( \frac{g_{12} g_{23}}{\Delta_{12}} \mright)^2 + \frac{2 g_{12} g_{13} g_{23}}{\Delta_{12} \Delta_{13}}
     ,
    \label{eq:E100=_evaluated}
\end{aligned}
\end{equation}
\end{widetext}
where we after the last equality sign have evaluated the diagrams with the rules in Eqs.~(\ref{eq:x})--(\ref{eq:square_contract}). The evaluated terms follow the order of appearance of the diagrams, as is manifest by the coupling strengths in each term. We have also in \eqref{eq:E100=_evaluated} introduced a dashed-diagram notation to account for the fact that diagrams of the kind in \eqref{eq:-x-} correct two different level repulsions such that
\begin{equation}
    \includegraphics[valign=c]{E100_-x-.pdf} = \includegraphics[valign=c]{E100_cx-.pdf} + \includegraphics[valign=c]{E100_-xc.pdf} \,,
    \label{eq:dash_notation}
\end{equation}
where $\includegraphics[valign=b, width=\baselineskip]{E100_cx-.pdf} = - (g_{12}^2 / \Delta_{12}) (g_{13} / \Delta_{13})^2$ and $\includegraphics[valign=b, width=\baselineskip]{E100_-xc.pdf} = - (g_{13}^2 / \Delta_{13}) (g_{12} / \Delta_{12})^2$.

With \eqref{eq:E100=_evaluated}, we find that the bare energy (trivial diagram) is corrected by three mechanisms: level repulsions from the energy levels of the states $\ket{010}$ (first round bracket in order of appearance) and $\ket{001}$ (second round bracket), and a three-loop mechanism (third round bracket). 

From the evaluated expressions, we note that each mechanism has a distinct common factor, e.g., $1 / \Delta_{12}$ for the level repulsion from the energy level of $\ket{010}$. Because of these common factors, which give the leading-order contribution from each mechanism, the three mechanisms can be varied to a great extent independently by varying several system parameters at the same time. For example, changing the sign $g_{12} \to -g_{12}$ flips the parity of the three-loop mechanism while keeping the level repulsions unchanged. The relative parameter independence between the mechanisms makes it meaningful to describe the energy corrections in terms of mechanisms instead of system parameters. The mechanism picture is thus a useful coarse-grained complement to describing the energy corrections in terms of the system parameters, with the advantage of being less complex. With the coarse-grained mechanisms, we reduce the complexity from the many terms in \eqref{eq:E100=_evaluated} down to only three mechanisms.

Continuing with the second-excitation energy $E_{\Delta, 110}$, we identify eight mechanisms that give energy corrections. We find these mechanisms similar to \eqref{eq:E100=_evaluated} and by additionally using the contraction in \eqref{eq:square_contract}:
\begin{widetext}
\begin{equation}
\begin{aligned}
    E_{\Delta, 110} &= 
    \includegraphics[valign=c, scale=0.89]{E110_x.pdf} \, 
    + 
    \mleft( 
    \includegraphics[valign=c, scale=0.89]{E110_-x.pdf} +
    \includegraphics[valign=c, scale=0.89]{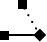} +
    \includegraphics[valign=c, scale=0.89]{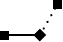} +
    \includegraphics[valign=c, scale=0.89]{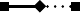} + 
    \includegraphics[valign=c, scale=0.89]{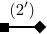}
    \mright)
    +
    \mleft( 
    \includegraphics[valign=c, scale=0.89]{E110_x-.pdf} +
    \includegraphics[valign=c, scale=0.89]{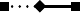} +
    \includegraphics[valign=c, scale=0.89]{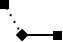} +
    \includegraphics[valign=c, scale=0.89]{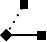} +
    \includegraphics[valign=c, scale=0.89]{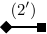} \,
    \mright)
    +
    \mleft( \, \includegraphics[valign=c, scale=0.89]{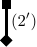} \mright)
    \\
    &+ \mleft( \,
    \includegraphics[valign=c, scale=0.89]{E110_Lx.pdf} +
    \includegraphics[valign=c, scale=0.89]{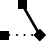} +
    \includegraphics[valign=c, scale=0.89]{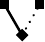} +
    \includegraphics[valign=c, scale=0.89]{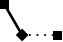} + 
    \includegraphics[valign=c, scale=0.89]{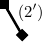}
    \mright) 
    +
    \mleft( 
    \includegraphics[valign=c, scale=0.89]{E110_xJ.pdf} +
    \includegraphics[valign=c, scale=0.89]{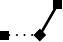} +
    \includegraphics[valign=c, scale=0.89]{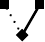} +
    \includegraphics[valign=c, scale=0.89]{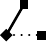} + 
    \includegraphics[valign=c, scale=0.89]{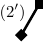}
    \mright)
    +
    \mleft( \, 
    \includegraphics[valign=c, scale=0.89]{E110_CIx.pdf} + 
    \includegraphics[valign=c, scale=0.89]{E110_CxI.pdf} + 
    \includegraphics[valign=c, scale=0.89]{E110_xCI.pdf} \, 
    \mright ) ,
    \label{eq:E110=_evaluated}
\end{aligned}
\end{equation}
\end{widetext}
where we have used the dashed-diagram notation from \eqref{eq:dash_notation} and again round brackets to separate the mechanisms (see \appref{app:diagram_evaluations} for the evaluation of the diagrams). Note that the first five brackets are level repulsions from the energy levels assigned to $\ket{200}$, $\ket{020}$, $\ket{002}$, $\ket{101}$, and $\ket{011}$, respectively. Of these level repulsions, only the repulsion from $\ket{002}$ is given by a single (contracted) second-order level repulsion. The remaining mechanisms are three-loop mechanisms (last bracket).

\subsection{Correlations in the mechanisms}
\label{sec:mechanism_correlations}

Above, we found that the energy corrections are grouped into 14 mechanisms (recall that $E_{\Delta, 010}$ has an equivalent diagram expansion to $E_{\Delta, 100}$). To understand the contributions to the ZZ coupling from these mechanisms, we could individually compare each of them, but this is impractical due to the sheer numbers. Instead, we opt for a more compact comparison based on the observation that the level repulsions are correlated with respect to specific detunings. We use the correlations to further coarse grain the picture from 14 mechanisms down to four correlated and one contracted level repulsions, and one total three-loop mechanism.

To give an example of one these correlations, we consider the level repulsion on the level of $\ket{100}$ from $\ket{010}$ in \eqref{eq:E100=_evaluated}; it has the leading factor $1/\Delta_{12}$. We compare this level repulsion to the one on $\ket{110}$ from $\ket{020}$, which has the leading factor $1/(\Delta_{12} - \alpha_2)$. The two level repulsions are correlated with respect to the detuning $\Delta_{12}$. They either align or counteract depending on the relation between $\Delta_{12}$ and $\alpha_2$, with strengths following the same functional dependency on $\Delta_{12}$. We recall that the level repulsion from the level of $\ket{010}$ comes with a negative contribution to the ZZ coupling due to the negative sign in \eqref{eq:ZZ}. If we take the negative contribution into account, the two level repulsions are negatively correlated.

We note that the correlations are directly distinguishable from the directions of the edges in the diagram expansions. This distinguishability is by design of the Hamiltonian graph in \figref{fig:Hamiltonian_graph} and then leveraged by the SW diagrammatics. In particular, for the correlation above, it is distinguished by noting that the first bracket in \eqref{eq:E100=_evaluated} and the second bracket in \eqref{eq:E110=_evaluated} consist of diagrams with the same horizontal edge in the subdiagram: $\includegraphics[valign=c]{E100_x-.pdf}$. Inspired by the common subdiagram, we use $\zeta_{\includegraphics[valign=c, scale=0.6]{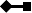}}$ to denote the contribution from the above correlated level repulsion to the ZZ coupling. The correlation is further manifest from the fact that the mentioned brackets include three identical diagrams if we ignore that the diagrams involve different states. Thus, when performing arithmetics with identical diagrams, we need to explicitly label the initial states.

Investigating the other repulsions, we find from the direction of the edges in the diagrams three more correlations. In close resemblance to $\zeta_{\includegraphics[valign=c, scale=0.6]{Subscript_e.pdf}}$, the level repulsions from the levels on $\ket{010}$ from $\ket{100}$, and on $\ket{110}$ from $\ket{200}$ are correlated with respect to the detuning $-\Delta_{12}$. We denote this contribution with $\zeta_{\includegraphics[valign=c, scale=0.6]{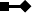}}$. We also find the two correlations: (1) the level repulsion on the level of $\ket{100}$ from $\ket{001}$ [second round bracket in \eqref{eq:E100=_evaluated}] is correlated with the repulsion on $\ket{110}$ from $\ket{011}$ [fifth bracket in \eqref{eq:E110=_evaluated}], and (2) the level repulsions on $\ket{010}$ from $\ket{001}$, and on $\ket{110}$ from $\ket{101}$ are correlated. These two correlations are with respect to the detunings $\Delta_{13}$ and $\Delta_{23}$, and we denoted their contributions with $\zeta_{\includegraphics[valign=c, scale=0.6]{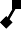}}$ and $\zeta_{\includegraphics[valign=c, scale=0.6]{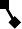}}$, respectively.

In total, we find four correlated level repulsions. Beyond these repulsions, we also have the (contracted) second-order level repulsion from $\ket{002}$ and the three-loop mechanisms. We denote the second-order level repulsion with $\zeta_{\includegraphics[valign=c, scale=0.6]{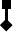}}$ and collect the total contribution from the three-loop mechanisms in $\zeta_3$. We choose to collect the three-loop mechanisms in a $\zeta_3$ to in \secref{sec:ec_predictions} compare their total contribution to the ZZ coupling with the correlated level repulsions. By then structuring the level repulsions in Eqs.~(\ref{eq:E000=})--(\ref{eq:E110=}) with respect to the correlations and then inserting them into \eqref{eq:ZZ}, we obtain:
\begin{equation}
    \zeta_\Delta \equiv 
    \zeta_{\includegraphics[valign=c, scale=0.6]{Subscript_w.pdf}}
    + \zeta_{\includegraphics[valign=c, scale=0.6]{Subscript_e.pdf}}
    + \zeta_{\includegraphics[valign=c, scale=0.6]{Subscript_nw.pdf}}
    + \zeta_{\includegraphics[valign=c, scale=0.6]{Subscript_ne.pdf}}
    + \zeta_{\includegraphics[valign=c, scale=0.6]{Subscript_n.pdf}}
    + \zeta_3,
    \label{eq:ZZ=}
\end{equation}
where the diagram expansions for each contribution to the ZZ coupling are:

\begin{widetext}
\begin{align}
    \zeta_{\includegraphics[valign=c, scale=0.6]{Subscript_w.pdf}} &=
    \mleft( 
    \underset{\ket{110}}{\includegraphics[valign=c]{E110_-x.pdf}} +
    \underset{\ket{110}}{\includegraphics[valign=c]{E110_-Lcx.pdf}} +
    \underset{\ket{110}}{\includegraphics[valign=c]{E110_-xJc.pdf}} +
    \underset{\ket{110}}{\includegraphics[valign=c]{E110_-x-c.pdf}} + 
    \underset{\ket{110}}{\includegraphics[valign=c]{E110_2a-x.pdf}}
    \mright) 
    - 
    \mleft( 
    \underset{\ket{010}}{\includegraphics[valign=c]{E010_-x.pdf}} +
    \underset{\ket{010}}{\includegraphics[valign=c]{E110_-Lcx.pdf}} +
    \underset{\ket{010}}{\includegraphics[valign=c]{E110_2a-x.pdf}} 
    \mright), \label{eq:ZZ=_w} \\
    \zeta_{\includegraphics[valign=c, scale=0.6]{Subscript_e.pdf}} &=
    \mleft( 
    \underset{\ket{110}}{\includegraphics[valign=c]{E110_x-.pdf}} +
    \underset{\ket{110}}{\includegraphics[valign=c]{E110_-cx-.pdf}} +
    \underset{\ket{110}}{\includegraphics[valign=c]{E110_Lcx-.pdf}} +
    \underset{\ket{110}}{\includegraphics[valign=c]{E110_xJc-.pdf}} + 
    \underset{\ket{110}}{\includegraphics[valign=c]{E110_x-2a.pdf}}
    \mright) 
    - 
    \mleft( 
    \underset{\ket{100}}{\includegraphics[valign=c]{E100_x-.pdf}} +
    \underset{\ket{100}}{\includegraphics[valign=c]{E100_cx-.pdf}} +
    \underset{\ket{100}}{\includegraphics[valign=c]{E110_x-2a.pdf}} 
    \mright), \label{eq:ZZ=_e} \\[1.5ex]
    \zeta_{\includegraphics[valign=c, scale=0.6]{Subscript_nw.pdf}} &=
    \mleft( 
    \underset{\ket{110}}{\includegraphics[valign=c]{E110_Lx.pdf}} +
    \underset{\ket{110}}{\includegraphics[valign=c]{E110_-cLx.pdf}} +
    \underset{\ket{110}}{\includegraphics[valign=c]{E110_LxJc.pdf}} +
    \underset{\ket{110}}{\includegraphics[valign=c]{E110_Lx-c.pdf}} +
    \underset{\ket{110}}{\includegraphics[valign=c]{E110_2aLx.pdf}}
    \mright) 
    - 
    \mleft( 
    \underset{\ket{010}}{\includegraphics[valign=c]{E010_x-.pdf}} +
    \underset{\ket{010}}{\includegraphics[valign=c]{E110_-cLx.pdf}} +
    \underset{\ket{010}}{\includegraphics[valign=c]{E110_2aLx.pdf}} 
    \mright), \label{eq:ZZ=_nw} \\[1.5ex]
    \zeta_{\includegraphics[valign=c, scale=0.6]{Subscript_ne.pdf}} &= 
    \mleft( 
    \underset{\ket{110}}{\includegraphics[valign=c]{E110_xJ.pdf}} +
    \underset{\ket{110}}{\includegraphics[valign=c]{E110_-cxJ.pdf}} +
    \underset{\ket{110}}{\includegraphics[valign=c]{E110_LcxJ.pdf}} +
    \underset{\ket{110}}{\includegraphics[valign=c]{E110_xJ-c.pdf}} + 
    \underset{\ket{110}}{\includegraphics[valign=c]{E110_xJ2a.pdf}}
    \mright) 
    - 
    \mleft( 
    \underset{\ket{100}}{\includegraphics[valign=c]{E100_-x.pdf}} +
    \underset{\ket{100}}{\includegraphics[valign=c]{E100_-xc.pdf}} +
    \underset{\ket{100}}{\includegraphics[valign=c]{E110_xJ2a.pdf}} 
    \mright), \label{eq:ZZ=_ne} \\[1.5ex]
    \zeta_{\includegraphics[valign=c, scale=0.6]{Subscript_n.pdf}} &= 
     \underset{\ket{110}}{\includegraphics[valign=c]{E110_xI2.pdf}} 
    , \qquad
    \zeta_3 = 
    \underset{\ket{110}}{\includegraphics[valign=c]{E110_CIx.pdf}} +
    \underset{\ket{110}}{\includegraphics[valign=c]{E110_CxI.pdf}} +
    \underset{\ket{110}}{\includegraphics[valign=c]{E110_xCI.pdf}} 
    -
    \underset{\ket{010}}{\includegraphics[valign=c]{E010_xCI.pdf}} -
    \underset{\ket{100}}{\includegraphics[valign=c]{E100_xCI.pdf}} .
    \label{eq:ZZ=_3}
\end{align}
\end{widetext}
Here, the correlated level repulsions are given by the first four lines. In Eqs.~(\ref{eq:ZZ=_w})--(\ref{eq:ZZ=_3}), we use the dashed notation from \eqref{eq:dash_notation} and explicitly label each diagram underneath with the initial state. As mentioned above, this labeling is required to distinguish between identical diagrams with different initial states. We refer to \appref{app:diagram_evaluations} for the evaluation of the diagram expansions in Eqs.~(\ref{eq:ZZ=_w})--(\ref{eq:ZZ=_3}).

Analogously to the level repulsions discussed in \secref{sec:configurations}, Eqs.~(\ref{eq:ZZ=_w})--(\ref{eq:ZZ=_3}) remain without additions of new mechanisms in setups with other superconducting qubits than transmons. This observation holds in setups where the excitation-conserving edges in \figref{fig:Hamiltonian_graph} couple the same states, e.g., in fluxonium architectures. With fluxonium, the main change, in addition to other energy-level configurations (recall \secref{sec:configurations}), is that the coupling strengths follow different functional dependencies compared to \figref{fig:Hamiltonian_graph}. To showcase the correlated and contracted level repulsions and the three-loop mechanisms in a specific case, we focus the remainder of the section on the three-transmon system.

\subsection{Predictions from the excitation-conserving mechanisms}
\label{sec:ec_predictions}

\begin{figure*}
    \centering
    \includegraphics[width=\textwidth]{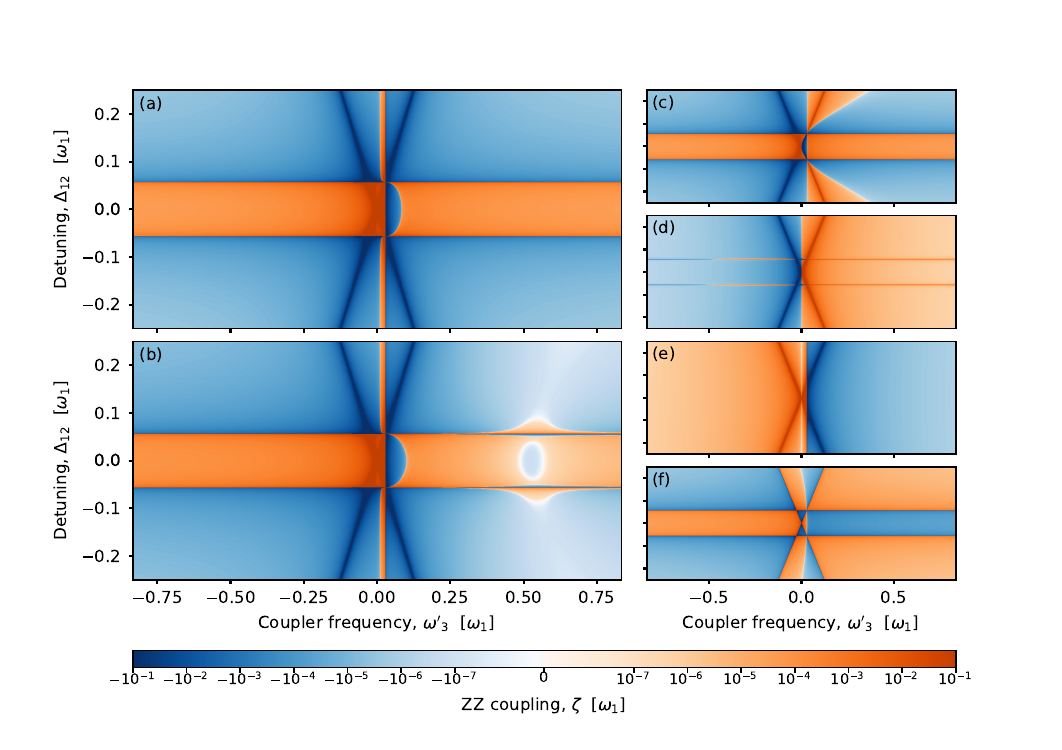}
    \caption{Analytical predictions for the static ZZ coupling from the excitation-conserving mechanisms. The predictions in (a--b) are gradually constructed from the level repulsions in (c--e) and the three-loop mechanisms in (f). The positive (negative) contributions to the ZZ coupling strength are represented with an orange (blue) gradient, while the white regions represent zero ZZ coupling. The plots are generated with the system parameters in units of $\omega_1$: $g_{13} = g_{23} = 75/4 \times 10^{-3}$, $g_{12} = g_{13} / 30 $, and $\alpha_1 = \alpha_2 = \alpha_3 = 3 g_{13}$ in the parameter space of the shifted coupler frequency relative to the mean qubit frequency $\omega'_3 = \omega_3 - (\omega_1 + \omega_2)/2$ and the qubit detuning $\Delta_{12} = \omega_1 - \omega_2$. (a) The contribution
    $
    \zeta_{\includegraphics[valign=c, scale=0.6]{Subscript_w.pdf}} +
    \zeta_{\includegraphics[valign=c, scale=0.6]{Subscript_e.pdf}} + 
    \zeta_{\includegraphics[valign=c, scale=0.6]{Subscript_nw.pdf}} +
    \zeta_{\includegraphics[valign=c, scale=0.6]{Subscript_ne.pdf}} +
    \zeta_{\includegraphics[valign=c, scale=0.6]{Subscript_n.pdf}} 
    $
    from level repulsions in (c--e). (b) The total contribution $\zeta_\Delta$ from all mechanisms in (c--f). (c) The correlated level repulsions assigned to $\ket{010}$, $\ket{100}$, $\ket{020}$, and $\ket{200}$
    , i.e., 
    $
    \zeta_{\includegraphics[valign=c, scale=0.6]{Subscript_w.pdf}} +
    \zeta_{\includegraphics[valign=c, scale=0.6]{Subscript_e.pdf}}
    $.
    (d) The correlated level repulsions assigned to $\ket{001}$, $\ket{011}$, and $\ket{101}$, i.e., 
    $
    \zeta_{\includegraphics[valign=c, scale=0.6]{Subscript_nw.pdf}} +
    \zeta_{\includegraphics[valign=c, scale=0.6]{Subscript_ne.pdf}}
    $.
    (e) The second-order level repulsion assigned to $\ket{002}$, i.e., $\zeta_{\includegraphics[valign=c, scale=0.6]{Subscript_n.pdf}}$. (f) The sum of the three-loop mechanisms $\zeta_3$. The panels (c--f) follow the order of appearance in Eqs.~(\ref{eq:ZZ=_w})--(\ref{eq:ZZ=_3}).}
    \label{fig:ZZ_landscape_EC_mechanisms}
\end{figure*}
Here, we predict the strength of the ZZ coupling between the transmon qubits in \figref{fig:three_qubit_circuit} from the correlated and contracted level repulsions (from here on: level repulsions) and the three-loop mechanisms in Eqs.~(\ref{eq:ZZ=_w})--(\ref{eq:ZZ=_3}). We especially focus on predicting all possible parameter regions with zero or strong ZZ coupling, for which we recall that there is a finite number of parameter regions defined from the 24 energy-level configurations discussed in \secref{sec:configurations}. In each parameter region, we evaluate the contributions of the level repulsions and the three-loop mechanisms. We use the resulting evaluations to sequentially construct the ZZ coupling in \figref{fig:ZZ_landscape_EC_mechanisms}. In the figure, we consider the coupling-strength regime $\abs{g_{12}} \ll \abs{g_{13}}, \abs{g_{23}}$, while the two regimes $\abs{g_{12}} \sim \abs{g_{13}}, \abs{g_{23}}$ and $\abs{g_{12}} \gg \abs{g_{13}}, \abs{g_{23}}$ are considered in \appref{app:ZZ_other_parameters}. The conclusions presented in this section are general for the three regimes unless otherwise stated. 

Focusing first on the separate level repulsions and three-loop mechanisms, we show their contributions to the ZZ coupling in \figpanels{fig:ZZ_landscape_EC_mechanisms}{c}{f}. We plot these contributions as functions of the shifted coupler frequency relative to the mean qubit frequency $\omega'_3 = \omega_3 - (\omega_1 + \omega_2)/2$ and the qubit detuning $\Delta_{12} = \omega_1 - \omega_2$. For the sake of brevity, we plot $\zeta_{\includegraphics[valign=c, scale=0.6]{Subscript_w.pdf}} + \zeta_{\includegraphics[valign=c, scale=0.6]{Subscript_e.pdf}}$ together in \figpanel{fig:ZZ_landscape_EC_mechanisms}{c}, which are both jointly correlated with $\Delta_{12}$. Likewise, we combine the $\Delta_{13}$- and $\Delta_{23}$-correlated level repulsions and plot $\zeta_{\includegraphics[valign=c, scale=0.6]{Subscript_nw.pdf}} + \zeta_{\includegraphics[valign=c, scale=0.6]{Subscript_ne.pdf}}$ in \figpanel{fig:ZZ_landscape_EC_mechanisms}{d}. These two repulsions have a joint correlation with respect to $\omega'_3$, but it is weaker than with respect to $\Delta_{13}$ and $\Delta_{23}$ independently.

We remark that the mechanism contributions have their strongest magnitudes in the dark blue and dark orange regions in \figref{fig:ZZ_landscape_EC_mechanisms}. These narrow dark regions are the poles $\Delta_{ij} = 0$ of the perturbative series in Eqs.~(\ref{eq:ZZ=_w})--(\ref{eq:ZZ=_3}). Thus, the dark regions correspond to the partition lines in \figref{fig:configurations}; they signal changes in the energy-level configurations. Note that not all partition lines have a corresponding dark region. These missing dark regions are explained by the fact that not all poles $\Delta_{ij} = 0$ in the second-excitation subspace are present when truncating the perturbation series to fourth order, e.g., the resonance between the $\ket{200}$ and $\ket{011}$ is missing. Taking into account that the perturbative series is not valid close to poles, we find that the level repulsions and the three-loop mechanisms, with few exceptions, do not change sign, i.e., color, between the poles. Hence for the predictions of the parameter regions with zero or strong ZZ coupling, we take a similar approach to the one in \secref{sec:intuitive_picture_predictions} for the regions bounded by the poles. Thus, we first consider where the mechanisms balance to result in a zero ZZ coupling and then how to maximally break this balance to create regions of strong ZZ coupling.

We examine the balance between the correlated level repulsions 
in \figpanels{fig:ZZ_landscape_EC_mechanisms}{c}{e}. Starting with the $\Delta_{12}$-correlated level repulsions in \figpanel{fig:ZZ_landscape_EC_mechanisms}{c}, we note that the main feature is a (blue) negative background with a (orange) positive horizontal band, sometimes referred to as the straddling regime \cite{kochChargeinsensitiveQubitDesign2007}. The positive band is given by $\abs{\Delta_{12}} < \abs{\alpha_1}, \abs{\alpha_2}$, corresponding to the configurations where the (bare) energy level of $\ket{110}$ is above the levels of $\ket{020}$ and $\ket{200}$ (see \secref{sec:configurations} for details). We note also the X-shaped feature that we attribute to the second-order level repulsions. Due to the second-order repulsions, the regions of \figpanel{fig:ZZ_landscape_EC_mechanisms}{c} remain with the same sign in the case of no qubit coupling $g_{12} = 0$.

Continuing to the other level repulsions in \figpanels{fig:ZZ_landscape_EC_mechanisms}{d}{e}, we note that the directions, i.e., signs, of these repulsions are mainly determined by the shifted coupler frequency $\omega'_3$. This simple dependence is expected from the fact that these repulsions mainly involve states with coupler excitations. Less expected however, we find that the direction in \figpanel{fig:ZZ_landscape_EC_mechanisms}{d} primarily is opposite to the one in \figpanel{fig:ZZ_landscape_EC_mechanisms}{e}. This difference implies that the level repulsions in the first-excitation subspace assigned to $\ket{001}$ dominate the repulsions in the second-excitation subspace assigned to $\ket{011}$ and $\ket{101}$. We explain this difference by observing in Eqs.~(\ref{eq:ZZ=_nw}) and (\ref{eq:ZZ=_ne}) that the repulsions in the second-excitation subspace have more correcting diagrams that dampen the repulsions. 

With the directions of the correlated repulsions in \figpanels{fig:ZZ_landscape_EC_mechanisms}{c}{e}, we reckon that there are several parameter regions where these repulsions counteract. Indeed in the coupling-strength regime $\abs{g_{12}} \ll \abs{g_{13}}, \abs{g_{23}}$, the $\Delta_{12}$-correlated level repulsions in \figpanel{fig:ZZ_landscape_EC_mechanisms}{c} are sufficiently weak to be balanced by the other level repulsions. We sum the level repulsions to plot $
\zeta_{\includegraphics[valign=c, scale=0.6]{Subscript_w.pdf}} +
\zeta_{\includegraphics[valign=c, scale=0.6]{Subscript_e.pdf}} + 
\zeta_{\includegraphics[valign=c, scale=0.6]{Subscript_nw.pdf}} +
\zeta_{\includegraphics[valign=c, scale=0.6]{Subscript_ne.pdf}} +
\zeta_{\includegraphics[valign=c, scale=0.6]{Subscript_n.pdf}} 
$ in \figpanel{fig:ZZ_landscape_EC_mechanisms}{a}. We find two parameter regions (white; for $\Delta_{12} > 0$) with zero joint contribution. For example, the region in the positive horizontal band has a zero ZZ coupling caused by the repulsions in Figs.~\hyperref[fig:ZZ_landscape_EC_mechanisms]{\ref*{fig:ZZ_landscape_EC_mechanisms}(c)} ($\zeta_{\includegraphics[valign=c, scale=0.6]{Subscript_w.pdf}} + \zeta_{\includegraphics[valign=c, scale=0.6]{Subscript_e.pdf}}$) and \hyperref[fig:ZZ_landscape_EC_mechanisms]{\ref*{fig:ZZ_landscape_EC_mechanisms}(d)} ($\zeta_{\includegraphics[valign=c, scale=0.6]{Subscript_nw.pdf}} + \zeta_{\includegraphics[valign=c, scale=0.6]{Subscript_ne.pdf}}$) aligning and then being counteracted with the repulsion in \figpanel{fig:ZZ_landscape_EC_mechanisms}{e} ($\zeta_{\includegraphics[valign=c, scale=0.6]{Subscript_n.pdf}}$).

Since \figpanel{fig:ZZ_landscape_EC_mechanisms}{a} includes the contributions from all level repulsions, we compare it to the predictions from the intuitive picture in \figref{fig:configurations}. We observe an excellent agreement between the two figures based on the signs, i.e., the regions with blue and orange colors, and also the white regions with zero contribution. Recall in \figref{fig:configurations} that the whole region is white in the presence of a predicted zero ZZ coupling. The white regions in \figref{fig:configurations} are hence exaggerated in comparison to \figpanel{fig:ZZ_landscape_EC_mechanisms}{a}, but the level repulsions in the intuitive and mechanism pictures predict the same two parameter regions for zero ZZ coupling. We observe that the two regions are near the maximum contribution from the level repulsion assigned to $\ket{002}$ in \figpanel{fig:ZZ_landscape_EC_mechanisms}{e}. In fact, not including this level repulsion in \figpanel{fig:ZZ_landscape_EC_mechanisms}{a} removes the two parameter regions with zero ZZ coupling. Based on the involved mechanisms, we say that these two regions have zero ZZ coupling of level-repulsion type.

Having considered the level repulsions, we turn to add the contribution from the three-loop mechanisms. We show the contribution $\zeta_3$ in \figpanel{fig:ZZ_landscape_EC_mechanisms}{f}. In contrast to the other mechanisms, the three-loop mechanisms are linear in the coupling strengths and proportional to $g_{12} g_{13} g_{23}$. By changing the sign of any of these coupling strengths, which is possible for floating transmons \cite{seteFloatingTunableCoupler2021a}, we can invert the sign of $\zeta_3$ and as a result the blue and orange colors in \figpanel{fig:ZZ_landscape_EC_mechanisms}{f}; the other mechanisms are invariant under these sign changes. Hence, for individual regions between poles, we are free to add the three-loop mechanisms as a positive or negative contribution to the sum of the other mechanisms in \figpanel{fig:ZZ_landscape_EC_mechanisms}{a}. 

For positive coupling strengths, we obtain the ZZ coupling in \figpanel{fig:ZZ_landscape_EC_mechanisms}{b} for the regime $\abs{g_{12}} \ll \abs{g_{13}}, \abs{g_{23}}$. We observe that the two regions (for $\Delta_{12} > 0$) with zero ZZ couplings of level-repulsion type from \figpanel{fig:ZZ_landscape_EC_mechanisms}{a} are marginally altered, while a new zero ZZ coupling region (white) appears in the horizontal band $\abs{\Delta_{12}} < \abs{\alpha_1}, \abs{\alpha_2}$ for larger $\omega'_3$. Since the new region is constructed from adding the three-loop mechanisms, we call this a zero ZZ coupling of three-loop type.

We also find a weak ZZ coupling region outside of the horizontal band $\abs{\Delta_{12}} < \abs{\alpha_1}, \abs{\alpha_2}$ for $\omega'_3$. However, note that the region outside of the horizontal band does not clearly appear between a positive (orange) and a negative (blue) region away from the poles. Therefore, we cannot infer from the intermediate-value argument as outlined in \secref{sec:truncation_scheme} that the region includes a point of zero ZZ coupling.

Shifting away from balancing the mechanisms, we consider how to align the correlated level repulsions and the loop mechanisms to construct the strongest possible ZZ coupling. Viewing \figpanels{fig:ZZ_landscape_EC_mechanisms}{c}{f}, we can align all the mechanisms in a positive direction for $\omega'_3 < 0$ except \figpanel{fig:ZZ_landscape_EC_mechanisms}{d} in the horizontal band. Approaching the pole at $\omega'_3 = 0$, the strengths of all mechanisms increase. Hence, we predict the strongest possible ZZ coupling to be achievable in the horizontal band when approaching $\omega'_3 \to 0^-$. We remark that this predicted parameter region is visibly strong in \figpanel{fig:ZZ_landscape_EC_mechanisms}{b} and analogous to the predicted region in \secref{sec:intuitive_picture_predictions}. However, the predicted region is in a non-perturbative parameter region, meaning that even if we can predict the placement, we are unable to quantitatively predict the strength of the ZZ coupling in this particular region.

\begin{figure*}
    \centering
    \includegraphics[width=\textwidth]{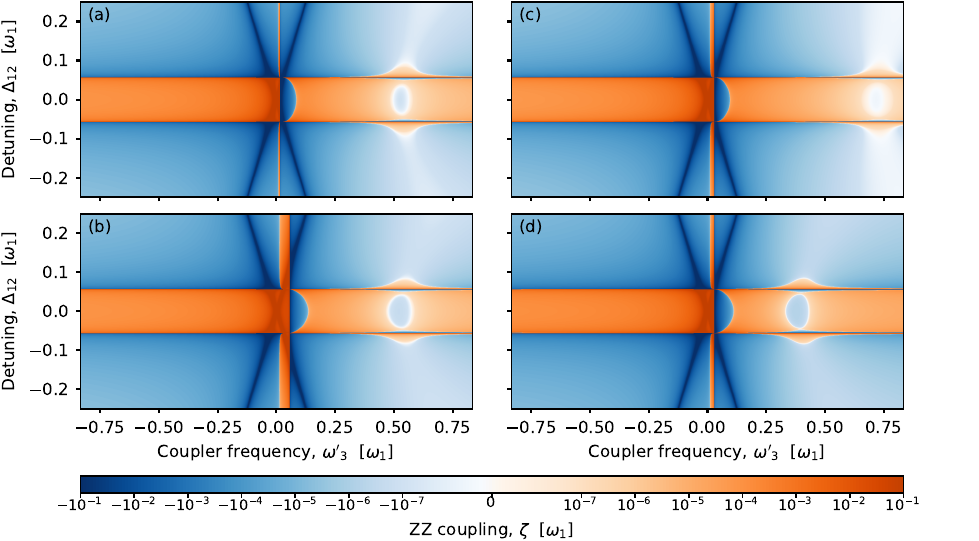}
    \caption{Variations of the analytical predictions in \figpanel{fig:ZZ_landscape_EC_mechanisms}{b} with respect to the coupler anharmonicity $\alpha_3$ in (a-b) and the qubit coupling strength $g_{12}$ in (c-d). (a) Decreased coupler anharmonicity $\alpha_3 \to \alpha_3 /2$. (b) Increased coupler anharmonicity $\alpha_3 \to 2 \alpha_3$. (c) Decreased qubit coupling strength $g_{12} \to 3 g_{12} / 4$. (d) Increased qubit coupling strength $g_{12} \to 4 g_{12} / 3$. The remaining system parameters are identical to \figref{fig:ZZ_landscape_EC_mechanisms}. Changes in $\alpha_3$ allows for control of the level-repulsion-type zero ZZ coupling, while changes in $g_{12}$ gives control of the three-loop-type zero ZZ coupling.}
    \label{fig:ZZ_landscape_EC_control}
\end{figure*}

We conclude that there are two alternatives to balance the level repulsions and the three-loop mechanisms in the 24 energy-level configurations from \secref{sec:configurations}. In the first alternative, the level repulsion assigned to $\ket{002}$ is the critical mechanism to achieve the balance, while the three-loop mechanisms play the same important role in the second alternative. We thus find two distinct types of regions with zero ZZ coupling. For the regions with strong ZZ coupling, we predict them to emerge around poles of the second-excitation subspace as given by \figref{fig:ZZ_landscape_EC_mechanisms}. In both instances of zero and strong ZZ coupling, we have encountered regions of interest where the perturbative approach used is not sufficient to answer all relevant questions. We will hence return to these regions in \secref{sec:numerical_predictions} with numerical methods.

\subsection{How to control the parameter regions of zero and strong ZZ coupling}
\label{sec:ZZ_coupling_control}

We consider how to control the features, e.g., their positions in frequency space, of the parameter regions with zero and strong ZZ coupling in \figpanel{fig:ZZ_landscape_EC_mechanisms}{b}. By having control of these features, the ZZ coupling can be engineered in the design process of a device. The features are controlled by choosing the anharmonicities $\alpha_i$ and the coupling strengths $g_{ij}$. In \figref{fig:ZZ_landscape_EC_control}, we show how the analytical predictions from the excitation-conserving mechanisms in \figpanel{fig:ZZ_landscape_EC_mechanisms}{b} vary with respect to changes in the coupler anharmonicity $\alpha_3$ and the qubit coupling strength $g_{12}$. These system parameters are excellent control parameters for the position of the parameter regions with zero ZZ coupling.

The frequency positions of the poles are controlled by the anharmonicities. Of the five lines of poles in \figpanel{fig:ZZ_landscape_EC_mechanisms}{b}, only the two poles $\Delta_{13} = 0$ and $\Delta_{23} = 0$, forming the X-shaped feature around $\omega'_3 = 0$, do not depend on the anharmonicities. The remaining three poles: $\Delta_{12} = - \alpha_1$, $\Delta_{12} =  \alpha_2$, and $\omega'_3 = - \alpha_3$, are controllable. We can use the anharmonicities $\alpha_1$ and $\alpha_2$ to vertically shift or change the size of the orange horizontal band in  \figpanel{fig:ZZ_landscape_EC_mechanisms}{b}. Similarly, the vertical pole at $\omega'_3 = - \alpha_3$ is horizontally shifted with the anharmonicity $\alpha_3$. We show the effect of changing $\alpha_3$ in \figpanels{fig:ZZ_landscape_EC_control}{a}{b}. We note that the orange vertical region increases in size with increasing anharmonicities, i.e., a more negative $\alpha_3$. Also, the level-repulsion-type zero ZZ coupling extends to larger $\omega'_3$. Compared to these two feature changes, the remaining features undergo only minor changes in comparison to \figpanel{fig:ZZ_landscape_EC_mechanisms}{b}. We understand the effect of varying $\alpha_3$ from the level repulsions in \figpanels{fig:ZZ_landscape_EC_mechanisms}{d}{e}. By increasing $\abs{\alpha_3}$, we separate the maximum of the level repulsion assigned to $\ket{002}$ in \figpanel{fig:ZZ_landscape_EC_mechanisms}{e} from the maximum in \figpanel{fig:ZZ_landscape_EC_mechanisms}{d}, causing the observed effect in \figpanels{fig:ZZ_landscape_EC_control}{a}{b}.

The position of the three-loop-type zero ZZ coupling is controllable by the qubit coupling strength $g_{12}$. By varying $g_{12}$, the region is shifted in the direction of $\omega'_3$. We show in \figpanels{fig:ZZ_landscape_EC_control}{c}{d} two variations of \figpanel{fig:ZZ_landscape_EC_mechanisms}{b} with respect to changes in $g_{12}$. For positive $g_{12}$, we note that the three-loop-type region horizontally shifts towards the center $\omega'_3 = 0$ with increasing $g_{12}$ until it collides and merges with the level-repulsion-type zero ZZ coupling. For decreasing $g_{12}$, the region shifts towards increasing $\omega'_3$. If we decrease $g_{12}$ below zero and into negative values, the three-loop-type zero ZZ coupling emerges on the negative half-plane $\omega'_3 < 0$. If we further decrease $g_{12}$, the region again approaches the center $\omega'_3 = 0$ but now from the opposite direction. The wrap-around of the region is a consequence of the fact that the three-loop mechanisms in \figpanel{fig:ZZ_landscape_EC_mechanisms}{f} change their signs with the sign of $g_{12}$. 

\begin{figure*}
    \centering
    \includegraphics[width=\textwidth]{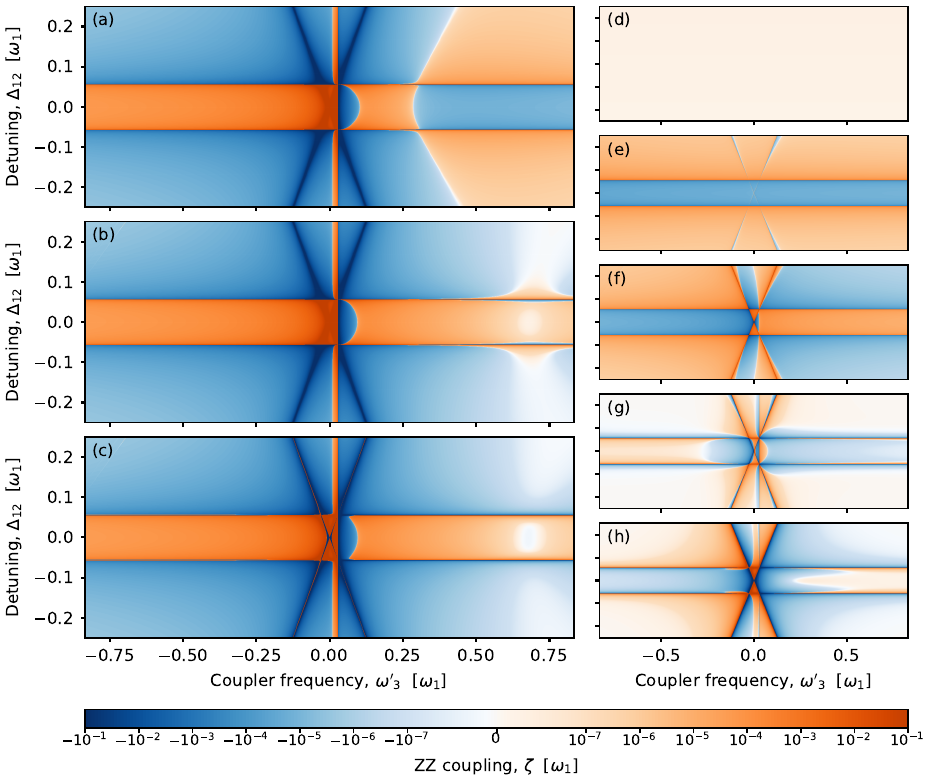}
    \caption{Addition of non-excitation-conserving and fifth-order contributions to the predictions in \figref{fig:ZZ_landscape_EC_mechanisms}. The predictions in (a-c) are constructed by progressively adding the second- to fifth-order non-excitation-conserving contributions in (d-g) and the fifth-order excitation-conserving contribution in (h) to the prediction from the excitation-conserving mechanisms in \figpanel{fig:ZZ_landscape_EC_mechanisms}{b}. (a) Addition up to third order of (d-e). (b) Addition up to fourth order of (d-f). (c) Addition up to fifth order of (d-h). (d) The second-order non-excitation-conserving contribution. (e) The third-order non-excitation-conserving contribution. (f) The fourth-order non-excitation-conserving contribution. (g) The fifth-order non-excitation-conserving contribution. (h) The fifth-order excitation-conserving contribution. The plots are generated with the same system parameters as in \figref{fig:ZZ_landscape_EC_mechanisms}. We remark that the plots are not generated from evaluating the diagram expansions up to fifth order but directly from evaluating the energy corrections given by the SW transformation up to fourth order in Eqs.~(\ref{eq:E2})--(\ref{eq:E4}) and in \eqref{eq:E5} for the fifth-order correction.}
    \label{fig:ZZ_landscape_NEC_mechanisms}
\end{figure*}

\subsection{Effect of non-excitation-conserving mechanisms and fifth-order corrections}
\label{sec:nec_corrections}

Here, we investigate how the non-excitation-conserving mechanisms and fifth-order corrections affect the predictions in \secref{sec:ec_predictions}. For this investigation, we again make good use of the diagram-expansion technique to probe the non-excitation-conserving mechanisms. We expect based on the estimates in \secref{sec:truncation_scheme} that the main contributions are from the second- and third-order diagram expansions, unless the fourth-order expansions include significantly more diagrams. To simplify the presentation, we only show diagrams including contributions to first order in $g_{ij} / \Sigma_{ij}$. By recalling the Hamiltonian graph in \figref{fig:Hamiltonian_graph}, the second- and third-order non-excitation-conserving diagram expansions are
\begin{align}
    E_{\Sigma, 000} &= \includegraphics[valign=c]{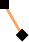} + \includegraphics[valign=c]{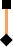} + \includegraphics[valign=c]{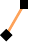} \, , \label{eq:E000+-} \\
    E_{\Sigma, 100} &= \includegraphics[valign=c]{O000_Lx.pdf} + \includegraphics[valign=c]{O000_xI.pdf} + \includegraphics[valign=c]{O000_xJ.pdf} + \includegraphics[valign=c]{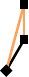} + \includegraphics[valign=c]{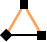} \, , \label{eq:E100+-} \\
    E_{\Sigma, 110} &= \includegraphics[valign=c]{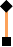} + \includegraphics[valign=c]{O000_Lx.pdf} + \includegraphics[valign=c]{O000_xI.pdf} + \includegraphics[valign=c]{O000_xJ.pdf} + \includegraphics[valign=c]{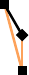} + \includegraphics[valign=c]{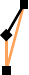} \nonumber \\
    &+ \includegraphics[valign=c]{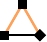} + \includegraphics[valign=c]{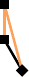} + \includegraphics[valign=c]{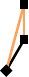} + \includegraphics[valign=c]{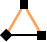} \, , \label{eq:E110+-}
\end{align}
where the orange solid lines are the non-excitation-conserving edges. We have here omitted the diagram expansion for $E_{\Sigma, 010}$ since it is equivalent to the one for $E_{\Sigma,100}$, similar to Eqs.~(\ref{eq:E010=})--(\ref{eq:E100=}). 

We give the fourth-order expansions in \appref{app:4th_order_non_EC_diagrams}. We find that the number of diagrams sharply increases from the third- to the fourth-order expansions. Considering only diagrams including contributions to first order in $g_{ij} / \Sigma_{ij}$, we find that the fourth-order expansions in Eqs.~(\ref{eq:E000+-4th})--(\ref{eq:E110+-4th}) involve 215 diagrams in total, which is an order of magnitude larger than the 23 diagrams in Eqs.~(\ref{eq:E000+-})--(\ref{eq:E110+-}) (including the equivalent diagrams for $E_{\Sigma,010}$). Thus, even if the individual fourth-order diagrams have a smaller contribution, the larger number of diagrams imply that they can amplify to have a total contribution comparable to the lower-order expansions. 

Figure~\ref{fig:ZZ_landscape_NEC_mechanisms} confirms that the fourth-order contributions are significant. The figure continues the gradual addition of contributions to the ZZ coupling that we began in \figref{fig:ZZ_landscape_EC_mechanisms}. In particular, the non-excitation conserving contributions from second to fifth order are given in \figpanels{fig:ZZ_landscape_NEC_mechanisms}{d}{g}, while \figpanel{fig:ZZ_landscape_NEC_mechanisms}{h} give the fifth-order excitation-conserving contribution. We note that the third-order contribution in \figpanel{fig:ZZ_landscape_NEC_mechanisms}{e} is comparable to the fourth-order contribution in \figpanel{fig:ZZ_landscape_NEC_mechanisms}{f}. The importance of the fourth-order contribution is clear when comparing its impact between the predictions in \figpanels{fig:ZZ_landscape_NEC_mechanisms}{a}{b}. In \figpanel{fig:ZZ_landscape_NEC_mechanisms}{a}, we have added the second- and third-order non-excitation-conserving contributions to the prediction in \figpanel{fig:ZZ_landscape_EC_mechanisms}{b}. We find that adding these contributions distinctly changes the right half-plane and removes the three-loop-type zero ZZ coupling. By then adding the fourth-order contribution in \figpanel{fig:ZZ_landscape_NEC_mechanisms}{b}, we recover the three-loop-type region. Compared to the prediction from the excitation-conserving mechanisms, the three-loop-type zero ZZ coupling is moved towards larger $\omega'_3$, and the ZZ coupling strength is overall shifted towards more positive values.

Since the fourth-order contribution from non-excitation-conserving mechanisms is notable, we also consider the effects from both fifth-order non- and excitation-conserving mechanisms. We derive the fifth-order contributions in \appref{app:5th_order_SWT} from the Schrieffer-Wolff transformation. The contributions are shown in \figpanels{fig:ZZ_landscape_NEC_mechanisms}{g}{h}, where we mainly observe weak regions (white). Still, the contributions are noticeable in the horizontal band and we observe, e.g, a minor transformation of the three-loop-type zero ZZ coupling in \figpanel{fig:ZZ_landscape_NEC_mechanisms}{c}. Hence, adding the fifth-order contributions yields no other change than minor transformations of the weakest regions, meaning that the main effects are already captured in the fourth-order contributions. 

\begin{figure}
    \centering
    \includegraphics[width=\columnwidth]{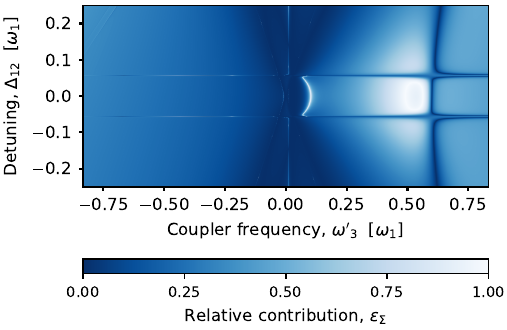}
    \caption{Relative ratio between the contributions from the excitation-conserving and non-excitation-conserving mechanisms up to fifth order. As given by \eqref{eq:contribution_ratio}, a ratio of 0 (blue) represents dominating excitation-conserving contributions. A ratio of 1 (white) represents dominating non-excitation-conserving contributions, while a ratio of 0.5 implies an equal contribution. The plot is generated with the same system parameters as in \figref{fig:ZZ_landscape_EC_mechanisms}.}
    \label{fig:nec_ratio}
\end{figure}

Having shown that the non-excitation-conserving contributions have a notable effect, we turn our focus towards their combined contribution. We compare the excitation-conserving and non-excitation-conserving contributions in \figref{fig:nec_ratio} by considering the relative contribution
\begin{equation}
    \varepsilon_\Sigma = \frac{\abs{\zeta_\Sigma}}{\abs{\zeta_\Sigma} + \abs{\zeta_\Delta}} .
    \label{eq:contribution_ratio}
\end{equation}
As expected, the non-excitation-conserving contribution is dominant around the regions predicted in \figref{fig:ZZ_landscape_EC_mechanisms} to have zero ZZ coupling. The center region around $\omega'_3 \approx 0$ is primarily dominated by the excitation-conserving contribution, while the regions for increasing $\abs{\omega'_3}$ have a larger ratio of non-excitation-conserving contribution. These ratios are expected since $\abs{\Delta_{ij}} \ll \Sigma_{ij}$ in the center region for all three combinations of $(i,j)$. We note that a typical value for the ratio in \figref{fig:nec_ratio} is around 0.25.

With a typical ratio around 0.25, we might expect new features to appear in \figpanel{fig:ZZ_landscape_NEC_mechanisms}{c} that are not present in \figpanel{fig:ZZ_landscape_EC_mechanisms}{b}. However, we do not observe any new larger features such as additional regions with zero ZZ coupling. Instead, including the non-excitation-conserving contribution primarily causes a transformation of already present features. We explain the lack of new features from observing that there are equivalences between the diagrams in the excitation-conserving and the non-excitation-conserving diagram expansions. To start from a concrete example, we enumerate the diagrams in the order they appear in Eqs.~(\ref{eq:E110=}) and (\ref{eq:E110+-}), and note diagram 21 in \eqref{eq:E110=} and diagram 8 in \eqref{eq:E110+-}. We deform the latter and evaluate the diagrams using \eqref{eq:tri} to explain how the two diagrams are equivalent:
\begin{align}
    \includegraphics[valign=c]{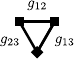} \ &\sim \ \raisebox{2pt}{\includegraphics[valign=c]{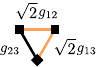}} \, , \\[1ex]
    2\frac{g_{12} g_{13} g_{23}}{\Delta_{13} \Delta_{23}} &\sim -4\frac{ g_{12}  g_{13} g_{23}}{\Sigma_{13} \Delta_{23}} + \mathcal{O}\mleft[ \mleft( \frac{\alpha_1}{\Sigma_{13}} \mright)^2 \mright] , \label{eq:equivalent_evaluation}
\end{align}
where we also have added the coupling strengths to the edges. The diagrams are equivalent in the sense that their evaluations are proportional under exchange of $\Delta_{ij} \leftrightarrow \Sigma_{ij}$ to first order in $\alpha_i / \Sigma_{ij}$. This equivalence holds for the majority of the three-loop diagrams, except for diagrams 7 and 10 in \eqref{eq:E110+-}. However, if $\Sigma_{12}$, $\Sigma_{13}$, and $\Sigma_{23}$ are comparable, which typically holds for transmons, diagrams 7 and 10 have equivalent excitation-conserving diagrams subject to exchanging $\Sigma_{13} \leftrightarrow \Sigma_{23}$.

We find similar equivalences for the second- and fourth-order diagram expansions. Only the fourth-order expansion includes non-excitation-conserving diagrams that do not have corresponding equivalent diagrams in the excitation-conserving expansions. The reason for this is that, e.g., the diagrams that transverse the excitation-conserving edges in the fourth-excitation subgraph include new edges with detunings different than the ones in the lowest three excitation subspaces. Hence, the non-equivalent diagrams in the fourth-excitation subgraph introduce new poles $\Delta_{ij} = \alpha'$ in the perturbative expansion, where $\alpha'$ is some combination of the anharmonicities not present in the lowest three excitation subspaces. The equivalent non-excitation-conserving diagrams have the important property that they preserve the pole structure of the excitation-conserving expansions. We note that the majority of the non-excitation-conserving diagrams have an equivalent excitation-conserving diagram.

The lack of new larger features is then explained by the observation that adding the non-excitation-conserving contributions primarily does not change the pole structure of the excitation-conserving expansions. Instead, adding the non-excitation-conserving contributions can be seen as further corrections of the excitation-conserving mechanisms. To see why the preserved pole structure does not add new larger features, we consider the curvature of the ZZ coupling $\partial_z \zeta$ with respect to any system parameter $z$. Inspired by \eqref{eq:equivalent_evaluation}, we consider the sum of two equivalent diagrams:  
\begin{equation}
    \zeta(z) = f(z) \mleft[ g_\Delta(z) + g_\Sigma(z) \mright],
\end{equation}
where $f(z)$ is the common factor between the diagrams, and $g_\Delta(z)$ and $g_\Sigma(z)$ represent the remainders. Using \eqref{eq:equivalent_evaluation} as a concrete example, then $z = \omega_3$, $f(\omega_3) = 2 g_{12} g_{13} g_{23}/\Delta_{23}$, $g_\Delta(\omega_3) = 1/ \Delta_{13}$, and $g_\Sigma(\omega_3) = -2/ \Sigma_{13}$. If $\abs{\partial_z g_\Delta(z)} \gg \abs{\partial_z g_\Sigma(z)}$ in a region around $z = z_0$, it holds that the curvature of the ZZ coupling is dominated by the excitation-conserving diagram around $z = z_0$:
\begin{equation}
    \partial_z \zeta(z) = \partial_z [f(z) g_\Delta(z)] + \partial_z f(z) g_\Sigma(z_0),
\end{equation}
where $g_\Sigma(z)$ is relative to $g_\Delta(z)$ constant in the region around $z = z_0$. Note that the condition $\abs{\partial_z g_\Delta(z)} \gg \abs{\partial_z g_\Sigma(z)}$ is strongly satisfied in the example of \eqref{eq:equivalent_evaluation}. Under this condition, the excitation-conserving expansion is a good predictor of the curvature, and by extension the qualitative features, of the ZZ coupling, while the non-excitation-conserving contributions are needed for quantitative predictions.

We conclude this section by summarizing the answers to the initial three questions:
\begin{itemize}
    \item Both level repulsion and the three-loop mechanism are relevant to predict the static ZZ coupling. Of these two mechanism types, there are 14 excitation-conserving instances that we coarse grain to five level repulsions, and a total three-loop mechanism within the mechanism picture. These are the primary mechanisms of the ZZ coupling.
    \item In general, no mechanism is dominant. The balance between the mechanisms depends on the energy-level configurations and coupling strengths. By balancing the  level repulsions and the three-loop mechanisms in all energy-level configurations, we find two types of zero ZZ coupling.
    \item The excitation-conserving contributions are sufficient to predict the qualitative features of the static ZZ coupling. The non-excitation-conserving contributions are required to give accurate quantitative predictions. 
\end{itemize}
We recall that these conclusions are a result of a perturbative treatment applied to an effective Hamiltonian. Consequently, the quantitative predictions are limited by the estimated truncation error of $2\pi \times \SI{100}{\kilo \hertz}$. As established in \secref{sec:truncation_scheme}, the predicted zero ZZ coupling regions are only reliable as long as the requirements for the intermediate-value theorem are fulfilled.
Opposite to the zero ZZ coupling predictions, the validity of the perturbation theory breaks down around its poles where we in fact predict regions of strong ZZ coupling. Hence in the non-perturbative regions around the poles, the perturbation theory is limited in its quantitative predictions of the largest ZZ coupling strengths. To complement our predictions for the static ZZ coupling, we turn our attention in the next section to numerically predicting the regions of zero and strong ZZ coupling. 

\section{Numerical predictions for the ZZ coupling}
\label{sec:numerical_predictions}
From the intuitive picture in \secref{sec:intuitive_picture} and the mechanism picture in \secref{sec:analytical_predictions}, we have predicted multiple parameter regions with zero or strong ZZ coupling for the three-transmon system. Some of these predictions are in regions where the quantitative accuracy of the used effective Hamiltonian and perturbation theory is insufficient. For example, the fourth-order perturbation theory in \secref{sec:ec_predictions} does not capture third- or higher-order level repulsions that may become relevant in non-perturbative regions. To complete the picture of the ZZ coupling beyond the identified primary mechanisms, we here remove previously used approximations and numerically predict the static ZZ coupling. We focus on the analytically predicted regions of zero or strong ZZ coupling and use exact diagonalization on the circuit Hamiltonian in \eqref{eq:circuit_Hamiltonian} to achieve arbitrary precision in the zero-ZZ-coupling regions. Pushing the predictions into the non-perturbative regions of strong ZZ coupling, it becomes a problem to identify which eigenenergies and eigenstates that form the ZZ coupling in \eqref{eq:ZZ}. We begin this section by defining this state-assignment problem and show how it can be solved by mapping it on the stable marriage problem \cite{galeCollegeAdmissionsStability1962}.

\subsection{The state-assignment problem}

To introduce the state-assignment problem, we recall for the ZZ coupling that we are interested in the eigenenergies of the computational states. Importantly, there is not a unique way to choose these computational states. In general, and in particular for transmons, we typically choose the ground and first-excited states due to better coherence properties. These states are well-defined in the case of uncoupled qubits and couplers. However, coupling the qubits and the couplers hybridizes the well-defined bare states, mixing them into eigenstates, i.e., dressed states, that are not localized in a particular qubit or coupler. For strong hybridization, it is not obvious to identify the states that can be considered to be the ground and first-excited states of the qubits. One option, and the one we will pursuit here, is to choose the dressed states that are the most similar to the computational bare states. We call the problem of assigning the dressed states to the bare states based on some similarity metric the state-assignment problem.

\begin{figure}
    \centering
    \hspace*{-10mm} 
    \includegraphics[width=0.9\columnwidth]{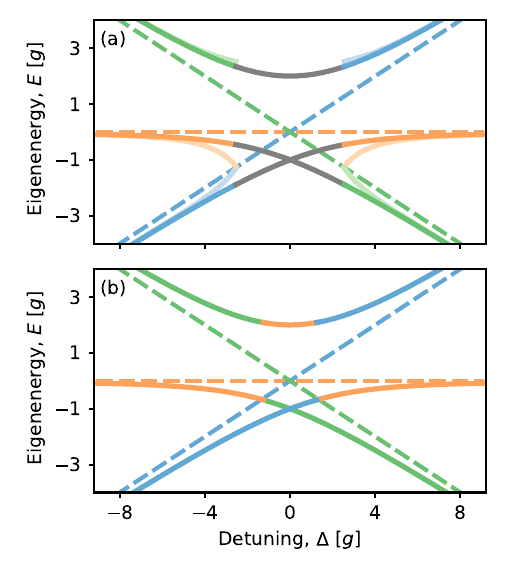}
    \caption{Energy spectrum of a three-level system as a function of the bare detuning in units of the coupling strength. The eigenenergies (solid lines) are colored after their assignment to the bare energies (dashed lines). In (a), the coloring follows the intuitive assignment based on the proximity to the bare energies outlined in the main text. Where the assignment is ambiguous, the eigenenergies are colored gray. The semi-transparent lines are the fourth-order approximations of the eigenenergies given by the SW transformation. The approximate energies are excluded where they strongly diverge from the eigenenergies. Note that the approximate energies diverge close to where the assignment becomes ambiguous. In (b), the coloring follows the state assignment given by the Gale--Shapley algorithm in \algref{alg:Gale-Shapley}. Here, the semi-transparent lines from the SW transformation have been removed for simplification.}
    \label{fig:state_assignment_problem}
\end{figure}

To illustrate, we consider a coupled three-level system, which is the smallest system where the state-assignment problem does not have a trivial solution. In particular, we consider a three-level system with the Hamiltonian
\begin{equation}
    H = \mqty( 
    -\Delta / 2 & g & g \\
    g & 0 & g \\
    g & g & \Delta / 2
    ),
    \label{eq:Hamiltonian_three_level_system}
\end{equation}
where $\Delta$ is the bare detuning between the lowest and highest energy levels, and $g$ is a coupling strength that we set to be the same between all pairs of states. We show in \figpanel{fig:state_assignment_problem}{a} the energy spectrum of \eqref{eq:Hamiltonian_three_level_system} near an avoided level crossing colored after an intuitive state assignment. Away from the avoided level crossing, the eigenenergies (solid lines) are pairwise in close proximity to the bare energies (dashed lines). Due to this proximity, we expect it to be reasonable to assign each eigenenergy to the closest bare energy. However, moving towards the avoided level crossing, the eigenenergies diverge from the bare energies and no longer form well-defined pairs. In this center region, it is not obvious how to consistently assign the dressed states to the bare states and a more rigorous method is needed. 

Note that we did not pay attention to the state-assignment problem with the analytical predictions in \secref{sec:analytical_predictions}. The reason behind this is that the problem primarily requires careful consideration in parameter regions similar to non-perturbative regions. To make this point explicit, \figpanel{fig:state_assignment_problem}{a} also shows perturbative approximations (semi-transparent lines) of the eigenenergies. The center region of the avoided level crossing is non-perturbative, as seen from the fact that the perturbative approximations diverge from the eigenenergies. It is in the same center region the state assignment is ambiguous. It is no coincidence that the non-perturbative region correlates with the difficulty of the state-assignment problem; both issues occur due to the hybridization of the mixed states. 

To conclude the introduction of the state-assignment problem, we state it more formally: given a set of $N$ bare states $\ket{i}$ and a set of $N$ eigenstates $\ket{I}$ that satisfy the eigenvalue problem $H \ket{I} = E \ket{I}$, find a bijective map $I \to i$ such that some similarity metric $S = S(I, i)$ is maximized for all $I$. For a Hermitian Hamiltonian $H$, there exists a unitary operator $U$ relating the bare states (lowercase letters) and the eigenstates (uppercase letters): $\ket{I} = U \ket{i}$.

\subsection{Mapping the state-assignment problem onto the stable marriage problem}

There are available solutions to the state-assignment problem in the literature. In particular, for the system of a transmon capacitively coupled to a driven resonator, two solutions \cite{shillitoDynamicsTransmonIonization2022a, gotoLabelingEigenstatesQubitcavity2024} have recently been proposed. In Ref.~\cite{shillitoDynamicsTransmonIonization2022a}, Shillito \emph{et al.} use a similarity metric based on a particular state overlap defined with creation and annihilation operators to recursively compute a state assignment. In Ref.~\cite{gotoLabelingEigenstatesQubitcavity2024}, Goto and Koshino replace the previous overlap and introduce a preprocessing step using a fix energy threshold. Here, we take a different approach and show that the state-assignment problem can be mapped onto the stable marriage problem \cite{galeCollegeAdmissionsStability1962}. This approach has the advantage that the solution to the latter is well studied and is guaranteed to have a stable solution. A detailed comparison of the different methods is beyond the scope of this paper, but is interesting for future work. Still, we note that the three methods differ with respect to the systems they are applied to and with regards to their implemented similarity metrics.

We rephrase the stable marriage problem from Ref.~\cite{galeCollegeAdmissionsStability1962}. The problem considers a community of $N$ Alices and $N$ Bobs. Each Alice and Bob has a ranked preference for the members in the other group. The problem is to marry all Alices and Bobs such that there are no Alice and Bob who prefer marrying each other instead of their current match. If there does not exist any such pair that prefer leaving their partners, the set of marriages is said to be stable.

To map the state-assignment problem to the stable marriage problem, we let the dressed states be the Alices and the bare states be the Bobs. Both sets of states are of equal size $N$. The missing requisite is then a ranked preference, i.e, a similarity metric, for each state in the two groups. We construct a preference matrix for the dressed states (Alices) of the bare states (Bobs) from the transformation rule $\ket{I} = U \ket{i}$. We take the absolute value of the overlap $\abs{ \braket{j}{I} } = \abs{\! \mel{j}{U}{i}}$ to be the preference for $\ket{I}$ to be assigned (married) to $\ket{j}$. The complementary preference matrix for $\ket{j}$ preferring $\ket{I}$ is then given by inverting the unitary transformation rule: $ \abs{ \braket{I}{j} } = \abs{\! \mel{I}{U^\dagger}{J} } $. The ranking is easily achieved by sorting, e.g., $\abs{\! \mel{j}{U}{i}}$ with respect to the index $j$. 

We note that the choice of the ranked preference is partially arbitrary in the sense that it needs to be chosen such that it is compatible with the observable under consideration. For example, the similarity metrics used in Refs.~\cite{shillitoDynamicsTransmonIonization2022a, gotoLabelingEigenstatesQubitcavity2024} can be viable options in the case of the driven system. For the case of the undriven three-transmon system and the ZZ coupling, we use the simpler preference $\abs{\! \mel{j}{U}{i}}$.

The solution to the stable marriage problem is given by the Gale--Shapley algorithm \cite{galeCollegeAdmissionsStability1962}. We reformulate the algorithm in \algref{alg:Gale-Shapley} as pseudocode in terms of the bare and dressed states. The Gale--Shapley algorithm is guaranteed to always find a stable solution within time $\mathcal{O}(N^2)$ \cite{galeCollegeAdmissionsStability1962}. We note that there in general exist more than one stable solution. \algref{alg:Gale-Shapley} gives the stable solution optimal for the dressed states, meaning that there does not exist any other stable solution where any dressed state is assigned to a bare state with higher preference. By exchanging the roles of the two sets of states, we can use \algref{alg:Gale-Shapley} to find the stable solution which is optimal for the bare states. The optimal solution for the bare states is not necessarily equal to the optimal solution for the dressed states. We prefer the optimal solution for the dressed states in the state-assignment problem since we prioritize that each dressed state is assigned to its most similar bare state. 

%%% BEGIN Gale-Shapley algorithm %%%%
\newcommand{\forcond}{$i=0$ \KwTo $n$}
\SetKwComment{Comment}{\# }{}%
\SetStartEndCondition{ }{}{}%
\SetKwProg{Fn}{def}{\string:}{}
\SetKwFunction{Range}{range}%%
\SetKw{KwTo}{in}\SetKwFor{For}{for}{\string:}{}%
\SetKwIF{If}{ElseIf}{Else}{if}{:}{elif}{else:}{}%
\SetKwFor{While}{while}{:}{fintq}%
\renewcommand{\forcond}{$i$ \KwTo\Range{$n$}}
\AlgoDontDisplayBlockMarkers\SetAlgoNoEnd\SetAlgoNoLine%
\DontPrintSemicolon%

\begin{algorithm}
\caption{The Gale--Shapley algorithm}\label{alg:Gale-Shapley}

\KwData{The ranked preference matrix $U$}
\KwResult{The assigned pairs of dressed and bare states} \;
Let every dressed and bare state be unassigned \;

\While{there are unassigned dressed states}{
    $d$ = Unassigned dressed state \;
    $b$ = Highest ranked bare state that $d$ has not proposed to \;
    \eIf{$b$ is unassigned}{
        Assign the pair $(d, b)$ \;
    }{
        \Comment*[h]{Some other $d'$ is assigned to $b$} \;
        \eIf{$b$ prefers $d$ to $d'$}{
            Assign $(d, b)$ \;
            Unassign $d'$ \;
        }{
            Keep $(d', b)$ assigned \;
        }
    }
}
\end{algorithm}
%%% END Gale-Shapley algorithm %%%%

To illustrate the use of the Gale--Shapley algorithm, we return to the energy spectrum of the three-level system in \figref{fig:state_assignment_problem}. The gray sections of the eigenenergies in the non-perturbative region in \figpanel{fig:state_assignment_problem}{a} can now be removed. We show the energy spectrum with assignments according to \algref{alg:Gale-Shapley} in \figpanel{fig:state_assignment_problem}{b}. Note that the assignments in the perturbative region are in agreement in \figpanels{fig:state_assignment_problem}{a}{b}.

\subsection{Numerical procedure}
\label{sec:numerical_procedure}

We implement the Gale--Shapley algorithm as a central element in our numerical procedure for computing the ZZ coupling. To go beyond the analytical description in \secref{sec:analytical_predictions}, we return to the more fundamental circuit Hamiltonian in  Eqs.~(\ref{eq:circuit_Hamiltonian})--(\ref{eq:interaction_Hamiltonian}). The numerical procedure is designed to perform an efficient and automatic exact diagonalization of the circuit Hamiltonian's low-energy subspace. The numerical efficiency is improved by projecting the Hilbert space into one with a limited number of excitations, while the automatic component is a direct result of the Gale--Shapley algorithm in \algref{alg:Gale-Shapley}. We outline the numerical procedure in this subsection and give further technical details in \appref{app:simulation_details}.

The numerical procedure has three main stages: (1) computing transmon eigenstates, (2) computing system eigenstates, and (3) state assignment. In stage (1), we consider the decoupled transmon Hamiltonians in \eqref{eq:transmon_Hamiltonian} separately. We represent the transmon Hamiltonians in the charge basis (see \secref{sec:circuit_hamiltonian}) and truncate each transmon subspace to $N$ states. The number of states $N$ is later determined from the numerical convergence of the computed ZZ coupling [see stage (2) for further outline]. We numerically diagonalize the decoupled transmon Hamiltonians to find the transmon eigenstates, which we refer to as the bare states of the system.

Having computed the bare states, we reintroduce in stage (2) the capacitive couplings to consider the fully coupled system. We represent the circuit Hamiltonian in the bare transmon basis as given by Eqs.~(\ref{eq:Mathieu_Hamiltonian_H0}) and (\ref{eq:Mathieu_Hamiltonian_V}). This representation results in a potentially large Hilbert space of dimension $N^3$. We reduce the dimensionality by projecting out the states with total excitation numbers larger than a maximum number of total excitations $M$, which is later determined together with $N$. The result is a reduction of the Hilbert space to $(M^3 + 6M^2 + 11M + 6)/6$ states, where the expression derives from the combinatorics of having a fixed number of excitations in three subspaces. The intuition behind this projection is drawn from the fact that the Hamiltonian graph in \figref{fig:Hamiltonian_graph} highlights that the matrix structure of the Hamiltonian follows the excitation subspaces. We note also from the perturbation theory in \secref{sec:analytical_predictions} that we expect the corrections to the eigenenergies to diminish with increasing total excitation numbers. We determine $M$ and $N$ by increasing the two cutoffs until the computed ZZ coupling converge to a value within the variations of the algorithmic precision given by the used diagonalization routine; see \appref{app:simulation_details} for details. The determined $M$ and $N$ are then used to numerically diagonalize the fully coupled system, from which we obtain the dressed transmon energies and states.

In stage (3), we use the Gale--Shapley algorithm to assign computational states from the dressed states. The corresponding dressed energies are the numerical eigenenergies we use to compute the ZZ coupling. The automatic assignment removes the otherwise tedious and error-prone task of manually assigning the computational states. Note that a new state assignment is computed for every instance of system parameters. 

\subsection{Predictions from exact diagonalization}
\label{sec:predictions_exact_diagonalization}

\begin{figure*}
    \centering
    \includegraphics[width=0.98\textwidth]{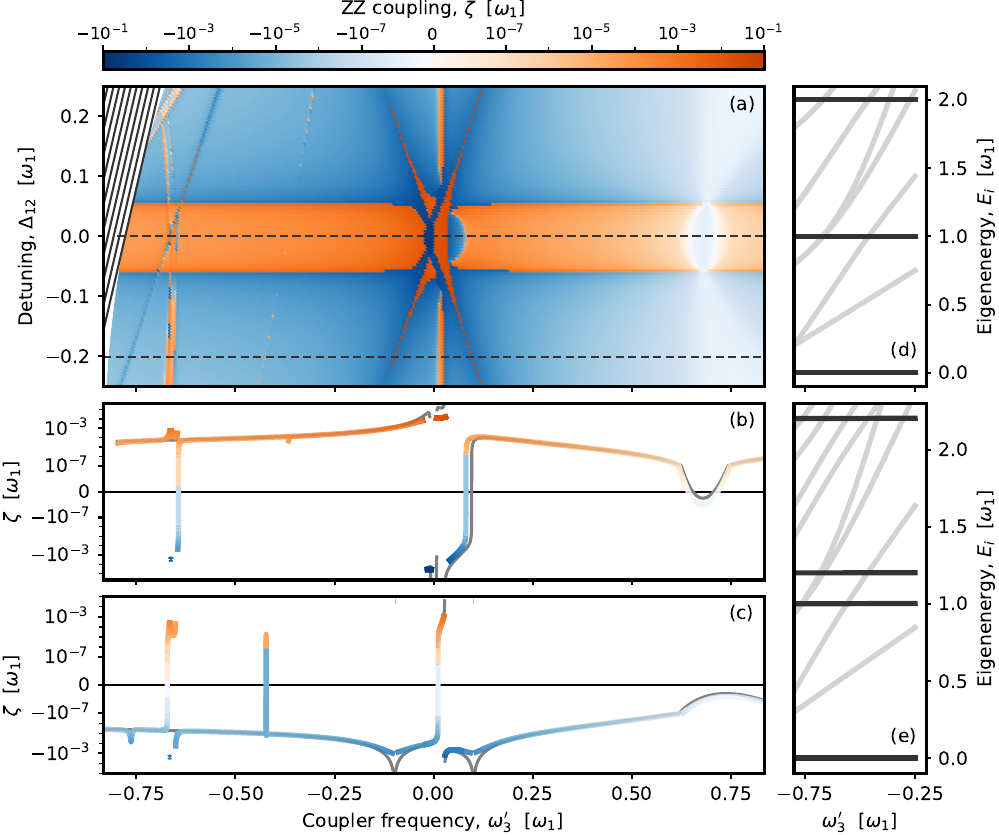}
    \caption{Numerical predictions for the static ZZ coupling from exact diagonalization of the circuit Hamiltonian. (a) The ZZ coupling strength as function of the bare coupler frequency $\omega'_3 = \omega_3 - (\omega_1 + \omega_2)/2$ and the bare qubit detuning $\Delta_{12} = \omega_1 - \omega_2$. All quantities are in units of $\omega_1$. Panel (a) includes two cross sections (dashed black lines) at $\Delta_{12} = 0$ and $\Delta_{12} = -0.2$ that go through the different regions of strong (dark orange or blue) and zero ZZ coupling (white).  These regions are clustered around: $\omega'_3 \approx -0.5$, $\omega'_3 \approx 0$, and $\omega'_3 \approx 0.7$. Note the excluded region (upper left corner; solid black lines) due to that all bare coupler frequencies are not achievable for the used system parameters. (b) The cross section at $\Delta_{12} = 0$ in (a). (c) The cross section at $\Delta_{12} = -0.2$ in (a). In (b-c), the cross section is colored identically to panel (a) following a logarithmic scale for $\abs{\zeta} > 10^{-7}$ and a linear scale for $\abs{\zeta} \leq 10^{-7}$. Note that the transition to the linear scale is the cause of the minor kinks in the curves around $\omega'_3 \approx 0.7$. The gray curves in panels (b-c) are the fifth-order analytical predictions given by the SW transformation, and the black horizontal line marks zero ZZ coupling. (d) For the cross section at $\Delta_{12} = 0$, the energy spectrum of the computational states (black) and states including coupler excitations (gray) that (anti-) cross the computational energy levels. The energy spectrum is given for the region around $\omega'_3 \approx -0.5$. (e) Same as panel (d) but for the cross section at $\Delta_{12} = -0.2$. Panels (d-e) highlight that the changes in the ZZ coupling around $\omega'_3 \approx -0.5$ occur due to anti-crossings with coupler energy levels. The ZZ coupling and the energy spectra are computed for fix charge energies $E_C^{i}$ and varying mutual charge energies $E_C^{ij}$. The $E_C^{i}$ are fixed such that the anharmonicities $\alpha_i = 225 / 4 \times 10^{-3} $ for $\omega_1 = \omega_2 = \omega_3$. The $E_C^{ij}$ are varied to fix the coupling strengths: $g_{13} = g_{23} = 75/4 \times 10^{-3}$ and $g_{12} = g_{13} / 30 $. The $\omega'_3$ and $\Delta_{12}$ are varied as a result of varying the Josephson energies $E_J^{(i)}$.}
    % Comment on what the colors represent in the different panels
    \label{fig:ZZ_landscape_numerical}
\end{figure*}

We here finalize our predictions of the ZZ coupling in the three-transmon system. The numerical predictions are shown in \figref{fig:ZZ_landscape_numerical}. In \figpanel{fig:ZZ_landscape_numerical}{a}, we give the ZZ coupling in the 24 energy-level configurations from \secref{sec:configurations}, making it comparable to the analytical predictions in, e.g., \figref{fig:ZZ_landscape_EC_mechanisms}. The black dashed lines in \figpanel{fig:ZZ_landscape_numerical}{a} represent two cross sections that cut through the predicted regions of zero and strong ZZ coupling. We use these cross sections to inspect the details of the ZZ coupling in \figpanels{fig:ZZ_landscape_numerical}{b}{c}, which also include the corresponding predicted ZZ coupling from the SW transformation (gray curves).

Having already computed Figs.~\ref{fig:ZZ_landscape_EC_mechanisms} and \ref{fig:ZZ_landscape_NEC_mechanisms}, it is intricate to generate \figpanel{fig:ZZ_landscape_numerical}{a} such that it is a precise comparison. The cause of this issue is that the effective and circuit Hamiltonian are parameterized differently. The effective Hamiltonian directly uses the bare system parameters while the circuit Hamiltonian is parameterized with the Josephson and charging energies. Hence in the numerical predictions, the bare system parameters are not directly accessible; they are computed from the circuit parameters making it difficult to generate \figref{fig:ZZ_landscape_numerical} with the exact same bare parameters as in Figs.~\ref{fig:ZZ_landscape_EC_mechanisms} and \ref{fig:ZZ_landscape_NEC_mechanisms}. 

To make the numerical bare parameters as similar as possible to the analytical bare parameters we proceed as follows: we first fix $E_J^{(1)}$ to set the energy scale $\omega_1 = 1$. We vary $E_J^{(2)}$ and $E_J^{(3)}$ to achieve the variations in $\omega'_3$ and $\Delta_{12}$. Then for $E_J^{(1)} = E_J^{(2)} = E_J^{(3)}$, i.e., at the center of \figpanel{fig:ZZ_landscape_numerical}{a}, we fix $E_C^{(i)}$ to obtain the same bare anharmonicities as in the analytical predictions. Note that the bare anharmonicities depend on the Josephson energies and thus deviate from the ones in the analytical predictions during the variations of the Josephson energies. These deviations are in general small in the transmon regime since $\alpha_i \approx - E_C^{(i)}$ to first order in the ratio $E_C^{(i)} / E_J^{(i)} \ll 1$~\cite{kochChargeinsensitiveQubitDesign2007}. For $E_C^{(ij)}$, we vary them with $E_J^{(i)}$ such that the coupling strengths $g_{ij}$ [recall \eqref{eq:effective_V}] are fixed identically to the analytical case. We refer the reader to \appref{app:simulation_details} for further details.

If needed, the parameter deviations for the bare anharmonicities can be mitigated by using the numerically computed bare parameters as the parameters of the effective model. Indeed, for the cross sections in \figpanels{fig:ZZ_landscape_numerical}{b}{c}, where we more thoroughly want to compare to the analytical predictions, we input the numerical bare parameters in the SW transformation. However, as already stated above, the parameter deviations are in general small in \figpanel{fig:ZZ_landscape_numerical}{a}. For the purpose of \figpanel{fig:ZZ_landscape_numerical}{a}, we find the minor system-parameter deviations to be acceptable.

Taking a bird's-eye view of \figpanel{fig:ZZ_landscape_numerical}{a}, we observe the same main features as in the analytical predictions in Figs.~\ref{fig:ZZ_landscape_EC_mechanisms} and \ref{fig:ZZ_landscape_NEC_mechanisms}. In particular, we again find the predicted two types of regions with zero ZZ coupling and the regions of strong ZZ coupling (cf.~\secref{sec:ec_predictions}). The agreement between the numerical and analytical predictions is visually excellent with respect to the placement in frequency space of the zero and strong ZZ coupling regions. We note that the regions are partially transformed when comparing \figpanel{fig:ZZ_landscape_numerical}{a} and \figpanel{fig:ZZ_landscape_NEC_mechanisms}{c}. For example, the three-loop-type zero ZZ coupling in \figpanel{fig:ZZ_landscape_numerical}{a} near $\omega'_3 \approx 0.7$ is more distinct than its counterpart in \figpanel{fig:ZZ_landscape_NEC_mechanisms}{c}. That this feature is partially transformed is unsurprising since the region supports a ZZ coupling with strength close to the analytical truncation error.

Beyond the similarities, the numerical predictions reveal two new features that are not shown by the analytics. Directing first our attention to the poles $\Delta_{13}, \Delta_{23} = 0$ (the X-shaped region around $\omega'_3 \approx 0$), we observe regions of strong ZZ coupling with opposite signs compared to, e.g, \figpanel{fig:ZZ_landscape_NEC_mechanisms}{c}. Being at the center of poles, it is expected that the perturbative approach does not capture these features. Carefully viewing the cross sections in \figpanels{fig:ZZ_landscape_numerical}{b}{c} shows that there are no continuous transitions to the regions with opposite signs. The discontinuous transitions, and in extension the opposite signs, are a result of changes in the state assignments (recall \figref{fig:state_assignment_problem} for discrete shifts due to the state assignment).

Continuing to the second new feature, we find around $\omega'_3 \approx -0.5$ several new line-shaped regions. In particular, we note the vertical and mainly positive (orange) regions between $\omega'_3 = -0.75$ and $\omega'_3 = -0.50$. Since the ZZ coupling partially changes signs in this region, there are potential subregions of zero ZZ coupling. Inspecting the cross section in \figpanel{fig:ZZ_landscape_numerical}{c} confirms that there are subregions of zero ZZ coupling outside the horizontal band ($\abs{\Delta_{12}} > \abs{\alpha_1}, \abs{\alpha_2})$. 

Figures~\hyperref[fig:ZZ_landscape_numerical]{\ref*{fig:ZZ_landscape_numerical}(d)-(e)} show that the line-shaped regions coincide with potential avoided level crossings. These crossings are between levels assigned to the computational states (black lines) and states including coupler excitations (gray lines). For instance, comparing \figpanel{fig:ZZ_landscape_numerical}{c} and \figpanel{fig:ZZ_landscape_numerical}{e}, we note that there are more line intersections in the energy spectrum than spiked regions in the cross section. Closer inspection shows that only the intersections with proper avoided level crossings have a corresponding spike in the cross section [not visible in \figpanel{fig:ZZ_landscape_numerical}{e}]. The other actual level crossings occur between levels assigned to states with different total-excitation parity (even or odd). Between these states, the coupling strengths are suppressed in line with the parity symmetry primarily manifest in the effective Hamiltonian instead of the circuit Hamiltonian (see \secref{sec:Hamiltonian_graph}). From these observations, we conclude that the second new feature is caused by avoided level crossings between the computational energy levels and levels with coupler excitations where both levels have the same total-excitation parity. This new mechanism gives a third type of zero ZZ coupling.

Furthermore, we explain why the line-shaped regions are not predicted by the perturbative approach in \secref{sec:nec_corrections} (or by the intuitive picture in \secref{sec:intuitive_picture_predictions}) by noting that the regions appear for low coupler frequencies. For example, assuming $\omega_1 = \omega_2$ gives a coupler frequency equivalent to $\omega_3 < \omega_1 / 2$. With these sufficiently low coupler frequencies, the excitation subgraphs in \figref{fig:Hamiltonian_graph} are no longer separated [recall assumption (3) for the intuitive picture in \secref{sec:nec_corrections}], resulting in increased level interactions between states in different excitation subgraphs. To capture these interactions, higher than the used fifth-order perturbation theory is needed. For instance, the third-order level repulsion between the levels assigned to $\ket{100}$ and $\ket{003}$ is first captured by a sixth-order SW transformation. It is possible that the line-shaped regions are present in higher-order perturbation theory, but such an investigation is beyond the scope of this paper.

We now turn our focus to the analytically predicted regions of zero and strong ZZ coupling. Figures~\hyperref[fig:ZZ_landscape_numerical]{\ref*{fig:ZZ_landscape_numerical}(b)-(c)} confirm the presence of the zero ZZ coupling of level-repulsion type around $\omega'_3 \approx 0$ and the three-loop-type zero ZZ coupling in the horizontal band ($\abs{ \Delta_{12} } < \abs{\alpha_1}, \abs{\alpha_2}$). Recall that the quantitative accuracy of the effective Hamiltonian and perturbation theory was not sufficient to predict with certainty that the region outside the horizontal band ($\abs{ \Delta_{12} } > \abs{\alpha_1}, \abs{\alpha_2}$) contained a zero ZZ coupling of three-loop type. From the numerics, we do not find any system parameters that give a zero ZZ coupling for the considered regions. This prediction is illustrated in \figpanel{fig:ZZ_landscape_numerical}{c} around $\omega'_3 \approx 0.7$, where the line given by numerics remains faint blue, i.e., always takes negative values and never intersects the zero axis (black line). Still, the faint blue color implies an exceedingly weak ZZ coupling in this region. We predict it to be possible to construct a ZZ coupling in this region on the scale of $\zeta \sim 2 \pi \times (1 \text{--} 10) \, \SI{}{\hertz} $ for conventional system parameters (e.g., given in \secref{sec:truncation_scheme}). We emphasize that this scale is several orders of magnitude lower than the estimated maximum average ZZ coupling of $\SI{100}{\kilo \hertz}$ needed for high-fidelity two-qubit gates in \secref{sec:estimate_ZZ}.

For the regions of strong ZZ coupling close to the poles of the perturbation theory, the general trend is that the analytical predictions overestimate (the absolute value of) the ZZ coupling. The trend is clear in both \figpanels{fig:ZZ_landscape_numerical}{b}{c}, where the gray lines given by the Schrieffer--Wolff transformation depart from the colored lines given by numerics. We find that the numerical and analytical predictions agree concerning the parameter region with the strongest ZZ coupling; this region is within the horizontal band ($\abs{ \Delta_{12} } < \abs{\alpha_1}, \abs{\alpha_2}$) while approaching $\omega'_3 \to 0^-$. For the system parameters used in \figref{fig:ZZ_landscape_numerical}, the region of strongest ZZ coupling exceeds $10^{-2} \omega_1$. Recalling \secref{sec:CZ}, this ZZ coupling is sufficient to implement a CZ gate with a gate time of $\SI{100}{\nano \second}$ given conventional system parameters.

To conclude, we further consider the implications for experimental implementations of adiabatic CZ gates. Figure \figpanelNoPrefix{fig:ZZ_landscape_numerical}{a} highlights that there are several frequency regions where points of zero ZZ coupling are continuously connected to points of strong ZZ coupling. For example, control of the coupler frequency is enough to follow the given cross sections that connect all three types of zero ZZ coupling with strong ZZ coupling points. These results show the existence of alternative frequency regions, but also pose the question of which of these regions that are ideal for implementation. There are several aspects to take into consideration and we give here a few examples. Most obviously, the gate time is affected by the maximally achievable ZZ coupling strength. We note that the difficulty may be different to adiabatically reach the different strong points, for example due to the number of energy levels involved in the different avoided level crossings. Regarding minimizing the gate times, we also emphasize that \figpanel{fig:ZZ_landscape_numerical}{a} gives the static ZZ coupling from states assigned by the Gale--Shapley algorithm. Under careful adiabatic control, it is possible that the strong ZZ coupling regions can be extended further than the state assignment suggests, giving even stronger ZZ coupling strengths.

In addition, the choice of frequency region affects other central properties if the three-transmon system is to be used as a component in a quantum processor. For example, the qubit detuning is both a variable in the charge-drive crosstalk and in the extent frequency crowding is a problem \cite{osmanMitigationFrequencyCollisions2023}. The coupler frequency can affect the coherence properties of the system; a low coupler frequency can make the coupler, and in extension the whole system, more susceptible to charge noise. The predictions based on the static analysis in this paper give the different frequency regions that are exciting options for the adiabatic CZ gate. To fully understand which of these regions that has the optimum properties for implementing an adiabatic CZ gate, we need to complement the static analysis with a thorough analysis of the dynamical properties.

\section{Conclusion and outlook}
\label{sec:Conclusions}

We have introduced a theoretical framework to comprehensively explain the emergence of static ZZ couplings between superconducting qubits. The resulting explanation was developed in three linked pictures that each emphasize different aspects of the origin of the ZZ coupling. The first picture demonstrates the emergence of ZZ coupling as a result of energy-level repulsions and their configurations. The second picture gives a perturbative view and refines the previous picture by dividing the ZZ coupling into its primary underlying mechanisms. The third picture removes all simplifications and approaches the problem numerically. Individually, the three complementary descriptions all show that level repulsion is one of the primary mechanisms producing the ZZ coupling. Through the second picture, we were also able to distinguish a three-loop mechanism from the level repulsions and showed that it is a second primary mechanism underlying the ZZ coupling.

We applied our framework to the setup of two fixed-frequency transmon qubits connected by a flux-tunable transmon coupler, which is the most common architecture in experiments with superconducting qubits. We found that this setup, with only weakly negative anharmonicities, supports a large parameter space that is split into 24 energy-level configurations. By exhausting every configuration, we predicted all parameter regions with weak and strong ZZ coupling that the primary mechanisms can create. Regions with weak ZZ coupling enable the operation of high-fidelity quantum gates and algorithms by eliminating a type of coherent error pervasive in quantum computers, while regions with strong ZZ coupling can be harnessed for implementing fast (and thus high-fidelity) CPHASE or CZ gates. From our understanding of the mechanisms, we could show both that the regions of weak ZZ coupling can be classified into three types, all accessible by current technology without any major redesign needed, and that there are no other regions beyond these where the ZZ coupling is zero.

For two types of weak-ZZ regions, we gave practical guidelines in \secref{sec:ZZ_coupling_control} on how to engineer the ZZ coupling in experiments. This control of the ZZ coupling is enabled by the understanding that the two types of regions are directly linked to either level repulsions from the coupler or the three-loop mechanisms. These mechanisms can be regulated through the coupler's anharmonicity and the direct qubit-qubit coupling, respectively.

The results we found for superconducting qubits in general, and the three-transmon setup in particular, were enabled by improvements of analytical and numerical methods. Our improvements of analytical methods for handling ZZ couplings mainly consist of a diagrammatic technique for perturbation theory using the Schrieffer-Wolff transformation. The technique is closely connected to a graph representation of the Hamiltonian describing the system at hand. The diagrams in the technique enable efficient bookkeeping of the many terms at higher perturbative orders, but also greatly enhance our understanding of the mechanisms responsible for the ZZ coupling by showing which processes contribute to it. Our numerical computations were aided by the application of the Gale--Shapley algorithm for stable matching to the problem of assigning eigenstates of our coupled three-transmon system to the bare states of the uncoupled system. This algorithm is crucial in non-perturbative regions where the system is strongly hybridized.

The implications of our results are manifold. The most obvious is the potential for using the multiple regions of zero ZZ coupling we found as operating points to increase the fidelity of both single- and two-qubit gates in all kinds of transmon-based superconducting quantum computers. Recent experimental implementations~\cite{liTunableCouplerRealizing2020a, stehlikTunableCouplingArchitecture2021a, sungRealizationHighFidelityCZ2021} have confirmed the existence and explored specific instances of these regions, e.g., for smaller detunings than anharmonicities, but we find the other regions to currently be unutilized. We expect that our results will expedite the experimental exploration of parameter regions with zero ZZ coupling. Similarly, we identified parameter regions for our three-transmon system where points with zero ZZ coupling were close to points with strong ZZ coupling. Although our results for the ZZ coupling here are only for static system configurations, it is expected that adiabatically tuning between such points would enable fast CZ and CPHASE gates after finding optimal parameters and pulse shapes.

Beyond individual single- and two-qubit gates, the improved understanding of the ZZ coupling in the multitude of parameter regions will affect the design of large-scale quantum processors. First of all, an architecture needs to be found where the qubit type, frequencies, and coupling strengths are allocated so that nearest-neighbor pairs of qubits can exploit the parameter regions for high-fidelity gates. Then, it will be necessary to consider whether ZZ couplings, and their related counterparts ZZZ and even higher-order Z$^\text{n}$ couplings, occur over a longer range than nearest neighbors in such a setup. The analytical and numerical methods with the picture of mechanisms introduced in this work can immediately be applied to Z$^\text{n}$ couplings in larger systems; the analytical methods can provide insights at a scale where full numerical calculations become too expensive. For example, a similar mechanism picture can be created for ZZZ couplings by extending the analysis to the third-excitation subgraph (recall the Hamiltonian graph in \figref{fig:Hamiltonian_graph}). Another tantalizing idea for design at a grander scale is to use other regions of the parameter space for ZZ coupling to implement quantum simulations or digital-analog quantum computing \cite{parra-rodriguezDigitalanalogQuantumComputation2020, garcia-de-andoinDigitalanalogQuantumComputation2024}, where one makes use of tunable ZZ interactions.

There are various possible directions for future research. For transmon setups, one degree of freedom that could be added to the description is the offset charges, which was neglected here since their impact on the lowest energies are exponentially suppressed. Still, they could influence static and dynamic properties, especially for adiabatic CZ and CPHASE gates. Similarly, one could also incorporate higher Josephson harmonics, which have been shown to affect properties of state-of-the-art superconducting qubits~\cite{Willsch2024}. Beyond transmons, which were the example qubits in this work, other types of superconducting qubits, e.g., fluxonium, or even semiconductor qubits, could be studied in detail. For fluxonium, the same mechanism picture is applicable but in other energy-level configurations (recall Section~\ref{sec:configurations} and \ref{sec:mechanism_correlations}). Hence, the framework introduced in this work paves the way to also understand the ZZ coupling in these other types of qubits, and its methods will facilitate the study of Z$^\text{n}$ couplings in quantum processors beyond a few qubits. The methods are easy to implement, are analytically and numerically efficient, and offer clear physical interpretations. We therefore foresee our framework becoming a prominent component in understanding and using ZZ to Z$^\text{n}$ couplings in available and future quantum computers. 

\begin{acknowledgments}

We thank Anuj Aggarwal, Jens Koch, Th\'{e}o S\'{e}pulcre, and Ariadna Soro for useful discussions. We acknowledge support from the Knut and Alice Wallenberg Foundation through the Wallenberg Centre for Quantum Technology (WACQT) and from the Horizon Europe programme HORIZON-CL4-2022-QUANTUM-01-SGA via the project 101113946 OpenSuperQPlus100. The computations were enabled by resources provided by the National Academic Infrastructure for Supercomputing in Sweden (NAISS), partially funded by the Swedish Research Council through grant agreement number 2022-06725. AFK is also supported by the Swedish Research Council (grant number 2019-03696) and the Swedish Foundation for Strategic Research (grant numbers FFL21-0279 and FUS21-0063).

\end{acknowledgments}

\appendix

\section{Normal ordering of the transmon Hamiltonian}
\label{app:normal_ordering}

In \secref{sec:anharmonic_oscillator}, we considered the anharmonic-oscillator approximation of the transmon Hamiltonian. In particular, we defined the creation- and annihilation-like operators $\Hat{a}_i$ and $\Hat{a}_i^\dagger$, which were used in the normal ordering  in \eqref{eq:normal_Hamiltonian}. In this appendix, we show how to obtain the normal ordering. Our motivation for normal ordering is to ensure that the anharmonic approximation is as close as possible to the initial transmon Hamiltonian. In the following, we simplify the notation from the main text by suppressing the indices, i.e., $\Hat{a}_i \to \Hat{a}$, that here do not serve any useful purpose. 

We let $\Hat{\phi}$ and $\Hat{n} \equiv -\i \partial_\phi$ be operators on the space of $2\pi$-periodic and square-integrable states, $L^2[-\pi, \pi)$. These operators, combined with Eqs.~(\ref{eq:a}) and (\ref{eq:a_dagger}), give the creation- and annihilation-like operators
\begin{align}
    \Hat{a} &= \frac{1}{\sqrt{2 \lambda}} \mleft( \Hat{\phi} + \lambda  \partial_\phi \mright), \\
    \Hat{a}^\dagger &= \frac{1}{\sqrt{2 \lambda}} \mleft( \Hat{\phi} - \lambda  \partial_\phi \mright), 
\end{align}
which fulfill the canonical commutation relation $[\Hat{a}, \Hat{a}^\dagger] f(\phi) = f(\phi)$ when acting from the left on any state $f(\phi) \in L^2[-\pi, \pi)$. We caution that $a$ and $a^\dagger$ are not proper operators in $L^2[-\pi, \pi)$, which is a consequence of the fact that $\Hat{\phi}: f(\phi) \to \phi f(\phi) \notin L^2[-\pi, \pi)$, which in turn is due to that $\phi$ is not a $2 \pi$-periodic function. Consequently, $\Hat{\phi}$, and extension $a$ and $a^\dagger$, are only operators in $L^2[-\pi, \pi)$ in terms of arguments to periodic functions, e.g., $\cos{\Hat{\phi}}$.

We recall the transmon Hamiltonian in \eqref{eq:transmon_Hamiltonian} to realize that the main obstacle is the normal ordering of $\cos{\Hat{\phi}}$. On the other hand, the normal ordering of $\Hat{n}^2$ is straightforward:
\begin{equation}
    \Hat{n}^2 = - \frac{1}{2 \lambda} \mleft[\Hat{a}^2 - 2 \Hat{a}^\dagger \Hat{a} + \mleft(\Hat{a}^{\dagger} \mright)^2 \mright] ,
    \label{eq:n_square_normal}
\end{equation}
where we have neglected constant terms that are irrelevant for the Hamiltonian. Unless otherwise stated, we neglect all constant terms from here on.

To normal order the cosine term, we use the fact that a function on operators is defined from its power series:
\begin{equation}
    \cos{ \mleft[ \sqrt{ \frac{\lambda}{2} } \mleft( \Hat{a} + \Hat{a}^\dagger \mright) \mright] } \equiv \sum_{n=0}^\infty \frac{(- \lambda / 2)^n}{(2n)!} \mleft(\Hat{a} + \Hat{a}^\dagger \mright)^{2n} .
\end{equation}
As such, we are interested in the normal ordering of the binomial $\mleft(\Hat{a} + \Hat{a}^\dagger \mright)^{2n}$ for $n \in \mathbb{Z}^+$, where $\mathbb{Z}^+$ is the natural numbers including zero. The normal ordering of the binomial is~\cite{mikhailovOrderingBosonOperator1983a}
\begin{equation}
    \mleft(\Hat{a} + \Hat{a}^\dagger \mright)^{2n} = \sum_{k=0}^n \sum_{i=0}^{2(n-k)} \frac{(2n)! (\Hat{a}^\dagger)^i \Hat{a}^{2(n-k)-i}}{2^k k! i! [2(n-k)-i]!} .
    \label{eq:operator_binomial}
\end{equation}
After some simplifications, we obtain a normal-ordered series expansion of the cosine term
\begin{multline}
    \cos{ \left[ \sqrt{ \frac{\lambda}{2} } \left( \Hat{a} + \Hat{a}^\dagger \right) \right] } = \\ \sum_{n=0}^\infty \sum_{k=0}^n \frac{(-\lambda)^n}{2^{n+k} k!} \sum_{i=0}^{2(n-k)} \frac{ (\Hat{a}^\dagger)^i \Hat{a}^{2(n-k)-i}}{i! [2(n-k)-i]!}.
    \label{eq:cos_normal_start}
\end{multline}
Note that there are multiple terms including common factors in operators $(\Hat{a}^\dagger)^i \Hat{a}^{2(n-k)-i}$. 

To collect terms with common factors, we perform a resummation of the series in \eqref{eq:cos_normal_start}. The resummation is achieved by identifying suitable changes of indices. To remove the bound $k \leq n$, we define new indices: $n+k=p$ and $n-k=q$ such that $p,q \in \mathbb{Z}^+$ and $q \leq p$, which yields
\begin{multline}
    \cos{ \left[ \sqrt{ \frac{\lambda}{2} } \left( \Hat{a} + \Hat{a}^\dagger \right) \right] } = \\ \sum_{\substack{p,q=0 \\ q \leq p }}^\infty \frac{(-\lambda)^{\frac{p-q}{2}}}{4^{\frac{p-q}{2}} \mleft( \frac{p-q}{2} \mright)!} \frac{(-\lambda)^q}{2^q} \sum_{i=0}^{2q} \frac{ (\Hat{a}^\dagger)^i \Hat{a}^{2q-i}}{i! (2q-i)!} .
\end{multline}
Here, we have factored out the first factor to help identify that it corresponds to the power series of an exponential function. From this observation, we change indices to $k = (p-q)/2$, to make the resummation of the exponential function evident:
\begin{multline}
    \cos{ \left[ \sqrt{ \frac{\lambda}{2} } \left( \Hat{a} + \Hat{a}^\dagger \right) \right] } = \\ \sum_{q=0}^\infty \underbrace{  \left( \sum_{k=0}^\infty \frac{(-\lambda)^k}{4^k k!} \right)}_{\exp{(-\lambda/4)}} \frac{(-\lambda)^q}{2^q} \sum_{i=0}^{2q} \frac{ (\Hat{a}^\dagger)^i \Hat{a}^{2q-i}}{i! (2q-i)!}.
    \label{eq:cos_normal_mid}
\end{multline}
We can cosmetically improve \eqref{eq:cos_normal_mid} by changing the indices to $m = i$ and $n = 2q - i$ (n.b.~not the same $n$ as above) such that $m + n = 2q$ is even:
\begin{multline}
    \cos{ \left[ \sqrt{ \frac{\lambda}{2} } \left( \Hat{a} + \Hat{a}^\dagger \right) \right] } = \\ \e^{-\lambda/4} \sum_{\substack{m,n\geq 0 \\ m+n ~\mathrm{ even}}}^\infty \left( -\frac{\lambda}{2} \right)^{\frac{m+n}{2}} \frac{ (\Hat{a}^\dagger)^m \Hat{a}^n}{m! n!} .
    \label{eq:cos_normal_end}
\end{multline}
We note that $m=n=0$ gives a constant term, which we neglect.

Having obtained \eqref{eq:cos_normal_end}, combining it with \eqref{eq:n_square_normal} gives the normal ordering of the transmon Hamiltonian. Still, we note that $\lambda$ is a free parameter that needs additional conditions to be constrained. We impose the canonical condition (cf.~the standard treatment of the harmonic oscillator in, e.g., Ref.~\cite{sakuraiModernQuantumMechanics2020}) that terms proportional to $\Hat{a}^2$ and $(\Hat{a}^\dagger)^2$ cancel in the Hamiltonian. Using \eqref{eq:n_square_normal} and \eqref{eq:cos_normal_end} for $m = 2$ and $n = 0$ then gives
\begin{equation}
    \frac{E_C}{E_J} = \frac{1}{2} 
 \left( \frac{\lambda}{2} \right)^2 \e^{-\lambda/4} .
\label{eq:lambda_condition}
\end{equation}
We find that \eqref{eq:lambda_condition} is a transcendental equation with at most three solutions for $E_C / E_J > 0$: one for $\lambda < 0$, one for $0 < \lambda < 8$, and one for $\lambda > 8$. Note that $\lambda = 8$ is the local maximum for the right-hand side in \eqref{eq:lambda_condition}.

We distinguish between these three solutions by first parameterizing the transmon Hamiltonian in terms of the bare harmonic oscillator frequency $\omega_0$ and the bare anharmonicity $\alpha_0$. We define $\omega_0$ as the parameter in front of $\Hat{a}^\dagger \Hat{a}$ in the Hamiltonian, which combined with \eqref{eq:lambda_condition} gives
\begin{equation}
    \omega_0 = E_J \lambda \e^{-\lambda/4} .
    \label{eq:omega_relation}
\end{equation}
Similarly, we define $\alpha_0 / 2$ as the parameter in front of $\Hat{a}^\dagger \Hat{a}^\dagger \Hat{a} \Hat{a}$ such that
\begin{equation}
    \alpha_0 = - E_C.
    \label{eq:alpha_relation}
\end{equation}
Consequently, from Eqs.~(\ref{eq:lambda_condition})--(\ref{eq:alpha_relation}), we obtain a simple algebraic expression for $\lambda$ in terms of $\omega_0$ and $\alpha_0$:
\begin{equation}
    \lambda = -\frac{8\alpha_0}{\omega_0} .
    \label{eq:lambda_relation}
\end{equation}
Then, to set $\lambda$ to one of the three solutions in the anharmonic-oscillator approximation, we recall that the transmon Hamiltonian should be similar to a weak anharmonic oscillator with a negative anharmonicity~\cite{kochChargeinsensitiveQubitDesign2007}. Using \eqref{eq:lambda_relation}, we therefore discard the two solutions bounded by $\lambda < 0$ and $ \lambda > 8$ and choose the small-$\lambda$ solution in the region $ 0 <\lambda / 8 < 1$.

To conclude, we finish the parameterization of the transmon Hamiltonian in $\omega_0$ and $\alpha_0$. By inserting Eqs.~(\ref{eq:omega_relation})--(\ref{eq:lambda_relation}) in Eqs.~(\ref{eq:n_square_normal}) and (\ref{eq:cos_normal_end}), we obtain the normal ordering of the transmon Hamiltonian from the main text [cf. \eqref{eq:normal_Hamiltonian}]:
\begin{equation}
    H = \omega_0 \Hat{a}^\dagger \Hat{a} + 2 \alpha_0 \hspace{-2.46mm} \sum_{m,n \in M} \mleft( \frac{4\alpha_0}{\omega_0} \mright)^{\frac{m+n-4}{2}} \frac{ (\Hat{a}^\dagger)^m \Hat{a}^n}{m! \, n!},
\end{equation}
where $M = \{ m, n \in \mathbb{Z}^+ \mid m+n \geq 4, \, \text{and} \ m+n \ \text{is even} \}$.

\section{Schrieffer--Wolff transformation of the normal-ordered transmon Hamiltonian}
\label{app:SWT_transmon}

Here, we compute the generator $S^{(i)}$ for the Schrieffer-Wolff transformation of the normal-ordered transmon Hamiltonian in \eqref{eq:normal_Hamiltonian}. The computation is performed to first order in $\alpha_0^{(i)} / \omega_0^{(i)}$. Similarly to the single-transmon case in \appref{app:normal_ordering} above, the index $i$ has no utility in this appendix and we therefore drop it here, i.e., $S^{(i)} \to S$, to simplify the notation.

In the main text, we noted that it is sufficient to split the normal-ordered Hamiltonian as given in Eqs.~(\ref{eq:harmonic_Hamiltonian}) and (\ref{eq:harmonic_interaction}) to compute the $S$ to first order in $\alpha_0 / \omega_0$. To explain this, we assume the anharmonic-oscillator approximation and note that in the harmonic-oscillator basis $\{\ket{m}\}_{m=0}^\infty$, the diagonal part of the normal-ordered Hamiltonian is
\begin{equation}
    H_\mathrm{bare} = \omega_0 \Hat{a}^\dagger \Hat{a} + 2 \alpha_0 \sum_{m = 2}^{\infty} \mleft( \frac{4\alpha_0}{\omega_0} \mright)^{m - 2} \frac{ (\Hat{a}^\dagger)^m \Hat{a}^m}{(m!)^2}.
    \label{eq:harmonic_Hamiltonian_app}
\end{equation}
Here, in contrast to \appref{app:normal_ordering}, $\Hat{a}$ and $\Hat{a}^\dagger$ are proper operators in $L^2(-\infty, \infty)$ under the assumption of the anharmonic-oscillator approximation.

Likewise, the off-diagonal part is
\begin{equation}
    H_\mathrm{int} = 2 \alpha_0 \hspace{-2.46mm} \sum_{m,n \in M'} \mleft( \frac{4\alpha_0}{\omega_0} \mright)^{\frac{m+n-4}{2}} \frac{ (\Hat{a}^\dagger)^m \Hat{a}^n}{m! \, n!},
    \label{eq:harmonic_interaction_app}
\end{equation}
where $M' = \{ m, n \in \mathbb{Z}^+ \mid m \neq n, \, m+n \geq 4 \ \text{and} \ m+n \ \text{is even} \}$. Note that the leading-order terms in $H_\mathrm{bare}$ and $H_\mathrm{int}$ correspond to Eqs.~(\ref{eq:harmonic_Hamiltonian}) and (\ref{eq:harmonic_interaction}). Using the full $H_\mathrm{bare}$ and $H_\mathrm{int}$, we recall \eqref{eq:S1} to write the matrix elements of the generator for $m \neq m'$:
\begin{equation}
    \mel{m}{S}{m'} = \frac{\mel{m}{H_\mathrm{int}}{m'}}{\Delta_{mm'}},
    \label{eq:S_app}
\end{equation}
where $\Delta_{mm'} = \mel{m}{H_\mathrm{bare}}{m} - \mel{m'}{H_\mathrm{bare}}{m'}$. Here, the numerator scales as $\mel{m}{H_\mathrm{int}}{m'} \propto \alpha_0[1 + \mathcal{O}(\alpha_0 / \omega_0)]$, while the denominator has the dependence $\Delta_{mm'} \propto \omega_0[1 + \mathcal{O}(\alpha_0 / \omega_0)] $. As such, only the leading-order terms in Eqs.~(\ref{eq:harmonic_Hamiltonian_app}) and (\ref{eq:harmonic_interaction_app}) have a first-order contribution to $S$.

Having established that the leading-order terms in Eqs.~(\ref{eq:harmonic_Hamiltonian_app}) and (\ref{eq:harmonic_interaction_app}) are sufficient to compute the generator, we compute $S$ by using \eqref{eq:S_app}. Rewriting \eqref{eq:S_app} on operator form gives
\begin{equation}
    [S, H_\mathrm{bare}] = - H_\mathrm{int}.
    \label{eq:S_app_operator}
\end{equation}
The ansatz given in \eqref{eq:SWT_transmon_generator}, which we reproduce here for convenience,
\begin{equation}
    S = 2 \alpha_0 \mleft( \frac{1}{4!} \frac{1}{4 \omega_0} (\Hat{a}^\dagger)^4 + \frac{1}{3!} \frac{1}{2 \omega_0} (\Hat{a}^\dagger)^3 \Hat{a} - \text{H.c.} \mright),
\end{equation}
satisfies \eqref{eq:S_app_operator} to first order in $\alpha_0 / \omega_0$ for $S$. This can be shown from noting that
\begin{align}
    [(\Hat{a}^\dagger)^n, \Hat{a}^\dagger \Hat{a}] &= -n (\Hat{a}^\dagger)^n , \\
    [\Hat{a}^n, \Hat{a}^\dagger \Hat{a}] &= n \Hat{a}^n .
\end{align}

\section{Evaluation of the excitation-conserving diagram expansions}
\label{app:diagram_evaluations}
In \secref{sec:mechanism_description}, we obtained the excitation-conserving diagram expansions for the eigenenergies $E_{\Delta, i}$ in Eqs.~(\ref{eq:E000=})--(\ref{eq:E110=}). We evaluated the diagrams for $E_{\Delta, 000}$ and $E_{\Delta, 100}$, but postponed the evaluations of $E_{\Delta, 010}$ and $E_{\Delta, 110}$. Here, we include the remaining evaluations obtained from the rules in Eqs.~(\ref{eq:x})--(\ref{eq:square_contract}). We group the evaluations below after $E_{\Delta, 010}$ and $E_{\Delta, 110}$, and according to the identified mechanisms. Lastly, we also insert the evaluations in Eqs.~(\ref{eq:ZZ=_w})--(\ref{eq:ZZ=_3}) to give analytical expressions for the excitation-conserving contributions to the ZZ coupling.

\paragraph{Evaluation of $E_{\Delta, 010}$}

Recall that $E_{\Delta, 010}$ is equivalent to $E_{\Delta, 100}$ under relabeling of the qubit indices $1 \leftrightarrow 2$. Hence, by relabeling \eqref{eq:E100=_evaluated}, we directly obtain the diagram expansion after contraction
\begin{equation}
\begin{aligned}
    E_{\Delta, 010} = \includegraphics[valign=c]{E110_x.pdf} 
    &+ \mleft( 
    \includegraphics[valign=c]{E110_-x.pdf} +
    \includegraphics[valign=c]{E110_-Lcx.pdf} +
    \includegraphics[valign=c]{E110_2a-x.pdf}
    \mright)
    \\
    &+ \mleft( \,
    \includegraphics[valign=c]{E110_Lx.pdf} +
    \includegraphics[valign=c]{E110_-cLx.pdf} +
    \includegraphics[valign=c]{E110_2aLx.pdf}
    \mright)
    + \mleft( \, 
    \includegraphics[valign=c]{E110_CIx.pdf} \, 
    \mright ),
\end{aligned}
\end{equation}
where the dashed-diagram notation is defined in \eqref{eq:dash_notation} and the brackets delimit the mechanisms. We note that the bare energy $\includegraphics[valign=c]{E110_x.pdf} = \omega_2$ is corrected by the three mechanisms: level repulsions from the energy levels of the states $\ket{100}$ (first bracket in order of appearance) and $\ket{001}$ (second bracket), and a three-loop mechanism (third bracket). 

We evaluate the diagrams generating the level repulsion from the energy level of the state $\ket{100}$:
\begin{align}
    \includegraphics[valign=c]{E110_-x.pdf} &= - \frac{g_{12}^2}{\Delta_{12}} \mleft[1 - \mleft( \frac{g_{12}}{\Delta_{12}} \mright)^2 \mright],
    \label{eq:E010=_evaluated_first} \\[1ex]
    \includegraphics[valign=c]{E110_-Lcx.pdf} &= \frac{g_{12}^2}{\Delta_{12}} \mleft( \frac{g_{23}}{\Delta_{23}} \mright)^2,
    \\[1ex]
    \includegraphics[valign=c]{E110_2a-x.pdf} &= - \frac{1}{\Delta_{12}} \left( \frac{g_{13} g_{23}}{\Delta_{23}} \right)^2.
\end{align}
Similarly, we have for the level repulsion assigned to $\ket{001}$: and for the three-loop mechanism:
\begin{align}
    \includegraphics[valign=c]{E110_Lx.pdf} &= \frac{g_{23}^2}{\Delta_{23}} \mleft[1 - \mleft( \frac{g_{23}}{\Delta_{23}} \mright)^2 \mright],
    \\[1ex]
    \includegraphics[valign=c]{E110_-cLx.pdf} &= - \frac{g_{23}^2}{\Delta_{23}} \mleft( \frac{g_{12}}{\Delta_{12}} \mright)^2,
    \\[1ex]
    \includegraphics[valign=c]{E110_2aLx.pdf} &= \frac{1}{\Delta_{23}} \mleft( \frac{g_{12} g_{13}}{\Delta_{12}} \mright)^2,
\end{align}
and for the three-loop mechanism:
\begin{equation}
    \includegraphics[valign=c]{E110_CIx.pdf} = - \frac{2 g_{12} g_{13} g_{23}}{\Delta_{12} \Delta_{23}}.
\end{equation}

\paragraph{Evaluation of $E_{\Delta, 110}$} Continuing to $E_{\Delta, 110}$, which was given in \eqref{eq:E110=_evaluated}, we recall that the corrections can be grouped into five level repulsions and additional three-loop mechanisms. The repulsions are individually assigned to the states $\ket{200}$, $\ket{020}$, $\ket{101}$, $\ket{011}$, and $\ket{002}$. These repulsions involve coupling strengths that are affected by the dressing of the capacitive couplings in \eqref{eq:dressed_charge_operator}. The dressed capacitive couplings give unnecessarily long expressions, so we introduce the notation
\begin{equation}
    \Tilde{g}_{ij}(k) \equiv g_{ij} \mleft(1 + \frac{\alpha_k}{2 \omega_k} \mright)
    \label{eq:g_correction}
\end{equation}
to shorten them. Using this shorthand notation, the evaluations for the level repulsion assigned to $\ket{200}$ are
\begin{align}
    \includegraphics[valign=c]{E110_-x.pdf} &= - \frac{ \mleft[ \sqrt{2} \Tilde{g}_{12}(1) \mright]^2 }{\Delta_{12} + \alpha_1} \mleft[1 - \mleft( \frac{\sqrt{2} \Tilde{g}_{12}(1)}{\Delta_{12} + \alpha_1} \mright)^2 \mright],
    \\[1ex]
    \includegraphics[valign=c]{E110_-Lcx.pdf} &= \frac{ \mleft[ \sqrt{2} \Tilde{g}_{12}(1) \mright]^2 }{\Delta_{12} + \alpha_1} \mleft( \frac{g_{23}}{\Delta_{23}} \mright)^2,
    \\[1ex]
    \includegraphics[valign=c]{E110_-xJc.pdf} &= \frac{ \mleft[ \sqrt{2} \Tilde{g}_{12}(1) \mright]^2 }{\Delta_{12} + \alpha_1} \mleft( \frac{g_{13}}{\Delta_{13}} \mright)^2,
    \\[1ex]
    \includegraphics[valign=c]{E110_-x-c.pdf} &= \frac{ \mleft[ \sqrt{2} \Tilde{g}_{12}(1) \mright]^2 }{\Delta_{12} + \alpha_1} \mleft( \frac{ \sqrt{2} \Tilde{g}_{12}(2) }{\Delta_{12} - \alpha_2} \mright)^2,
    \\[1ex]
    \includegraphics[valign=c]{E110_2a-x.pdf} &= - \frac{1}{\Delta_{12} + \alpha_1} \mleft( \frac{\sqrt{2} \Tilde{g}_{13}(1) g_{23}}{\Delta_{23}} \mright)^2.
\end{align}

Second, we have for the repulsion assigned to $\ket{020}$:
\begin{align}
    \includegraphics[valign=c]{E110_x-.pdf} &= \frac{ \mleft[ \sqrt{2} \Tilde{g}_{12}(2) \mright]^2 }{\Delta_{12} - \alpha_2} \mleft[1 - \mleft( \frac{\sqrt{2} \Tilde{g}_{12}(2)}{\Delta_{12} - \alpha_2} \mright)^2 \mright],
    \\[1ex]
    \includegraphics[valign=c]{E110_-cx-.pdf} &= - \frac{ \mleft[ \sqrt{2} \Tilde{g}_{12}(2) \mright]^2 }{\Delta_{12} - \alpha_2} \mleft( \frac{  \sqrt{2} \Tilde{g}_{12}(1) }{\Delta_{12} + \alpha_1} \mright)^2,
    \\[1ex]
    \includegraphics[valign=c]{E110_Lcx-.pdf} &= - \frac{ \mleft[ \sqrt{2} \Tilde{g}_{12}(2) \mright]^2 }{\Delta_{12} - \alpha_2} \mleft( \frac{g_{23}}{\Delta_{23}} \mright)^2,
    \\[1ex]
    \includegraphics[valign=c]{E110_xJc-.pdf} &= - \frac{ \mleft[ \sqrt{2} \Tilde{g}_{12}(2) \mright]^2 }{\Delta_{12} - \alpha_2} \mleft( \frac{g_{13}}{\Delta_{13}} \mright)^2,
    \\[1ex]
    \includegraphics[valign=c]{E110_x-2a.pdf} &= \frac{1}{\Delta_{12} - \alpha_2} \mleft( \frac{\sqrt{2} g_{13} \Tilde{g}_{23}(2) }{\Delta_{13}} \mright)^2.
\end{align}

Third, the evaluations for the level repulsions assigned to $\ket{101}$ are:
\begin{align}
    \includegraphics[valign=c]{E110_Lx.pdf} &= \frac{g_{23}^2}{\Delta_{23}} \mleft[1 - \mleft( \frac{g_{23}}{\Delta_{23}} \mright)^2 \mright],
    \\[1ex]
    \includegraphics[valign=c]{E110_-cLx.pdf} &= -\frac{g_{23}^2}{\Delta_{23}} \mleft( \frac{ \sqrt{2} \Tilde{g}_{12}(1) }{\Delta_{12} + \alpha_1} \mright)^2,
    \\[1ex]
    \includegraphics[valign=c]{E110_LxJc.pdf} &= -\frac{g_{23}^2}{\Delta_{23}} \mleft( \frac{ g_{13} }{ \Delta_{13} } \mright)^2,
    \\[1ex]
    \includegraphics[valign=c]{E110_Lx-c.pdf} &= -\frac{g_{23}^2}{\Delta_{23}} \mleft( \frac{ \sqrt{2} \Tilde{g}_{12}(2) }{\Delta_{12} - \alpha_2} \mright)^2,
    \\[1ex]
    \includegraphics[valign=c]{E110_2aLx.pdf} &= \frac{1}{\Delta_{23}} \mleft( \frac{g_{12} g_{13}}{\Delta_{13}} - \frac{ 2 \Tilde{g}_{12}(1) \Tilde{g}_{13}(1) }{\Delta_{12} + \alpha_1} \mright)^2.
\end{align}

Fourth, we have for the repulsion assigned to $\ket{011}$:
\begin{align}
    \includegraphics[valign=c]{E110_xJ.pdf} &= \frac{g_{13}^2}{\Delta_{13}} \mleft[1 - \mleft( \frac{g_{13}}{\Delta_{13}} \mright)^2 \mright],
    \\[1ex]
    \includegraphics[valign=c]{E110_-cxJ.pdf} &= -\frac{g_{13}^2}{\Delta_{13}} \mleft( \frac{ \sqrt{2} \Tilde{g}_{12}(1) }{\Delta_{12} + \alpha_1} \mright)^2,
    \\[1ex]
    \includegraphics[valign=c]{E110_LcxJ.pdf} &= -\frac{g_{13}^2}{\Delta_{13}} \mleft( \frac{ g_{23} }{ \Delta_{23} } \mright)^2, 
    \\[1ex]
    \includegraphics[valign=c]{E110_xJ-c.pdf} &= -\frac{g_{13}^2}{\Delta_{13}} \mleft( \frac{ \sqrt{2} \Tilde{g}_{12}(2) }{\Delta_{12} - \alpha_2} \mright)^2,
    \\[1ex]
    \includegraphics[valign=c]{E110_xJ2a.pdf} &= \frac{1}{\Delta_{13}} \mleft( \frac{2 \Tilde{g}_{12}(2) \Tilde{g}_{23}(2) }{\Delta_{12} - \alpha_2} + \frac{g_{12} g_{23}}{\Delta_{23}} \mright)^2.
\end{align}

Lastly for the repulsions, the repulsion assigned to the state $\ket{002}$ is purely of second order and given by the contracted diagram
\begin{equation}
    \includegraphics[valign=c]{E110_xI2a.pdf} = - \frac{1}{2 \omega'_3 + \alpha_3} \mleft(\frac{\sqrt{2} g_{13} \Tilde{g}_{23}(3) }{\Delta_{13}} + \frac{\sqrt{2} \Tilde{g}_{13}(3) g_{23}}{\Delta_{23}} \mright)^2.
\end{equation}
Here, $\omega'_3 \equiv \omega_3 - (\omega_1 + \omega_2) / 2$ is the shifted coupler frequency relative to the mean qubit frequency.

In addition to the repulsions, $E_{\Delta, 110}$ includes corrections from three three-loop mechanisms. They are evaluated using \eqref{eq:tri} to:
% Equations are split to clean up white spaces.
\begin{align}
    \includegraphics[valign=c]{E110_CIx.pdf} &= - \frac{4 \Tilde{g}_{12}(1) \Tilde{g}_{13}(1) g_{23}}{(\Delta_{12} + \alpha_1) \Delta_{23}},
    \\[1ex]
    \includegraphics[valign=c]{E110_CxI.pdf} &= \frac{2 g_{12} g_{13} g_{23}}{\Delta_{13} \Delta_{23}},
\end{align}
and
\begin{align} 
    \includegraphics[valign=c]{E110_xCI.pdf} = \frac{4 \Tilde{g}_{12}(2) g_{13} \Tilde{g}_{23}(2)}{(\Delta_{12} - \alpha_2) \Delta_{13}} .
    \label{eq:E110=_evaluated_last} 
\end{align}

% \begin{align}
%     \includegraphics[valign=c]{E110_CIx.pdf} &= - \frac{4 \Tilde{g}_{12}(1) \Tilde{g}_{13}(1) g_{23}}{(\Delta_{12} + \alpha_1) \Delta_{23}},
%     \\[1ex]
%     \includegraphics[valign=c]{E110_CxI.pdf} &= \frac{2 g_{12} g_{13} g_{23}}{\Delta_{13} \Delta_{23}},
%     \\[1ex]
%     \includegraphics[valign=c]{E110_xCI.pdf} &= \frac{4 \Tilde{g}_{12}(2) g_{13} \Tilde{g}_{23}(2)}{(\Delta_{12} - \alpha_2) \Delta_{13}},
%     \label{eq:E110=_evaluated_last} .
% \end{align}

\paragraph{Evaluation of ZZ-coupling contributions} With Eqs.~(\ref{eq:E010=_evaluated_first})--(\ref{eq:E110=_evaluated_last}), and the previous evaluation in \eqref{eq:E100=_evaluated} in hand, we obtain the corresponding analytical expressions for Eqs.~(\ref{eq:ZZ=_w})--(\ref{eq:ZZ=_3}) by direct insertion:

\begin{widetext}
\begin{align}
    \zeta_{\includegraphics[valign=c, scale=0.6]{Subscript_w.pdf}} =
    &- \mleft(
    \frac{ \mleft[ \sqrt{2} \Tilde{g}_{12}(1) \mright]^2 }{\Delta_{12} + \alpha_1} 
    \mleft[
    1 
    - \mleft( \frac{\sqrt{2} \Tilde{g}_{12}(1)}{\Delta_{12} + \alpha_1} \mright)^2 
    - \mleft( \frac{g_{23}}{\Delta_{23}} \mright)^2
    - \mleft( \frac{g_{13}}{\Delta_{13}} \mright)^2
    - \mleft( \frac{ \sqrt{2} \Tilde{g}_{12}(2) }{\Delta_{12} - \alpha_2} \mright)^2
    \mright]
    + \frac{1}{\Delta_{12} + \alpha_1} \mleft( \frac{\sqrt{2} \Tilde{g}_{13}(1) g_{23}}{\Delta_{23}} \mright)^2 \mright)
    \nonumber \\ 
    &+ \mleft( \frac{g_{12}^2}{\Delta_{12}} 
    \mleft[
    1 
    - \mleft( \frac{g_{12}}{\Delta_{12}} \mright)^2 
    - \mleft( \frac{g_{23}}{\Delta_{23}} \mright)^2
    \mright]
    + \frac{1}{\Delta_{12}} \left( \frac{g_{13} g_{23}}{\Delta_{23}} \right)^2
    \mright),
    \label{eq:ZZ=_w_evaluated} \\[3ex]
    \zeta_{\includegraphics[valign=c, scale=0.6]{Subscript_e.pdf}} = &+
    \mleft(
    \frac{ \mleft[ \sqrt{2} \Tilde{g}_{12}(2) \mright]^2 }{\Delta_{12} - \alpha_2} 
    \mleft[
    1 
    - \mleft( \frac{\sqrt{2} \Tilde{g}_{12}(2)}{\Delta_{12} - \alpha_2} \mright)^2 
    - \mleft( \frac{  \sqrt{2} \Tilde{g}_{12}(1) }{\Delta_{12} + \alpha_1} \mright)^2
    - \mleft( \frac{g_{23}}{\Delta_{23}} \mright)^2
    - \mleft( \frac{g_{13}}{\Delta_{13}} \mright)^2
    \mright]
    + \frac{1}{\Delta_{12} - \alpha_2} \mleft( \frac{\sqrt{2} g_{13} \Tilde{g}_{23}(2) }{\Delta_{13}} \mright)^2
    \mright) \nonumber \\ 
    &- \mleft(
    \frac{g_{12}^2}{\Delta_{12}} \mleft[1 - \mleft( \frac{g_{12}}{\Delta_{12}} \mright)^2 - \mleft( \frac{g_{13}}{\Delta_{13}} \mright)^2 \mright] + \frac{1}{\Delta_{12}} \mleft( \frac{g_{13} g_{23}}{\Delta_{13}} \mright)^2
    \mright),
    \label{eq:ZZ=_e_evaluated} \\[3ex]
    \zeta_{\includegraphics[valign=c, scale=0.6]{Subscript_nw.pdf}} = &-
    \mleft(
    \frac{g_{23}^2}{\Delta_{23}} 
    \mleft[
    \mleft( \frac{ \sqrt{2} \Tilde{g}_{12}(1) }{\Delta_{12} + \alpha_1} \mright)^2
    + \mleft( \frac{ g_{13} }{ \Delta_{13} } \mright)^2
    + \mleft( \frac{ \sqrt{2} \Tilde{g}_{12}(2) }{\Delta_{12} - \alpha_2} \mright)^2
    \mright]
    - \frac{1}{\Delta_{23}} \mleft( \frac{g_{12} g_{13}}{\Delta_{13}} - \frac{ 2 \Tilde{g}_{12}(1) \Tilde{g}_{13}(1) }{\Delta_{12} + \alpha_1} \mright)^2
    \mright) \nonumber \\
    &+ \mleft(
    \frac{g_{23}^2}{\Delta_{23}} 
    \mleft[
    \mleft( \frac{g_{12}}{\Delta_{12}} \mright)^2
    \mright]
    - \frac{1}{\Delta_{23}} \mleft( \frac{g_{12} g_{13}}{\Delta_{12}} \mright)^2
    \mright),
    \label{eq:ZZ=_nw_evaluated} \\[3ex]
    \zeta_{\includegraphics[valign=c, scale=0.6]{Subscript_ne.pdf}} = &-
    \mleft(
    \frac{g_{13}^2}{\Delta_{13}} 
    \mleft[
    \mleft( \frac{ \sqrt{2} \Tilde{g}_{12}(1) }{\Delta_{12} + \alpha_1} \mright)^2
    + \mleft( \frac{ g_{23} }{ \Delta_{23} } \mright)^2
    + \mleft( \frac{ \sqrt{2} \Tilde{g}_{12}(2) }{\Delta_{12} - \alpha_2} \mright)^2
    \mright]
    - \frac{1}{\Delta_{13}} \mleft( \frac{2 \Tilde{g}_{12}(2) \Tilde{g}_{23}(2) }{\Delta_{12} - \alpha_2} + \frac{g_{12} g_{23}}{\Delta_{23}} \mright)^2
    \mright) \nonumber \\
    &+ \mleft(
    \frac{g_{13}^2}{\Delta_{13}} 
    \mleft[
    \mleft( \frac{g_{12}}{\Delta_{12}} \mright)^2
    \mright]
    - \frac{1}{\Delta_{13}} \mleft( \frac{g_{12} g_{13}}{\Delta_{12}} \mright)^2
    \mright),
    \label{eq:ZZ=_ne_evaluated} \\[3ex]
    \zeta_{\includegraphics[valign=c, scale=0.6]{Subscript_n.pdf}} = 
     &- \frac{1}{2 \omega'_3 + \alpha_3} \mleft(\frac{\sqrt{2} g_{13} \Tilde{g}_{23}(3) }{\Delta_{13}} + \frac{\sqrt{2} \Tilde{g}_{13}(3) g_{23}}{\Delta_{23}} \mright)^2
    \label{eq:ZZ=_n_evaluated} \\[3ex]
    \zeta_3 = &- \frac{4 \Tilde{g}_{12}(1) \Tilde{g}_{13}(1) g_{23}}{(\Delta_{12} + \alpha_1) \Delta_{23}}
    + \frac{2 g_{12} g_{13} g_{23}}{\Delta_{13} \Delta_{23}} + \frac{4 \Tilde{g}_{12}(2) g_{13} \Tilde{g}_{23}(2)}{(\Delta_{12} - \alpha_2) \Delta_{13}} + \frac{2 g_{12} g_{13} g_{23}}{\Delta_{12} \Delta_{23}} - \frac{2 g_{12} g_{13} g_{23}}{\Delta_{12} \Delta_{13}} .
    \label{eq:ZZ=_3_evaluated}
\end{align}
\end{widetext}
Here, in Eqs.~(\ref{eq:ZZ=_w_evaluated})--(\ref{eq:ZZ=_ne_evaluated}), the outer round brackets encapsulating each line correspond to the original round brackets in Eqs.~(\ref{eq:ZZ=_w})--(\ref{eq:ZZ=_ne}) that delimit the different level repulsions. Note that the second-order contributions in Eqs.~(\ref{eq:ZZ=_nw}) and (\ref{eq:ZZ=_ne}) have canceled, which is a result of the fact that the level repulsions assigned to $\ket{101}$ and $\ket{011}$ are independent of the anharmonicities to leading order.

\begin{figure*}
    \centering
    \includegraphics[width=\textwidth]{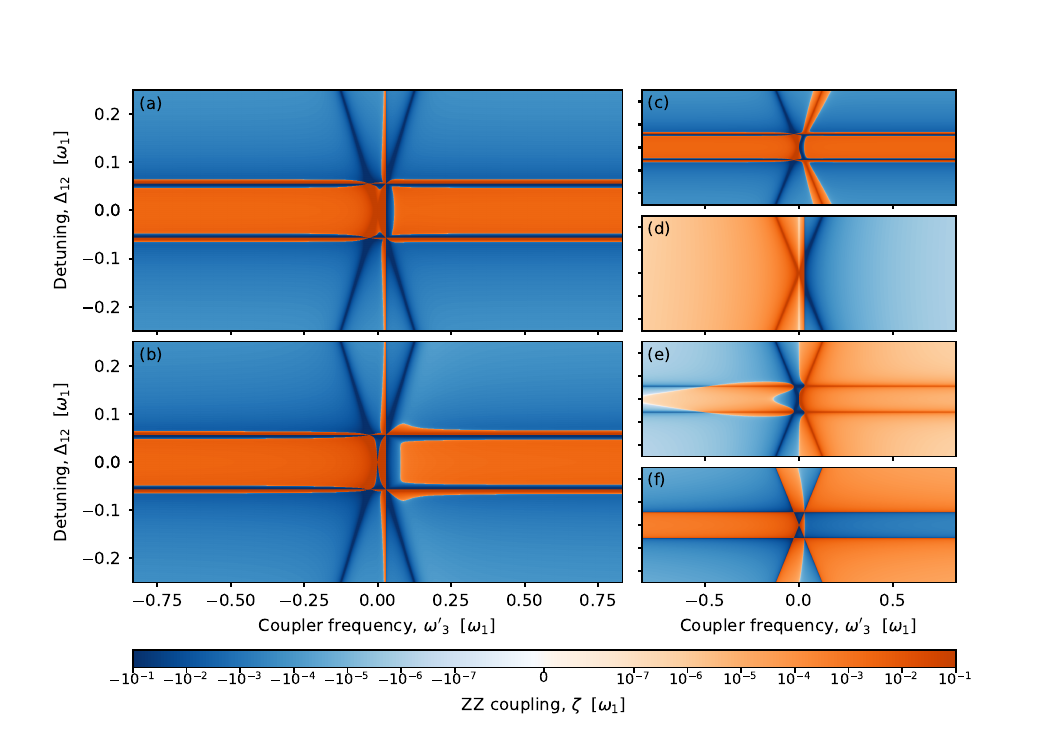}
    \caption{Variation of \figref{fig:ZZ_landscape_EC_mechanisms} in the regime $\abs{g_{12}} \sim \abs{g_{13}}, \abs{g_{23}}$ with an order of magnitude larger qubit coupling strength $g_{12} \to 10 g_{12}$. The plots are therefore generated with the system parameters (in units of $\omega_1$): $g_{13} = g_{23} = 75/4 \times 10^{-3}$, $\alpha_1 = \alpha_2 = \alpha_3 = 3 g_{13}$, and importantly $g_{12} = g_{13} / 3$. As in the figure in the main text, the positive (negative) contributions to the ZZ coupling strength are represented with an orange (blue) gradient, while the white regions represent zero ZZ coupling in the parameter space of the shifted coupler frequency relative to the mean qubit frequency $\omega'_3 = \omega_3 - (\omega_1 + \omega_2)/2$ and the qubit detuning $\Delta_{12} = \omega_1 - \omega_2$. (a) The contribution from level repulsions in (c--e). (b) The total contribution from all mechanisms in (c--f). (c) The correlated level repulsions assigned to: $\ket{010}$, $\ket{100}$, $\ket{020}$, and $\ket{200}$. (d) The second-order level repulsion assigned to $\ket{002}$. (e) The correlated level repulsions assigned to $\ket{001}$, $\ket{011}$, and $\ket{101}$. (h) The sum of the three-loop mechanisms. (c--f) correspond in order of appearance to the square brackets in \eqref{eq:ZZ=}.}
    \label{fig:ZZ_landscape_EC_mechanisms_case_equal}
\end{figure*}

\section{The ZZ coupling in other coupling-strength regimes}
\label{app:ZZ_other_parameters}

\begin{figure*}
    \centering
    \includegraphics[width=\textwidth]{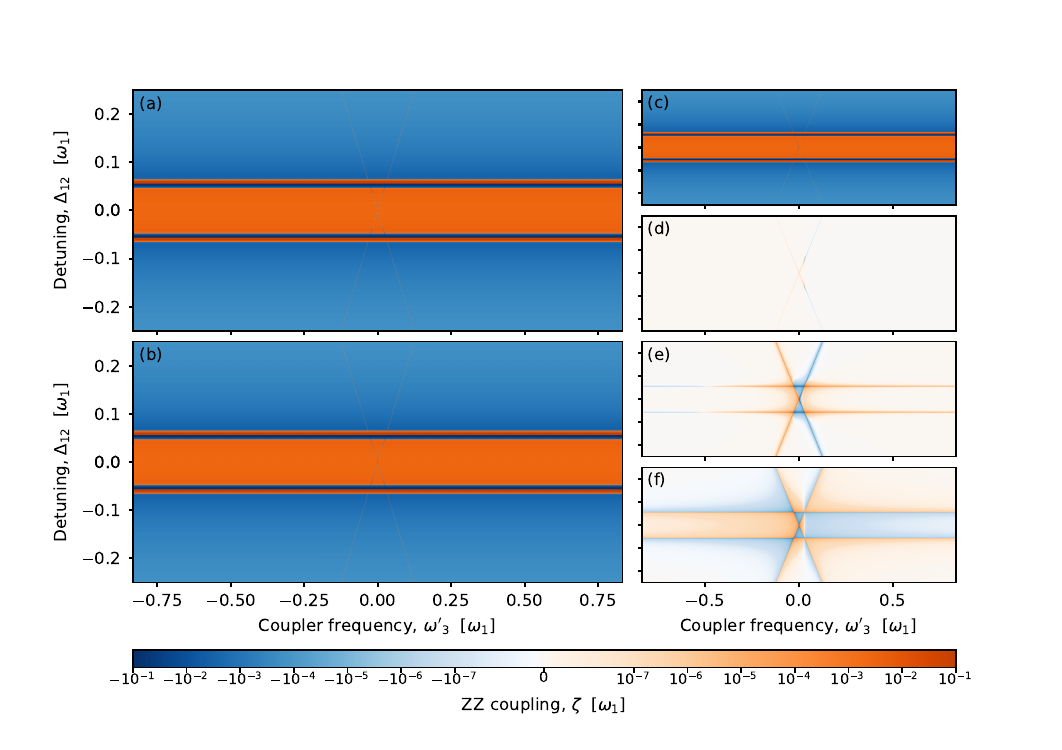}
    \caption{Variation of \figref{fig:ZZ_landscape_EC_mechanisms} in the regime $\abs{g_{12}} \gg \abs{g_{13}}, \abs{g_{23}}$. The plots are generated identically to \figref{fig:ZZ_landscape_EC_mechanisms_case_equal} with the only change that the coupling strengths to the coupler have been reduced to $g_{13} = g_{23} = g_{12} / 30$. See the caption for \figref{fig:ZZ_landscape_EC_mechanisms_case_equal} for further details.}
    \label{fig:ZZ_landscape_EC_mechanisms_case_larger}
\end{figure*}

In \figref{fig:ZZ_landscape_EC_mechanisms}, we presented the static ZZ coupling in the coupling-strength regime $\abs{g_{12}} \ll \abs{g_{13}}, \abs{g_{23}}$. We recreate that figure in this appendix in the two coupling-strength regimes $\abs{g_{12}} \sim \abs{g_{13}}, \abs{g_{23}}$ and $\abs{g_{12}} \gg \abs{g_{13}}, \abs{g_{23}}$. These different regimes are limits of the characteristics of the coupling between the assigned qubits in the three-transmon system in \figref{fig:three_qubit_circuit}. In the limit $\abs{g_{12}} \ll \abs{g_{13}}, \abs{g_{23}}$, the qubit coupling is mainly mediated by the coupler, which we call a coupler-like coupling. The qubit coupling in the limit $\abs{g_{12}} \gg \abs{g_{13}}, \abs{g_{23}}$ is mainly channeled through the direct capacitive coupling, giving a capacitive-like coupling. The regime $\abs{g_{12}} \sim \abs{g_{13}}, \abs{g_{23}}$ is a mixture of the coupler- and capacitive-like regimes where neither is dominant. For applications in quantum processors where the third transmon is used as a coupler, the coupler-like regime considered in the main text is the most common one.

Considering first the regime $\abs{g_{12}} \sim \abs{g_{13}}, \abs{g_{23}}$, we recreate in \figref{fig:ZZ_landscape_EC_mechanisms_case_equal} the analytical predictions for the static ZZ coupling from \figref{fig:ZZ_landscape_EC_mechanisms}. The new figure is identically generated from the diagram expansion in \eqref{eq:ZZ=}, with the only exception that we have tenfold $g_{12}$ compared to the qubit coupling strength in the main text, to $g_{12} = g_{13} / 3 = g_{23} / 3$. We show the contributions to the static ZZ coupling from the correlated level repulsions and three-loop mechanisms in \figpanels{fig:ZZ_landscape_EC_mechanisms_case_equal}{c}{f}.

We note few changes in the features of the correlated level repulsions and three-loop mechanisms. One difference compared to \figpanels{fig:ZZ_landscape_EC_mechanisms}{c}{f} is that the strengths have increased for the $\Delta_{12}$-correlated repulsion in \figpanel{fig:ZZ_landscape_EC_mechanisms_case_equal}{c} and the three-loop mechanisms in \figpanel{fig:ZZ_landscape_EC_mechanisms_case_equal}{f}. The increased strengths are expected since the contributions are proportional to $g_{12}^2$ and $g_{12}$, respectively. Another, more noticeable, difference is that the correlated level repulsions in \figpanel{fig:ZZ_landscape_EC_mechanisms_case_equal}{e} now have a region of positive contribution for $\omega'_3$. 

These changes carry over to the total contribution to the ZZ coupling in \figpanel{fig:ZZ_landscape_EC_mechanisms_case_equal}{b}. We note that the total contribution is now more dominated by the level repulsions, as seen from the close similarity between Figs.~\hyperref[fig:ZZ_landscape_EC_mechanisms]{\ref*{fig:ZZ_landscape_EC_mechanisms_case_equal}(a)} and \hyperref[fig:ZZ_landscape_EC_mechanisms]{\ref*{fig:ZZ_landscape_EC_mechanisms_case_equal}(b)}. Still in this coupling-strength regime, there exist regions of zero ZZ coupling. Even with the increased contribution from the $\Delta_{12}$-correlated repulsion, the two other correlated level repulsions are strong enough to counteract it. On the other hand, the three-loop mechanisms are not sufficient to create an isolated region with a three-loop-type zero ZZ coupling. Instead, the three-loop-type zero ZZ coupling has merged with the level-repulsion-type ones. 

Continuing to the capacitive-like regime $\abs{g_{12}} \gg \abs{g_{13}}, \abs{g_{23}}$, we show an instance of it in \figref{fig:ZZ_landscape_EC_mechanisms_case_larger}. We transition to this regime from the previous one in \figref{fig:ZZ_landscape_EC_mechanisms_case_equal} by keeping the qubit coupling strength fixed and decreasing the coupling strengths to the coupler such that $g_{13} = g_{23} = g_{12} / 30$ (recall that $g_{12} = g_{13} / 30$ in the main text). We observe a significant decrease in the contributions from the correlated level repulsions in \figpanels{fig:ZZ_landscape_EC_mechanisms_case_larger}{d}{e}, which is expected since the repulsions are proportional to either $g_{13}^2$ or $g_{23}^2$. Consequently, the main contribution to the ZZ coupling in \figpanel{fig:ZZ_landscape_EC_mechanisms_case_larger}{b} is from the $\Delta_{12}$-correlated repulsion in \figpanel{fig:ZZ_landscape_EC_mechanisms_case_larger}{c}. We note that the $\Delta_{12}$-correlated repulsion has lost the X-shaped feature present in the other coupling-strength regimes. We understand this change from the fact that the involved second-order level repulsions are suppressed in the capacitive-like regime.

With the $\Delta_{12}$-correlated repulsion being the main contribution in the capacitive-like regime, the other correlated level repulsions and the three-loop mechanisms are not sufficiently strong to counteract this main repulsion. We find as a consequence that both the level-repulsion- and the three-loop-type zero ZZ couplings are not present in the capacitive-like regime.

\section{Fourth-order non-excitation-conserving diagram expansions}
\label{app:4th_order_non_EC_diagrams}

In this appendix, we give the fourth-order diagram expansions that include non-excitation-conserving edges. As will become evident, these expansions include a large number of diagrams. Therefore, we choose to minimize the presentation by only showing diagrams that include contributions to first order in $g_{ij} / \Sigma_{ij}$, i.e., terms of the form $(g_{ij} / \Sigma_{ij}) \times (g_{kl} / \Delta_{kl})^n$, where $n$ is a positive integer. Despite ignoring the higher-order non-excitation-conserving contributions, it will be clear that the fourth-order expansions include exceedingly more diagrams than the corresponding third-order ones in Eqs.~(\ref{eq:E000+-})--(\ref{eq:E110+-}).

In the expansions below, we follow the notation from the Hamiltonian graph in \figref{fig:Hamiltonian_graph} and denote the (non-) excitation-conserving edges with (orange) black edges. The fourth-order expansion for the ground-state energy is
\begin{widetext}
\begin{equation}
\begin{aligned}
    E_{\Sigma, 000}^{(4)} = 
    &\mleft( 
    \includegraphics[valign=c]{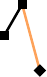} +
    \includegraphics[valign=c]{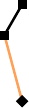} +
    \includegraphics[valign=c]{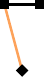} +
    \includegraphics[valign=c]{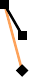} +
    \includegraphics[valign=c]{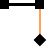} +
    \includegraphics[valign=c]{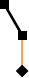} +
    \includegraphics[valign=c]{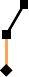} +
    \includegraphics[valign=c]{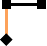} +
    \includegraphics[valign=c]{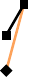} +
    \includegraphics[valign=c]{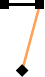} +
    \includegraphics[valign=c]{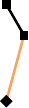} +
    \includegraphics[valign=c]{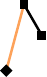}
    \mright) \\
    +
    &\mleft(
    \includegraphics[valign=c]{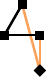} +
    \includegraphics[valign=c]{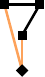} +
    \includegraphics[valign=c]{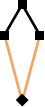} +
    \includegraphics[valign=c]{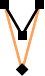} +
    \includegraphics[valign=c]{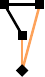} +
    \includegraphics[valign=c]{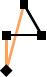}
    \mright)
    +
    \mleft(
    \includegraphics[valign=c]{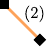} +
    \includegraphics[valign=c]{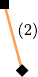} +
    \includegraphics[valign=c]{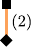} +
    \includegraphics[valign=c]{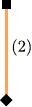} +
    \includegraphics[valign=c]{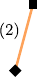} +
    \includegraphics[valign=c]{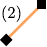}
    \mright) ,
\end{aligned}
\label{eq:E000+-4th}
\end{equation}
\end{widetext}
where we have ordered the individual diagrams in brackets following their kinds defined in Eqs.~(\ref{eq:x})--(\ref{eq:x-2}). For instance, the last two brackets include the four-loop mechanisms and the second-order level repulsions, respectively. In total, we find that the expansion includes 24 diagrams.

The number of diagrams doubles in the expansions for the first-excitation energies. We recall that the expansions for $E_{\Sigma, 010}^{(4)}$ and $E_{\Sigma, 100}^{(4)}$ are equivalent under relabeling of the qubits $1 \leftrightarrow 2$ in Eqs.~(\ref{eq:effective_H0}) and (\ref{eq:effective_V}). We thus only show the expansion for $E_{\Sigma, 100}^{(4)}$: 
\begin{widetext}
\begin{equation}
\begin{aligned}
    E_{\Sigma, 100}^{(4)} = 
    &\mleft( 
    \includegraphics[valign=c]{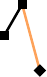} +
    \includegraphics[valign=c]{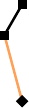} +
    \includegraphics[valign=c]{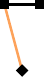} +
    \includegraphics[valign=c]{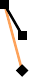} +
    \includegraphics[valign=c]{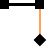} +
    \includegraphics[valign=c]{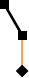} +
    \includegraphics[valign=c]{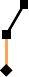} +
    \includegraphics[valign=c]{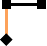} +
    \includegraphics[valign=c]{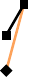} +
    \includegraphics[valign=c]{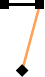} +
    \includegraphics[valign=c]{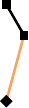} +
    \includegraphics[valign=c]{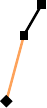} +
    \includegraphics[valign=c]{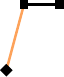} +
    \includegraphics[valign=c]{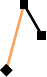}
    \mright. \\
    + &\hphantom{\Biggl( } \mleft.
    \includegraphics[valign=c]{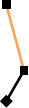} +
    \includegraphics[valign=c]{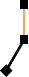} +
    \includegraphics[valign=c]{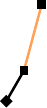} +
    \includegraphics[valign=c]{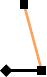} +
    \includegraphics[valign=c]{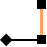} +
    \includegraphics[valign=c]{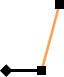}
    \mright)
    +
    \mleft(
    \includegraphics[valign=c]{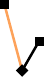} +
    \includegraphics[valign=c]{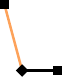} +
    \includegraphics[valign=c]{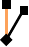} +
    \includegraphics[valign=c]{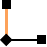} +
    \includegraphics[valign=c]{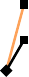} +
    \includegraphics[valign=c]{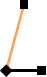}
    \mright) \\
    + 
    &\mleft(
    \includegraphics[valign=c]{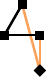} +
    \includegraphics[valign=c]{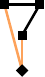} +
    \includegraphics[valign=c]{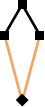} +
    \includegraphics[valign=c]{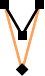} +
    \includegraphics[valign=c]{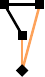} +
    \includegraphics[valign=c]{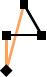} +
    \includegraphics[valign=c]{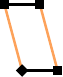} +
    \includegraphics[valign=c]{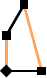} +
    \includegraphics[valign=c]{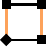} +
    \includegraphics[valign=c]{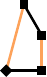} +
    \includegraphics[valign=c]{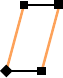} +
    \includegraphics[valign=c]{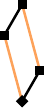}
    \mright. \\
    + &\hphantom{\Biggl( } \mleft.
    \includegraphics[valign=c]{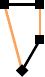} +
    \includegraphics[valign=c]{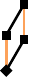} +
    \includegraphics[valign=c]{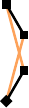} +
    \includegraphics[valign=c]{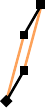} +
    \includegraphics[valign=c]{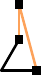}
    \mright)
    +
    \mleft(
    \includegraphics[valign=c]{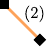}
    \includegraphics[valign=c]{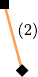} +
    \includegraphics[valign=c]{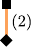} +
    \includegraphics[valign=c]{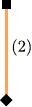} +
    \includegraphics[valign=c]{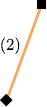} +
    \includegraphics[valign=c]{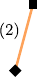} +
    \includegraphics[valign=c]{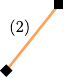} +
    \includegraphics[valign=c]{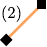}
    \mright) ,
\end{aligned}
\label{eq:E100+-4th}
\end{equation}
\end{widetext}
where we once again have collected the individual diagrams in brackets after their kinds. In comparison with \eqref{eq:E000+-4th}, \eqref{eq:E100+-4th} includes additional corrections to the first-order level repulsions, which form the second bracket (in order of appearance). We note that the last bracket with the second-order repulsions includes no additional contributions to the second-order repulsions assigned to the states in the first-excitation subgraph in \eqref{eq:E100=} . The additional contributions appear first at lowest order $(g_{ij} / \Sigma_{ij})^2$, which we neglect here. Adding up, the expansion in \eqref{eq:E100+-4th} has 51 diagrams, thus amounting to 102 diagrams for the first-excitation energies.

Lastly, we consider the fourth-order non-excitation-conserving expansion for $E_{\Sigma, 110}$. Again, we find almost a doubling of the number of diagrams. The expansion consists of 89 diagrams:
\begin{widetext}
\begin{equation}
\begin{aligned}
    E_{\Sigma, 110}^{(4)} = 
    &\mleft( 
    \includegraphics[valign=c]{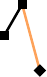} +
    \includegraphics[valign=c]{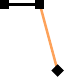} +
    \includegraphics[valign=c]{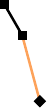} +
    \includegraphics[valign=c]{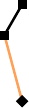} +
    \includegraphics[valign=c]{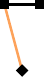} +
    \includegraphics[valign=c]{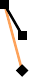} +
    \includegraphics[valign=c]{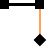} +
    \includegraphics[valign=c]{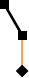} +
    \includegraphics[valign=c]{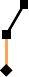} +
    \includegraphics[valign=c]{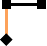} +
    \includegraphics[valign=c]{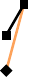} +
    \includegraphics[valign=c]{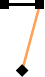} +
    \includegraphics[valign=c]{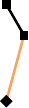} +
    \includegraphics[valign=c]{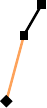}
    \mright. \\
    + &\hphantom{\Biggl( } \mleft.
    \includegraphics[valign=c]{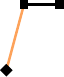} +
    \includegraphics[valign=c]{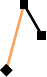} +
    \includegraphics[valign=c]{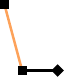} +
    \includegraphics[valign=c]{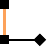} +
    \includegraphics[valign=c]{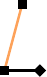} +
    \includegraphics[valign=c]{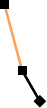} +
    \includegraphics[valign=c]{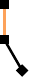} +
    \includegraphics[valign=c]{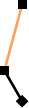} +
    \includegraphics[valign=c]{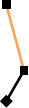} +
    \includegraphics[valign=c]{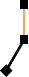} +
    \includegraphics[valign=c]{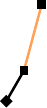} +
    \includegraphics[valign=c]{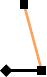} +
    \includegraphics[valign=c]{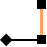}
    \mright. \\
    + &\hphantom{\Biggl( } \mleft.
    \includegraphics[valign=c]{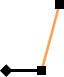}
    \mright)
    +
    \mleft(
    \includegraphics[valign=c]{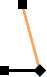} +
    \includegraphics[valign=c]{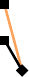} +
    \includegraphics[valign=c]{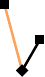} +
    \includegraphics[valign=c]{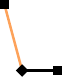} +
    \includegraphics[valign=c]{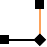} +
    \includegraphics[valign=c]{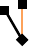} +
    \includegraphics[valign=c]{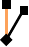} +
    \includegraphics[valign=c]{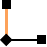} +
    \includegraphics[valign=c]{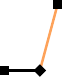} +
    \includegraphics[valign=c]{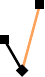} +
    \includegraphics[valign=c]{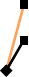}
    \mright. \\
    + &\hphantom{\Biggl( } \mleft.
    \includegraphics[valign=c]{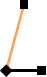} +
    \includegraphics[valign=c]{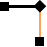} +
    \includegraphics[valign=c]{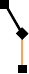} +
    \includegraphics[valign=c]{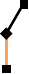} +
    \includegraphics[valign=c]{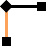}
    \mright)
    +
    \mleft(
    \includegraphics[valign=c]{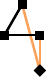} +
    \includegraphics[valign=c]{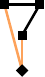} +
    \includegraphics[valign=c]{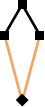} +
    \includegraphics[valign=c]{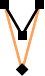} +
    \includegraphics[valign=c]{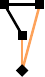} +
    \includegraphics[valign=c]{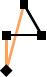} +
    \includegraphics[valign=c]{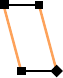}
    \mright. \\
    + &\hphantom{\Biggl( } \mleft.
    \includegraphics[valign=c]{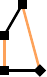} +
    \includegraphics[valign=c]{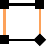} +
    \includegraphics[valign=c]{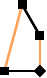} +
    \includegraphics[valign=c]{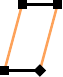} +
    \includegraphics[valign=c]{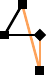} +
    \includegraphics[valign=c]{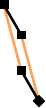} +
    \includegraphics[valign=c]{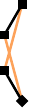} +
    \includegraphics[valign=c]{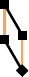} +
    \includegraphics[valign=c]{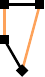} +
    \includegraphics[valign=c]{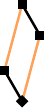} +
    \includegraphics[valign=c]{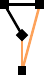} +
    \includegraphics[valign=c]{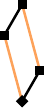} +
    \includegraphics[valign=c]{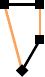}
    \mright. \\
    + &\hphantom{\Biggl( } \mleft.
    \includegraphics[valign=c]{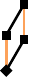} +
    \includegraphics[valign=c]{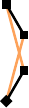} +
    \includegraphics[valign=c]{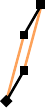} +
    \includegraphics[valign=c]{C1104th_66.pdf} +
    \includegraphics[valign=c]{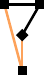} +
    \includegraphics[valign=c]{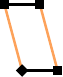} +
    \includegraphics[valign=c]{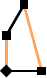} +
    \includegraphics[valign=c]{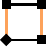} +
    \includegraphics[valign=c]{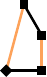} +
    \includegraphics[valign=c]{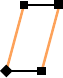} +
    \includegraphics[valign=c]{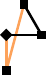} +
    \includegraphics[valign=c]{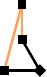} +
    \includegraphics[valign=c]{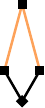}
    \mright. \\
    + &\hphantom{\Biggl( } \mleft.
    \includegraphics[valign=c]{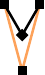} +
    \includegraphics[valign=c]{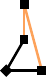} +
    \mright)
    +
    \mleft(
    \includegraphics[valign=c]{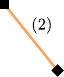} \hspace{-1mm} +
    \includegraphics[valign=c]{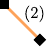} + \hspace{-1mm}
    \includegraphics[valign=c]{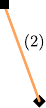} + \hspace{-1mm}
    \includegraphics[valign=c]{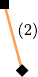} +
    \includegraphics[valign=c]{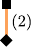} +
    \includegraphics[valign=c]{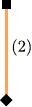} +
    \includegraphics[valign=c]{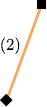} \hspace{-2mm} +
    \includegraphics[valign=c]{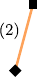} \hspace{-1mm} + \hspace{-1mm}
    \includegraphics[valign=c]{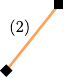}  \hspace{-2mm} +
    \includegraphics[valign=c]{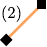} +
    \includegraphics[valign=c]{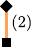}
    \mright) .
\end{aligned}
\label{eq:E110+-4th}
\end{equation}
\end{widetext}
All in all, we find that the fourth-order non-excitation-conserving expansions add up to at least 215 diagrams. Note that Eqs.~(\ref{eq:E000+-4th})--(\ref{eq:E110+-4th}) only show the diagrams that contribute to first order in $g_{ij} / \Sigma_{ij}$.

\section{The fifth-order energy corrections}
\label{app:5th_order_SWT}
In this appendix, we derive the fifth-order corrections to the bare energies from the Schrieffer-Wolff transformation. Starting from the eigenproblem in \eqref{eq:bare_eigenproblem}, we impose that $S_1$ ($S_2)$ diagonalizes $H$ to first (second) order in $V$ to its transformed form $H'$. The generators are then given by Eqs.~(\ref{eq:S1}) and (\ref{eq:S2}). The fact that $H'$ is diagonal to second order in $V$ implies that the lowest-order off-diagonal elements in $H'$ are at least of third order. It then holds that the generator $S_3$ is of third order, further implying that it does not contribute to the fifth-order energy corrections. Hence, it is sufficient to consider the fifth-order expansion of the left-hand side in \eqref{eq:bare_eigenproblem}:
\begin{equation}
\begin{aligned}
    \mathrm{LHS}^{(5)} &= \frac{1}{2!} [S_2,[S_2,[S_1,H_0]]] \\
    &+ \frac{1}{3!} [S_2,[S_1,[S_1,[S_1,H_0]]]] \\
    &+ \frac{1}{5!} [S_1,[S_1,[S_1,[S_1,[S_1,H_0]]]]] \\
    &+ \frac{1}{2!} [S_2,[S_2,V]] + \frac{1}{2!} [S_2,[S_1,[S_1,V]]] \\
    &+ \frac{1}{4!} [S_1,[S_1,[S_1,[S_1,V]]]] .
\end{aligned}
\end{equation}
Here, the expansion is a result of the Baker-Campbell-Hausdorff lemma \cite{sakuraiModernQuantumMechanics2020}
\begin{equation}
    \e^A B \e^{-A} \equiv \sum_{n=0}^\infty \frac{1}{n!} [(A)^n, B],
    \label{eq:BCH_identity}
\end{equation}
where $A$ and $B$ are any two operators, and the nested commutator is defined by
\begin{align}
[(A)^0,B] &\equiv B, \\
[(A)^n,B] &\equiv \underbrace{[A,\dotsb[A,[A}_{n \text { times }}, B]] \dotsb], \text{ for } n > 0 .
\end{align}
Using Eqs.~(\ref{eq:S1}) and (\ref{eq:S2}), we obtain the fifth-order energy corrections
\begin{equation}
\begin{aligned}
    E_{5, i} = &\mel{i}{\frac{1}{30} [S_1,[S_1,[S_1,[S_1,V]]]] }{i} \\
    + &\mel{i}{\frac{1}{3} [S_2,[S_1,[S_1,V]]]}{i} .
\end{aligned}
\label{eq:E5}
\end{equation}

\section{Details and convergence of numerical computations}
\label{app:simulation_details}

In Sections~\ref{sec:numerical_procedure} and \ref{sec:predictions_exact_diagonalization}, we outlined the procedure used to numerically compute the ZZ coupling in the three-transmon system. We recall that the procedure uses the circuit Hamiltonian in Eqs.~(\ref{eq:circuit_Hamiltonian})--(\ref{eq:interaction_Hamiltonian}), which is parameterized with the charging (Josephson) energies $E^{(i)}_C$ ($E^{(i)}_J$) of each transmon $i$, and the mutual charging energies $E^{(ij)}_C$. The circuit Hamiltonian is then expressed on matrix form using the charge states. Here, we provide complementary details for the numerical computations, such as the numerical values used in \secref{sec:predictions_exact_diagonalization} for the charging and Josephson energies. We also present the numerical convergence of the ZZ coupling as function of the cutoffs $N$ (number of states in each transmon subspace before projection) and $M$ (the maximum number of total excitations).

We want that the ZZ coupling to be numerically evaluated for the same bare system parameters as those used in the analytical evaluations in \secref{sec:analytical_predictions}. As discussed in \secref{sec:predictions_exact_diagonalization}, it is somewhat of a technical challenge to numerically obtain the already used analytical bare parameters due to the different parameterizations, especially as a consequence of the varying $\Delta_{12}$ and $\omega'_3$. (Recall that $\Delta_{12} = \omega_1 - \omega_2$ is the qubit detuning and $\omega'_3 = \omega_3 - (\omega_1 + \omega_2) / 2$ is the shifted coupler frequency.) We align the numerical bare parameters to sufficient precision with the bare quantities specified in \figref{fig:ZZ_landscape_numerical} by setting $E_J^{(1)} \approx 2.797$ and $E_C^{(i)} \approx 0.04947$ in units of $\omega_1$. We vary $E_J^{(2)}$ and $E_J^{(3)}$ to achieve the spans of $\omega'_3$ and $\Delta_{12}$ in \figref{fig:ZZ_landscape_numerical}. $E_C^{(ij)}$ is dependently adjusted to maintain constant coupling strengths. For example, we have $E_C^{(13)} = E_C^{(23)} \approx 0.003705$ and $E_C^{(12)} \approx 0.0001235$ for $E_J^{(1)} = E_J^{(2)} = E_J^{(3)} \approx 2.797$.

The numerical bare parameters are computed by separately solving for the eigenenergies and -states for the decoupled transmon Hamiltonians in \eqref{eq:transmon_Hamiltonian}. We use the C++ library Eigen~\cite{eigenweb} as our eigenvalue-problem solver. The same solver is used for computing the eigenenergies and -states for the coupled three-transmon system. We use $N = 27$ charge states for each transmon when solving the eigenvalue problem for the uncoupled system, and then bare eigenstates with a maximum $M = 14$ total excitations in the case of the coupled system.

% More significant digits
% 2.797205135277043
% 0.04946934825638417
% 0.003704893442571813
% 0.00012349644808572712

\begin{figure}
    \centering
    \includegraphics[width=\columnwidth]{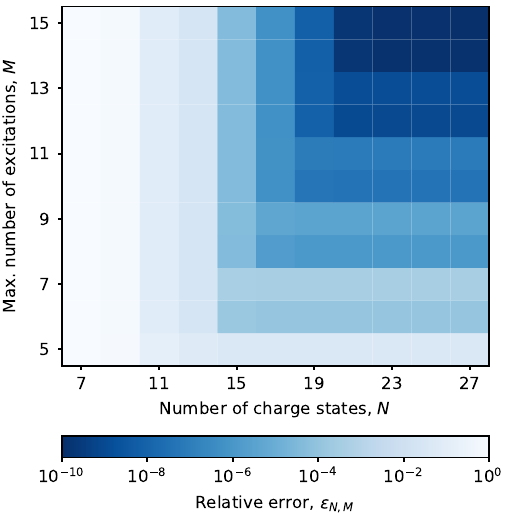}
    \caption{Relative error of the numerically computed ZZ coupling as a function of the number of charge states in each transmon subspace and the maximum total excitation. The relative error is represented with a blue-to-white gradient where blue (white) represents a small (large) error. The relative error has been evaluated uniformly over the bare frequency space in \figref{fig:ZZ_landscape_numerical} using $51 \times 51$ samples. Note that the color gradient has a lower bound resulting in that all relative errors below $10^{-10}$ are given the darkest blur color. For instance, this lower bound determines the coloring of $N_\text{max} = 27$ and $M_\text{max} = 15$ where the relative error is zero by definition.}
\label{fig:convergence}
\end{figure}

We determine $N = 27$ and $M = 14$ from the numerical convergence of the ZZ coupling. As a metric for the numerical convergence, we monitor the relative error
\begin{equation}
    \varepsilon_{N, M} = \frac{\sum_{\omega'_3, \Delta_{12}} \abs{\zeta_{N, M} (\omega'_3, \Delta_{12}) - \zeta_{N_\text{max}, N_\text{max}} (\omega'_3, \Delta_{12}) }}{\sum_{\omega'_3, \Delta_{12}} \abs{\zeta_{N_\text{max}, N_\text{max}} (\omega'_3, \Delta_{12}) }},
\end{equation}
where $\zeta_{N, M} (\omega'_3, \Delta_{12})$ is the ZZ coupling numerically evaluated at the bare frequencies $(\omega'_3, \Delta_{12})$ using $N$ charge states and bare eigenstates with maximum $M$ total excitations. The sums are over the set of evaluated bare frequencies, and $N_\text{max}$ and $M_\text{max}$ are the maximum $N$ and $M$ used. We use $\zeta_{N_\text{max}, \text{max}}$ in the relative error as the reference that we compare all other instances to. Using $N_\text{max} = 27$ and $M_\text{max} = 15$, we obtain in \figref{fig:convergence} the evolution of the relative error as a function of $N$ and $M$. We note the dark blue four-by-two rectangle in the upper right corner ($N \geq 21$ and $M \geq 14$) with a relative error $\varepsilon_{N, M} \approx 10^{-10}$ showing numerical convergence. Hence, we expect all $N$ and $M$ in the four-by-two rectangle to yield converged results. Compared to $M_\text{max} = 15$, we choose $M = 14$ to reduce the computational requirements. On the other hand, we use $N = N_\text{max} = 27$ since the computational load is less sensitive to $N$ in comparison to $M$.

%%%%%%%%%%%%%%%%%%%%%%%%%%%%%%%%%%%%%%%%%%%%%%%

\bibliography{ZZ_refs}

\end{document}